# Instability of steady flows in helical pipes

*A. Yu. Gelfgat*

*School of Mechanical Engineering, Faculty of Engineering, Tel-Aviv University, Ramat Aviv, Tel-Aviv 69978, Israel. e-mail: gelfgat@tau.ac.il*

**Abstract**

A parametric numerical study of three-dimensional instability of steady flows in a helical pipe of arbitrary curvature and torsion is carried out. The computations are performed by a numerical approach verified against independent experimental and numerical results in a previous study. The results are reported as dependences of the critical Reynolds number, critical wavenumber and the critical frequency on the dimensionless pipe curvature and torsion. A multiplicity of different disturbance modes becoming most unstable at different values of the governing parameters, is observed. Patterns of the most unstable modes are reported and classified. Different routes to instability including viscous and inviscid mechanisms, locally developing boundary and mixing layers, interaction between the Dean vortices and the through flow are described.

# 1. Introduction

Instabilities of pressure gradient driven flows in helical pipes have been studied since the pioneer works of White (1929) and Taylor (1929). The fully developed steady flow in a helical pipe appears to be noticeably more complicated than in a straight one, since the centrifugal and Coriolis forces induce additional spanwise vortical motion known as the Dean vortices. The Dean vortices necessarily appear in the helical pipe flow at any, even very small, pressure drop (Dean, 1927). These vortices effectively mix either heat or mass without any need of additional mixing means, which resulted in their wide usage in various applications (see, e.g., Vadhisth et al, 2008). Contrarily to flows in straight ducts, circular or rectangular, the helical pipe flow does not allow for an analytical solution for the steady base flow state, so that the numerical modeling is called for, even at low and moderate values of the Reynolds number. In this manuscript, we focus on examination of stability of calculated steady flows and computation of critical parameters at which the primary transition from steady to oscillatory flow takes place.

A particular interest in instabilities of the helical pipe flows is connected with the fact that unlike the straight circular pipe having no linear stability limit (Salven et al., 1980 ), the helical pipe flows become linearly unstable at finite Reynolds numbers already at very small curvatures. Moreover, the computed and measured critical Reynolds numbers are very close starting from dimensionless curvature 0.01 (Canton et al., 2016; Gelfgat, 2019), which shows that the transition takes place owing to the linear instability.

The helical pipe flow at low and moderate Reynolds numbers was considered as three-dimensional for a long time, until Germano (1982) showed that a two-dimensional formulation is possible in a specially tailored system of curvilinear orthogonal coordinates. Since then, computations of this flow became affordable (see, e.g., Yamamoto et al 1994; Hüttl & Friedrich, 2000; Gelfgat et al 2003; Nobari & Malvandi, 2013; Totorean et al. 2016; and references therein). Formulation in the Germano (1982) coordinates is applied also in this study, which makes the base state dependent on only two coordinates, and periodicity of perturbations along the third coordinate is assumed. Calculation of steady base flow and the linear stability analysis is carried out using the numerical approach described in Gelfgat (2007), which is quite general for a two-dimensional steady base flow subject to three-dimensional infinitesimal disturbances that are periodic in one spatial direction.

Several researchers have studied instability onset in different helical pipes experimentally (White, 1929; Taylor , 1929; Sreenivasan & Strykowski, 1983; Yamamoto et al, 1998; Webster & Humphrey, 1993; Kühnen et al 2014, 2015).  Most of the numerical studies of stability of helical pipes flows addressed either only two-dimensional disturbances (Yamamoto et al., 1998), or only toroidal geometry (Webster & Humphrey, 1997; Di Piazza & Ciofalo, 2011; Canton et al, 2016), which is a limit case of a helical pipe with the zero torsion. In our recent paper (Gelfgat, 2019) we reviewed all these results and compared between them. It was shown that the results obtained using the present numerical technique agree well with the numerical results of Canton et al (2016) computed for a toroidal pipe, as well as with the experimental results of Kühnen et al (2014, 2015) obtained for the toroidal pipes and the helical pipes with small torsion. All the other experimental and numerical results cited above exhibit a considerable scatter and the reported critical Reynolds numbers are considerably larger than those reported in later experiments of Kühnen et al (2014, 2015). In Gelfgat (2019) we also could partially reproduce the experimental results of Yamamoto et al (1998) obtained for large torsions, much larger than those applied in other experimental studies.

In this paper we base on the convergence and validation studies reported in Gelfgat (2019) and focus on parametric studies of instabilities of the helical pipe flows. These flows are defined by three governing parameters, the Reynolds number $Re$, the dimensionless curvature $\varepsilon$, and the dimensionless torsion $\tau$ of the pipe. As mentioned above, the linear stability of this flow in a toroidal pipe ($\tau = 0$) was studied by Canton et al (2016) and the results were verified against the experiments of Kühnen et al (2014), and later numerically by Gelfgat (2019). Several examples of stability results for a non-zero torsion were presented in Gelfgat (2019), however no systematic study where the torsion and the curvature are varied independently was ever published. In the first part of this study we report the stability results for the dimensionless pipe curvature and torsion in the ranges $0.01 \leq \varepsilon \leq 0.6$, and $0 \leq \tau/\varepsilon \leq 5$. To the best of the author's knowledge, such a parametric study is carried out for the first time.

Study of the linear stability of the helical pipe flow revealed that the primary instability sets in at the critical Reynolds number as a transition to a three-dimensional oscillatory flow. The instability is characterized additionally by the oscillation frequency $\omega_{cr}$, the wavenumber $k_{cr}$ that defines periodicity along the pipe centerline, and the most unstable perturbation represented by the leading eigenvector of the linearized governing equations. It is a common place nowadays that in the course of linear stability study, one observes several most unstable disturbances (perturbation modes) that replace each other with variation of the governing parameters. Several examples of that for convective and rotating flows can be found in Gelfgat & Bar-Yoseph (2004). In the considered ranges of the curvature and torsion, this study revealed 13 different most unstable perturbation modes replacing each other in the plane $(\varepsilon, \tau)$.

In the second part of this study, we make an attempt to classify the computed perturbation modes, to describe their features, and to offer, at least hypothetically, an explanation for possible physical mechanisms that lead to onset of instability. Slightly supercritical three-dimensional flows are visualized in the cross-pipe planes using the divergence free projection approach of Gelfgat (2016). We observe that in different flow configurations, the instability sets in either in both Dean vortices, or only inside one of them, sometimes altering the whole vortex and sometimes only in a boundary layer. In some cases we observe instabilities arising in locally developing viscous boundary layers, while in other cases an inviscid instability of local mixing layer configurations. The number of possible most unstable modes grows with the increase of the pipe curvature. This multiplicity of different patterns necessitate a large amount of graphical representations, which are supplied by animation files. The latter helps to understand the structure of most unstable perturbations, and to visualize slightly supercritical flows.

## 2. Coordinate system

The pipe centerline is a helical curve defined parametrically as

$$\boldsymbol{R}_0(t) = \{x(t), y(t), z(t)\} = \{c \cdot cos(t), c \cdot sin(t), bt\}, \tag{1}$$

where $c$ is the radius of the helix, $2\pi b$ is distance between coils (see Fig. 1), and $t$ is a parameter. The curvature and the torsion of the helical curve are defined as

$$\kappa = \frac{c}{b^2+c^2}, \quad \tau = \frac{b}{b^2+c^2}, \tag{2}$$

respectively. In the following we use also their ratio $\lambda = \tau/\kappa = b/c$.

Apparently, for a flow in an infinite helical pipe, all the positions at the curve centerline, as well as orthogonal to the centerline cross-sections are equivalent, so that one would expect to observe the same flow in every cross-section. In the case of the zero torsion, the helical pipe turns into a torus, so that the same local polar coordinates can be defined in every cross-section (Gelfgat et al, 2003; Canton et al, 2016). However, keeping this formulation for a non-zero torsion results in non-orthogonal coordinates, Wang (1981). The orthogonality can be restored using the coordinates of Germano (1982), who proposed to rotate the position of $\theta = 0$ of the polar angle along the pipe centerline as (assuming the torsion $\tau$ is a constant)

$$\xi = \theta - \int_{s_0}^{s} \tau ds = \theta - \tau(s - s_0) \tag{3}$$

The resulting coordinate system $(r, \xi, s)$ is orthogonal. The Lamé coefficients of these coordinates are $H_r = 1$, $H_\xi = r$, $H_s = 1 + krsin(\xi)$. Note that for the constant torsion

$$\frac{d}{ds} = \frac{\partial}{\partial s} + \frac{\partial}{\partial \xi}\frac{\partial \xi}{\partial s} = \frac{\partial}{\partial s} - \tau\frac{\partial}{\partial \xi} = \frac{\partial}{\partial s} - \kappa\lambda\frac{\partial}{\partial \xi} \tag{4}$$

In the coordinates $(r, \xi, s)$, we can assume that the three fluid velocity components and the pressure are independent on the position at the pipe centerline, i.e., independent on $s$, so that $\partial/\partial s = 0$. Thus, we arrive to a two-dimensional formulation for the velocity and the pressure of the base flow dependent only on $r$ and $\xi$.

### 3. Governing equations

#### 3.1. Base flow

Consider a flow of incompressible fluid in a helical pipe of the inner radius $a$, radius of the coil $c$, and a constant distance between the coils equal to $2\pi b$. The pipe is sketched in Fig. 1a. Figures 1b and 1c help to understand the arrangement of the coordinate system for a helix (Fig. 1b) and for a particular case of torus (Fig. 1c). In the plane occupied by the torus centreline we define a system of polar coordinates $(R, \phi)$, which produces also a system of the Cartesian coordinates $X = R\cos(\phi), Y = R\sin(\phi)$. The latter is not shown in the figure, but its location is obvious. The tangent to the centreline is defined by $\boldsymbol{T} = d\boldsymbol{R}/ds = -\boldsymbol{e}_X \sin(\phi) + \boldsymbol{e}_Y \cos(\phi)$, and the normal $\boldsymbol{N} = R\, d\boldsymbol{T}/ds = -\boldsymbol{e}_X \cos(\phi) - \boldsymbol{e}_Y \sin(\phi)$. These vectors are shown in Fig. 1c. Now, in the normal to the centreline cross-section of the torus, we define the local Cartesian system so that the axes $x$ and $y$ are parallel to the vectors $\boldsymbol{N}$ and $\boldsymbol{T}$, respectively. This brings us to the system of coordinates introduced by Wang (1981) and then used by Germano (1982). It is depicted in Fig. 1b, and is a usual right handed system if it is observed from the opposite side of the sheet.

The flow is created by a pressure gradient, which is constant along the pipe centreline

$$\frac{dP}{ds} = G = const, \tag{5}$$

and is governed by the continuity and momentum equations. The flow is defined by the three dimensionless parameters, which are dimensionless curvature $\varepsilon = a\kappa$, torsion to curvature ratio $\lambda$, and the Reynolds number $Re_d = \overline{U}d/\nu$, where $d = 2a$ is the pipe diameter, $\nu$ is the kinematic viscosity, and $\overline{U}$ is the flow mean velocity. The Reynolds number sometimes is replaced by the Dean number $De_d = Re_d\sqrt{\varepsilon}$.

The above definition of the Reynolds (Dean) number requires mean velocity value, which is convenient for experimental studies. In a numerical study, the mean velocity can be found only after calculation of the flow. Since it is not known a priori, its use in the problem formulation causes a certain inconvenience. Thus, to use this traditional formulation, a non-linear problem making dimensionless $\overline{U}$ equal to unity was solved in Canton et al. (2016). To make an alternative and more convenient non-dimensionalization, we use the pressure gradient based scales introduced in Gelfgat et al. (2003, 2019). Assuming that the pressure gradient $G$ is known, we define the scales of length, time, velocity and pressure as $a, (\rho a/G)^{1/2}, (Ga/\rho)^{1/2}$, and $Ga$. The resulting system of the dimensionless continuity and momentum equations reads

$$\nabla \cdot \boldsymbol{v} = 0, \quad \frac{\partial \boldsymbol{v}}{\partial t} + (\boldsymbol{v} \cdot \nabla)\boldsymbol{v} = -\frac{1}{H_s}\boldsymbol{e}_s - \nabla p + \frac{1}{R_G}\Delta \boldsymbol{v} \tag{6,7}$$

where the dimensionless parameter $R_G = (Ga^3/\rho\nu^2)^{1/2}$ replaces the Reynolds number. The equations (6,7) are solved together with the no-slip condition

$$\boldsymbol{v}(r = a, \xi, s) = 0. \tag{8}$$

After the flow is computed, its dimensionless mean velocity $\bar{V}_G$ can be easily obtained. Then the dimensional mean velocity is $\overline{U} = \bar{V}_G(Ga/\rho)^{1/2}$, and the resulting Reynolds number is calculated as $Re_d = 2a\,\overline{U}/\nu = 2a\,\bar{V}_G(Ga/\rho)^{1/2}/\nu = 2\bar{V}_G R_G$.

The friction factor is defined as ($L$ is the pipe length)

$$f = \frac{\Delta p}{\rho \overline{U}^2/2}\left(\frac{2a}{L}\right) = 4\frac{\Delta p}{L}\frac{a}{\rho \bar{V}_G^2(Ga/\rho)} = \frac{4}{\bar{V}_G^2} \tag{9}$$

For a "two-dimensional" flow independent on the coordinate $s$, the continuity equation in the helical coordinates reads

$$\nabla \cdot \boldsymbol{v} = \frac{1}{rH_s}\left\{\frac{\partial}{\partial r}[rH_s v_r] + \frac{\partial}{\partial \xi}[H_s v_\xi] - \varepsilon\lambda r\frac{\partial v_s}{\partial \xi}\right\} = 0. \tag{10}$$

It can be satisfied by introducing a function $\psi$ as

$$v_r = \frac{1}{rH_s}\frac{\partial[H_s\psi]}{\partial \xi}, \qquad v_\xi - \frac{\varepsilon\lambda r}{H_s}v_s = -\frac{1}{H_s}\frac{\partial[H_s\psi]}{\partial r}, \tag{11}$$

so that

$$v_r\boldsymbol{e}_r + \left[v_\xi - \frac{\varepsilon\lambda r}{H_s}v_s\right]\boldsymbol{e}_\xi = rot[\psi\boldsymbol{e}_s]. \tag{12}$$

The function $\psi$ can be interpreted as the stream function of the field $\left(v_r, v_\xi - \frac{\varepsilon\lambda r}{H_s}v_s, 0\right)$. In the following, we call $\psi$ pseudo – streamfunction. The velocity component $v_s$ is called centerline velocity, and part of the flow described by the pseudo – streamfunction is called cross-section flow. The Dean vortices corresponding to the negative and positive values of $\psi$ are called "negative vortex" and "positive vortex", respectively. In the plane chosen for plots, the motion in negative vortices is counter-clockwise, while the motion in positive vortices is clockwise. Also, in all plots and animations below the blue color corresponds to the smallest values of depicted functions, and the red color to the largest values. Thus, negative and positive vortices can be distinguished by the blue or the red color in their cores.

The momentum equations are written in general orthogonal coordinates as in Kochin et al. (1954)

$$\frac{\partial v_1}{\partial t} + \frac{v_1}{H_1}\frac{\partial v_1}{\partial x_1} + \frac{v_2}{H_2}\frac{\partial v_2}{\partial x_2} + \frac{v_3}{H_3}\frac{\partial v_3}{\partial x_3} + \frac{v_1 v_2}{H_1 H_2}\frac{\partial H_1}{\partial x_2} + \frac{v_1 v_3}{H_1 H_3}\frac{\partial H_1}{\partial x_3} - \frac{v_2^2}{H_1 H_2}\frac{\partial H_2}{\partial x_1} - \frac{v_3^2}{H_1 H_3}\frac{\partial H_3}{\partial x_1} = -\frac{1}{H_1}\frac{\partial p}{\partial x_1} +$$

$$+ \frac{1}{R_G}\frac{1}{H_2 H_3}\left\{ -\frac{\partial}{\partial x_2}\left[\frac{H_3}{H_1 H_2}\frac{\partial(H_2 v_2)}{\partial x_1}\right] + \frac{\partial}{\partial x_2}\left[\frac{H_3}{H_1 H_2}\frac{\partial(H_1 v_1)}{\partial x_2}\right] + \frac{\partial}{\partial x_3}\left[\frac{H_2}{H_1 H_3}\frac{\partial(H_1 v_1)}{\partial x_3}\right] - \frac{\partial}{\partial x_3}\left[\frac{H_2}{H_1 H_3}\frac{\partial(H_3 v_3)}{\partial x_1}\right]\right\}. \quad (13)$$

Here the indices 1, 2 and 3 stay for $r$, $\xi$ and $s$, respectively. Two other equations are obtained by cyclic permutations of the indices. These equations contain mixed second derivatives, which may cause a certain inconvenience and loss of accuracy at the discretization stage. To avoid this, the mixed derivatives are eliminated using the continuity equation (10). The resulting set of momentum equations is detailed in the Appendix.

*3.2. Linearized stability problem*

The linear stability of calculated steady flows is studied assuming three-dimensional infinitesimally small disturbances that are periodic along the pipe centerline direction $s$. The perturbations were represented in the form $\{\widetilde{\boldsymbol{v}}(r,\xi), \tilde{p}(r,\xi)\} exp[\sigma t + iks]$, where $\sigma$ is the complex time increment, $k$ is the wavenumber along the centerline and infinitesimally small perturbation amplitude is denoted by tilde. The linearization procedure is standard, except for derivatives in the $s-$ direction, for which eq. (4) must be replaced, for the dimensionless variables, by

$$\frac{d}{ds} = -\varepsilon\lambda\frac{\partial}{\partial \xi} + ik , \quad (14)$$

when the derivative of a disturbance is evaluated. Thus, e.g., the continuity equation for the velocity disturbances becomes

$$\nabla \cdot \widetilde{\boldsymbol{v}} = \frac{1}{rH_s}\left\{\frac{\partial}{\partial r}[rH_s\tilde{v}_r] + \frac{\partial}{\partial \xi}[H_s\tilde{v}_\xi] - \varepsilon\lambda r\frac{\partial \tilde{v}_s}{\partial \xi} + ik\tilde{v}_s\right\} = 0. \quad (15)$$

The corresponding momentum equations are listed in Appendix A. The linear stability problem reduces to the generalized eigenvalue problem

$$\sigma \mathbf{B}(\widetilde{\boldsymbol{v}}, \tilde{p})^T = \mathbf{J}(\widetilde{\boldsymbol{v}}, \tilde{p})^T , \quad (16)$$

where **J** is the complex Jacobian matrix and **B** is the diagonal matrix such that its diagonal elements corresponding to the time derivatives of $\tilde{v}$ are equal to one, while the elements corresponding to $\tilde{p}$ and the boundary conditions are zeros. Obviously, $det\mathbf{B} = 0$, so that the generalized eigenproblem (16) cannot be transformed into a standard one and is treated in the sift-and-inverse formulation

$$(\mathbf{J} - \sigma_0 \mathbf{B})^{-1}\mathbf{B}(\tilde{v},\tilde{p})^T = \vartheta(\tilde{v},\tilde{p})^T, \quad \vartheta = \frac{1}{\sigma - \sigma_0}. \tag{17}$$

where $\sigma_0$ is a complex shift.

## 4. Numerical technique and test calculations

The continuity and momentum equations were discretized on staggered grids using central finite differences with linear interpolation between the nodes where necessary. The Newton iteration was applied for calculation of steady flows. Application of the Newton method is identical to Gelfgat (2007), and is based on the LU decomposition of the sparse Jacobian matrix with further direct solution for the Newton corrections.

Study of stability of an *s*-independent two-dimensional steady flows for a given set of the governing parameters proceeds in the following way. For a fixed value of the wavenumber $k$ we vary the complex shift $\sigma_0$, calculating each time 10-20 eigenvalues closest to the shift, until the eigenvalue $\sigma$ of the eigenproblem (16) having the largest real part is found. This process is repeated for different values of the wavenumber $k$ until the eigenvalue $\hat{\sigma}$ having the largest real part for all real wavenumbers $k$ is computed. This eigenvalue is called leading. Apparently, $Real[\hat{\sigma}] = \max_{k}\{Real[\sigma(k)]\} > 0$ means instability of the steady flow. Our purpose is to find the critical value of the number $R_{G,cr}$, dependent on $\varepsilon$ and $\lambda$, at which $Real[\hat{\sigma}(k_{cr})] = 0$, where $k_{cr}$ is the critical wavenumber, at which the above equality holds. In all the calculations described below the imaginary part of $\hat{\sigma}(k_{cr})$ was non-zero. It estimates the frequency of appearing oscillations, is called critical frequency, and is denoted as $\hat{\omega}_{cr} = Im[\hat{\sigma}(k_{cr})]$. The corresponding eigenvector of (16) is called leading. It defines the most unstable perturbation of the base state. Its amplitude, to within multiplication by a constant, represents the amplitude of an oscillatory flow resulting from the instability onset.

For a given pair of the geometrical governing parameters $\varepsilon$ and $\lambda$, the result of the stability study is defined by the critical values $R_{G,cr}$, $k_{cr}$, $\hat{\omega}_{cr}$ and the leading eigenvector. The critical Reynolds number then can be calculated as $Re_{cr} = 2\bar{V}_G R_{G,cr}$, and the dimensionless critical frequency scaled by $(2a)/\bar{U}$ is $\omega_{cr} = 4\hat{\omega}_{cr}R_{G,cr}/Re_{cr}$. Since the perturbation at the critical point is proportional to $exp[i\hat{\omega}_{cr}t + ik_{cr}s] = exp[ik_{cr}(s - ct)]$, where $c = -\hat{\omega}_{cr}/k_{cr}$ is the phase speed of the developing traveling wave. Assuming $k_{cr} > 0$, the wave propagates downstream when $\hat{\omega}_{cr} < 0$ and $c > 0$, and upstream if $\hat{\omega}_{cr} > 0$ and $c < 0$.

The eigenproblem (17) is solved by the Arnoldi method. The shift-and-inverse formulation is provided by the ARPACK package of Lechouq et al. (1998). Following Gelfgat (2007), we calculate LU decomposition of the complex matrix $(\mathbf{J} - \sigma_0\mathbf{B})^{-1}$, so that calculation of the next Krylov vector for the Arnoldi method is reduced to one backward and one forward substitution. It should be noted that the Jacobian matrices for the Newton iteration and the stability analysis are different, since the latter contains the terms depending on the wavenumber $k$ that can also be complex. The Jacobian matrices were calculated directly from the numerical schemes. The corresponding parts of the code were verified by numerical differentiation of the equations' right hand sides.

The numerical approach was tested in Gelfgat (2019) in the following way. The steady state and stability problems were solved using momentum equations formulated in the form of Eq. (13) and in the modified form presented in the Appendix. Additionally, comparison with the results of Yamamoto (1994) was presented. The calculated steady flows were very close. The flow features reported contained the flow rates, friction factors, and minimal and maximal values of the pseudo stream function. Their Richardson extrapolations done using the grids of $50 \times 100$ and $100 \times 200$ nodes coincided up to the third decimal place. Convergence of the critical parameters was found to be strongly dependent on the curvature and the torsion. Nevertheless, refining the grids up to 300×600 nodes, we could establish the convergence up to the second decimal place in the worst case.

To gain an additional validation of the results, we consider a limiting case of zero torsion, which brings us to the flow in a toroidal pipe considered by Catton et al (2016). This case was considered in general Germano coordinates, as well as in the cylindrical coordinates. The toroidal boundary was treated by the immersed boundary method. This allowed us to compare our results obtained by the two independent approaches and compare them additionally with the independent calculations of Canton et al (2016). Unfortunately, it can be done only for the zero torsion. The critical parameters of Canton et al (2016) and those calculated in Germano and on cylindrical coordinates coincided within the second decimal place. All the details can be found in Gelfgat (2019).

To validate the calculated steady states against experimental measurements, the friction factors measured in De Amicis et al (2014) were compared with the calculated ones (Gelfgat, 2019).

It was shown that, as expected, the friction factors coincide at small Reynolds numbers, and diverge at the larger ones when the flow becomes turbulent. Here we report a similar comparison (Fig. 2), for which we used the experimental data of Cioncolini & Santini (2006). In Fig. 2 we observe that the measured and calculated friction factors are close up to $Re \leq 8000$, and start to diverge at larger values. To illustrate how the flow pattern changes with the increase of the Reynolds number, several calculated steady flows are included in Fig. 2 as inserts. In these frames, as well as in all the figures below, it is assumed that the central axis of the helix (Fig. 1) is located on the left hand side from every pipe cross-section plotted, so that the inner pipe boundary corresponds to the left point of a plot border. For more information on the steady flow patterns the reader is referred to Gelfgat (2019) and references therein.

Comparing the calculated critical Reynolds numbers with the experimentally measured ones (Gelfgat, 2019), we obtained a good agreement with the recent experiments of Kühnen et al (2014, 2015) done for $\varepsilon < 0.1$ and $\lambda < 0.05$. Several earlier experiments that studied the instability onset for similar pipe curvatures and small torsions (White, 1929; Taylor, 1929; Sreenivasan & Strykowski, 1983; Webster & Humphrey, 1993) found noticeably larger critical Reynolds numbers, and possibly missed the primary instability. A possible overestimation of the critical Reynolds number can be observed by comparison of Fig. 2 and the results reported below. The critical Reynolds number calculated for the parameters of Fig. 2 by the above described linear stability analysis is approximately 3525, while judging by the friction factor dependence (Fig. 2), it can be estimated to be above 8000.

## 5. Results

### 5.1. Stability diagrams

Study of the primary instability of the helical pipe flow was carried out for $0 < \varepsilon \leq 0.6$ and $0 \leq \lambda \leq 5$. The study can be extended to larger curvatures and torsions, however, such values are unusual for most of applications, as well as for most of academic studies. The computations were carried out on the uniform grid of 100×200 nodes. Several characteristic points on the stability diagram reported below were verified with calculations on the uniform 200×400 and finer grids to ensure that the results coincide at least within the two decimal places, i.e., the disagreement between results obtained on the two grids is less than 1%.

The critical points corresponding to transition from steady to oscillatory flow are reported in Fig. 3. Figure 3a shows dependence of the critical Reynolds number on the torsion to curvature ratio for all the values of curvature considered. The flows are stable below and on the left hand side of the curves, and are unstable above and on the right hand side of them. The corresponding critical frequencies and critical wavenumbers are shown in Figs. 3b and 3c.

The results reported in Fig. 3 need several additional comments. Firstly, as explained above, $\omega_{cr} < 0$ means that the instability sets in as a traveling wave propagating downstream. It follows from Figs. 3b and 3c that for $\lambda \leq 1$ all the most critical disturbances are downstream propagating waves. Only at larger torsions $\varepsilon > 0.2$, and not at all the curvatures considered, the instability sets in as a upstream propagating wave.

Secondly, the linearized stability problem is invariant for the replacement $k \to -k$ and $\omega \to -\omega$. For each positive value of $k_{cr}$ we found a single eigenvalue with zero real part. No multiple eigenvalues were observed. Also, the real part of leading eigenvalues always crossed the zero axes when the Reynolds number was varied to find the critical point. It allows us to conclude that at calculated critical points $d\sigma/dRe \neq 0$. Since the wavenumber $k$ can attain positive and negative values, the whole stability problem has a pair of complex conjugated eigenvalues $(0, \pm\omega_{cr})$ that correspond to the wavenumbers $\pm k_{cr}$, and their eigenvectors are complex conjugate functions representing the same travelling wave. Thus, we conclude that the steady – oscillatory transition takes place as a Hopf bifurcation (Hassard et al., 1981).

Thirdly, we observe several most unstable modes that replace each other when the curvature and the torsion are varied. These modes are represented by separate lines in Figs. 3b and 3c. A replacement of an unstable mode by another one causes more or less noticeable breaks of the $Re_{cr}(\lambda)$ curves shown in Fig. 3a. Note that an existence of many most unstable modes is quite usual for parametric stability studies (see, e.g., Gelfgat & Bar-Yoseph, 2004, and references therein).

In Figs.3b and 3c the different most unstable eigenmodes are numbered according to their appearance in the calculations. The same number is attained to modes with similar eigenvector patterns. Table 1 summarizes the curvature values and intervals of $\lambda$ where all the numbered eigenmodes are observed. The third column contains a short description of the observed features of the corresponding disturbance. The two last columns of Table 1 contain links to animations, in which time dependencies of perturbation patterns and slightly supercritical flows are visualized to help understand the descriptions given in Section 5.3.

*5.2. Visualization of slightly supercritical flows*

Recalling that all the disturbances are proportional to $exp(ik_{cr}s + i\omega_{cr}t)$, we note that for a single-frequency time-periodic flow, oscillations in time at fixed $s$ are similar to oscillations along the spatial period $s_0 \leq s \leq s_0 + 2\pi/k_{cr}$ at fixed time. Therefore, in the following we visualize only oscillations in time at an arbitrary fixed location $s$.

Each most unstable eigenmode can be plotted as a time-dependent function, or just as its absolute value showing the distribution of the oscillations amplitude. In this section we discuss how oscillations of the Dean vortices, i.e., oscillations of the secondary cross-pipe flow can be visualized. To do this we use the visualization method described in Gelfgat (2016), and calculate a divergence free projection of velocity on the plane $(r, \xi)$. For the base flow, which is independent on $s$, this projection is given by Eq. (12) and is $\boldsymbol{V}_{2D} = (v_r, v_\xi - \varepsilon \lambda r v_s/H_s, 0)$, $div\boldsymbol{V}_{2D} = 0$. Since there is no analytical expression for the divergence free projection of an arbitrary three-dimensional disturbance, it is calculated by the iterative SIMPLE-like procedure described in Gelfgat (2016). As a result, the disturbance in the fixed plane $s = s_0$ is expressed as $\widetilde{\boldsymbol{v}} = \widetilde{\boldsymbol{v}}_{2D} + \nabla \varphi$, where $\widetilde{\boldsymbol{v}}_{2D} = (\tilde{v}_{r,2D}, \tilde{v}_{r,2D}, 0)$, and $\nabla \cdot \widetilde{\boldsymbol{v}}_{2D} = 0$. The vector $\widetilde{\boldsymbol{v}}_{2D}$ is the divergence free projection of the three-dimensional disturbance vector on a plane $(r, \xi)$. It can be represented as $\widetilde{\boldsymbol{v}}_{2D} = rot\widetilde{\boldsymbol{\psi}}_{2D}$, where the vector potential $\widetilde{\boldsymbol{\psi}}_{2D}$ has a form $\widetilde{\boldsymbol{\psi}}_{2D} = (0,0, \tilde{\psi}_{2D})$, so that $\tilde{\psi}_{2D}$ is an analog of a two-dimensional stream function. Now a slightly supercritical oscillatory flow can be visualized using

$$v_s = V_s(r, \xi) + \epsilon Real[\tilde{v}_s(r, \xi) exp(ik_{cr}s + i\omega_{cr}t)] \tag{18}$$

$$\Psi = \psi(r, \xi) + \epsilon Real[\tilde{\psi}_{2D}(r, \xi) exp(ik_{cr}s + i\omega_{cr}t)] \tag{19}$$

where $V_s$ and $\tilde{v}_s$ are the centerline $s$-components of the base flow and the disturbance, respectively, $\psi$ is the pseudo – streamfunction of the base flow defined in Eq. (11), and $\Psi$ is the pseudo – streamfunction of the divergence free projection of the slightly oscillatory flow state. The amplitude $\epsilon$ is a function of the super-criticality $(Re - Re_{cr})$ and cannot be found within the linear stability approach. Its finding requires either a non-linear analysis of bifurcation, or a fully three-dimensional time-dependent solution, both of which are beyond the scope of this study. Here we are interested only in qualitative visualization of patterns of slightly supercritical flows. For this purpose, we choose the value of $\epsilon$ so that the amplitude of the second terms of Eqs. (18) and (19) do not exceed 10% of the amplitude of the first terms, while the oscillations of both the centerline velocity and the cross-section flow are visible.

An example of this visualization is presented in Figs. 4 and 5 and the corresponding animations. Color plots in Fig. 4 show snapshots of the leading perturbation of the three velocity components distanced by a quarter of the time oscillation period. The isolines of the centerline velocity and the pseudo stream function of the base flow are shown by lines. This figure shows that the velocities $v_s$ and $v_r$ are perturbed anti-symmetrically with respect to the pipe diameter $\xi = 0, \pi$, while the perturbation of $v_\xi$ is symmetric. However, these perturbation patterns are not very helpful in understanding of how the flow changes in a slightly supercritical regime. The answer to this question is presented in Fig. 5, where snapshots of the functions defined in Eqs. (18,19) are shown. In this figure and the corresponding animation we observe slight oscillations of the centerline velocity, and oscillations of the Dean vortices that oscillate in a counter phase. Below, based on the described visualization of disturbances and slightly supercritical flows, we make an attempt to classify the eigenmodes and to gather some more understanding in the processes leading to the onset of different instability modes. For this purpose all the 13 modes observed are illustrated in Figs. 6-31 in the same way as it was done in Figs. 4 and 5.

*5.3. Classification and description of the unstable eigenmodes*

In this section we make an attempt to classify the most unstable modes and to discuss which physical mechanisms can be responsible for onset of the above described instabilities and for the appearance of self-sustained oscillations of supercritical states. Clearly, such descriptions are mostly qualitative and sometimes even speculative, but nevertheless, we are making an attempt to provide some more insight into this question. First we note that results obtained for instabilities of inviscid vortices (Godeferd et al, 2001; Chomaz et al, 2010; Carnevale et al., 2016), as well as for inviscid vortex pairs (Billant, 1999; Roy et al, 2008; Leweke et al, 2016), in an unbounded domain, cannot be applied for the bounded viscous flow considered here. Moreover, in the present helical pipe configuration, the flow through the pipe and the secondary Dean vortices are already interconnected in the base flow, contrarily to inviscid vortices superimposed with axial flow (e.g., Roy et al, 2008) or with background rotation (e.g., Godeferd et al, 2001; Gargan-Shingles, 2016).

To examine which terms of the linearized equations contribute or do not contribute to the instability onset, we eliminate them individually and monitor the leading eigenvalue and pattern of the leading eigenvector. This simple computational experiment, successfully applied for example in Gelfgat (2011) and later studies, allows us to focus only on the terms responsible for the disturbances growth and to avoid discussing the irrelevant ones. In the text below we call it "zeroing terms numerical experiment".

We start from the limiting case of zero torsion, $\tau = \lambda = 0$, which corresponds to the toroidal pipe (Kühnen et al, 2014; Canton et al, 2016). In this case the isolines of the centerline velocity are always symmetric with respect to the diameter line $= 0, \pi$, while the Dean vortices are always antisymmetric. The *s*-component of flow vorticity and the pseudo stream function $\psi$ are also antisymmetric, so that, e.g., $\psi(\xi) = -\psi(2\pi - \xi)$. The instability can break this symmetry or preserve it, which allows us to make the first distinction between the eigenmodes. Thus, the instability at $\varepsilon = 0.01, \lambda = 0$ breaks the symmetry, as is seen from Fig. 5. This instability mode is denoted as mode 1. Note that the perturbations $\tilde{v}_s$ and $\tilde{v}_r$ are antisymmetric and break the symmetry of the corresponding symmetric components $v_s$ and $v_r$ of the base flow (Fig. 4). The base flow component $v_\xi$ is antisymmetric, so that its perturbation is symmetric and breaks its antisymmetry.

The above mentioned interconnection between the primary centerline and the secondary Dean vortex flow helps to understand the oscillatory mechanism of mode 1 (Figs. 4 and 5). An increase of the centerline velocity caused by a positive value of perturbation in the lower part of the upper left frame of Fig. 4 (the first period quarter) increases the centrifugal force, which leads to an intensification of the positive clockwise Dean vortex in the right frame of the first period quarter in Fig. 5. As a result, the zero pseudo streamline is deformed inwards to the negative vortex, so that the positive vortex occupies a slightly larger volume than the negative one. The intensification of the positive vortex causes larger energy losses in the part of the flow region it occupies, which slows down the through flow there. The latter is observed at the third quarter of the oscillation period (Fig. 4), where perturbation of the centerline velocity in the lower part becomes negative, while attaining a large positive value in the upper part. This leads to the intensification of the negative vortex, followed by decrease of the centerline velocity in the corresponding part of the flow region. Finally, the oscillations of both vortices and of the centerline velocity become self-sustained.

The zeroing terms numerical experiment, performed for the parameters of Figs. 4 and 5, shows that to obtain a similar eigenvalue and eigenvector one can leave in the perturbed momentum equations (A4)-(A6) the terms $ikV_s\boldsymbol{u}/H_s$, $\left(2V_\xi u_\xi/r + 2\varepsilon \sin(\xi)V_s u_s/H_s\right)\boldsymbol{e}_r$, $2\varepsilon\cos(\xi)V_s u_s \boldsymbol{e}_\xi/H_s$, and $\frac{\varepsilon}{H_s}V_s\left(u_\xi \cos(\xi) + u_r \sin(\xi)\right)\boldsymbol{e}_s$, and to zero all the others. The first vector term shows that advection of disturbances of all the three velocity components along the pipe centerline is necessary for the instability onset. The next two terms show that perturbation of the centerline velocity $u_s$ affects the radial and circumferential velocities via the centrifugal and Coriolis forcing. An additional centrifugal forcing comes from the circumferential velocity perturbation. The last term is the Coriolis forcing responsible for affecting the centerline velocity perturbation by two other perturbed components.

Unfortunately, considering other instability modes, we cannot point to a similar simple mechanism of self-sustained oscillations.

At $\varepsilon = 0.02, \lambda = 0$ the instability mode 1 is replaced by another one preserving the symmetry that persists until $\varepsilon = 0.1$. This symmetry preserving instability, denoted as mode 3, is illustrated in Figs. 6 and 7. Contrary to the previous case, the perturbations $\tilde{v}_s$ and $\tilde{v}_r$ are symmetric, while the perturbation $\tilde{v}_\xi$ is antisymmetric (Fig. 6). The Dean vortices oscillate in phase (Fig. 7), so that the zero pseudo streamline separating the vortices remains non-deformed. It is easy to verify that when the centerline velocity increases, the intensity of vortices, measured by the maximal and minimal values of the pseudo streamfunction, decreases and vice versa. This rules out the oscillations mechanism described above, where increase of centerline velocity and intensification of the Dean vortices takes place simultaneously. More likely, we observe here an exchange of energy between the through flow and the Dean vortices. Another qualitative difference in both cases can be seen by comparing the snapshots of the oscillatory flow (cf. Figs. 5 and 7, and corresponding animations). In the case of mode 1 (Fig. 5), the zero pseudo streamline oscillates, and an intensification of one vortex is followed by a weakening of the second one, so that the vortices affect each other. In the case of mode 3, the zero pseudo streamline at $\lambda = 0$ (Fig. 7) remains non-deformed, and the vortices grow and diminish simultaneously, so that no interaction between the vortices is observed. Several additional numerical experiments showed that if to zero the terms $u_r \frac{\partial V_r}{\partial r} \boldsymbol{e}_r$, $\frac{u_r}{r}\frac{\partial (rV_\xi)}{\partial r} \boldsymbol{e}_\xi$, and $\left(V_r \frac{\partial u_s}{\partial r} + \frac{V_\xi}{r}\frac{\partial u_s}{\partial \xi}\right)\boldsymbol{e}_s$ the eigenvalue and the eigenvector remain close to those computed via the full equations. The unimportance of the last term shows that advection of the centerline velocity perturbation along the base flow Dean vortices does not affect the instability. The possibility to neglect two previous terms means that advection of the mean flow $V_r$ and $V_\xi$ components along the *r*-direction does not affect the instability, while this is not true regarding the ξ-direction. Advection in the *s*-direction of all the flow components, base and perturbed, remains important. This brings us to the problem of stability of a pair of vortices superimposed with a through flow (Roy et al, 2008; Nagarathinam et al, 2015). However, as it was already mentioned, in the considered problem the vortices and the through flow are interconnected, and the flow is viscous, which makes the results obtained in the above studies inapplicable. Thus, the perturbation patterns reported in Roy et al (2008) are qualitatively different from those depicted in Fig. 6, as well as from all the others perturbation patterns reported below.

Continuing discussion and description of mode 3, we notice that the base flow component $V_\xi$, changing between $\pm 0.12$, is noticeably larger than $V_r$ changing between -0.04 and +0.02, so that advection along the ξ-direction is expected to be dominant. Also, the maximal and minimal values of the perturbation $u_s$ are located in the areas where the isolines of the center line velocity $V_s$ are strongly curved (Fig. 6). One-dimensional profiles of $V_s$ in this region have inflection points, which indicates on a possibility of an inviscid instability mechanism. Obviously, this criterion remains only an indication and cannot be applied for the considered flow.

Another possibility follows from relatively large values of $V_\xi$ and the localization of the perturbation $u_\xi$ near the upper and lower borders (rightmost frames in Fig 6). The isolines of $V_\xi$, not shown in above figures, exhibit a boundary layer near these parts of the border. Some characteristic profiles $V_\xi(r)$ plotted via the point where the base circumferential velocity attains its maximum are shown in Fig. 32 for this and other modes. The plotted examples correspond to some of the cases shown in Figs. 4-31 where a developed boundary layer of $V_\xi$ was observed, so that the circumferential velocity steeply grows from the boundary point inside the pipe. Thus, in the case of mode 3, and other cases when the perturbation $u_\xi$ is localized near or inside the boundary layer of $V_\xi$, we can assume instability of above boundary layer that also may interact with other destabilizing flow features.

Starting from $\varepsilon = 0.2$ and keeping $\lambda = 0$, we observe the symmetry breaking instability again, however it leads to a qualitatively different pattern of the slightly supercritical flow (Figs. 8 and 9, and the corresponding animations). This disturbance mode is denoted as mode 6. The perturbations of all the velocity components are localized near the Dean vortices "centers", where the pseudo stream function attains the minimal and maximal values. Oscillations of the Dean vortices are noticeable only near their "centers", while far from there the vortices remain almost steady. Examining the corresponding animations, listed in Table 1, we observe that the perturbations are advected along the pseudo streamlines from the regions of relatively large cross-flow velocities, where the streamlines are close, to the regions where the pseudo stream function attain its largest and smallest values, so that the cross flow velocities there are small due to the small derivatives $\partial\psi/\partial r$ and $\partial\psi/\partial \xi$. In these regions the disturbances fully dissipate. This mode of instability persists in the toroidal pipe ($\lambda = 0$) up to $\varepsilon = 0.6$.

The above observations of the leading perturbation patterns characteristic for the symmetric flows in the toroidal pipe, allow us to introduce some features that will help us to classify disturbances of non-symmetric flows at non-zero torsion. Thus, we can distinguish them by weak or strong oscillations of the zero pseudo streamline separating two Dean vortices, where strong oscillations will correspond to the symmetry breaking mode 1, while weak oscillations will be compared either with the symmetry preserving mode 3 or the symmetry breaking mode 6. We can examine whether the Dean vortices oscillate in close phases, similar to the symmetry preserving mode, or at noticeably different phases, similarly to the symmetry breaking modes. The perturbations can be characterized by their distribution over the whole bulk of the vortices or localization in certain flow regions.

Thus, quite expected changes of the perturbations and slightly supercritical flows are observed along the parts of neutral curves corresponding to modes 1 and 3 (not shown in the figures, only in the animations, see Table 1). For example, at $\varepsilon = 0.01, \lambda = 2$ the Dean vortices oscillate in counter phases and are similar to those depicted in Fig. 5. At $\varepsilon = 0.03, \lambda = 1$, the vortices oscillations are in close phases and are similar to those shown in Fig. 7. However, at larger curvatures, the supercritical flow changes qualitatively along the same eigenmode branch, as is shown in Figs. 10 and 11 for $\varepsilon = 0.3, \lambda = 1$, which correspond to mode 6. With the increase of torsion, the negative vortex becomes more intensive than the positive one. As a result, the instability sets in only inside the stronger negative vortex, located in the upper part of the pseudo streamlines plot, as is seen in Fig. 10. As a result, the flow oscillations are seen only in the upper part of the plots in Fig. 11, while the lower parts remain almost unchanged along the oscillation period.

An additional important observation follows from comparison of Figs. 8 and 9 with Figs. 10 and 11 plotted for the case of a non-zero torsion. When the base flow symmetry is broken due to a non-zero torsion, instability sets in in the stronger counter clockwise rotating vortex, while the weaker clockwise vortex remains almost non-perturbed. This leads us to a conclusion that interaction of the two Dean vortices play no role in the onset of this mode of instability. Since the perturbations attain their largest values far from the boundaries, we cannot assume here a boundary layer instability. At the same time, we observe again an inflection point at the $V_s(r)$ profile that passes through the minimum of the pseudo stream function, in the neighborhood of which we observe large perturbations of all the three velocity components. Thus, we can assume here an inviscid instability of the through flow, which appears in both vortices in the antisymmetric configuration (Figs. 8 and 9), and only in a stronger counter clockwise vortex in a non-symmetric case (Figs. 10 and 11).

All the most critical eigenmode branches described above start at $\lambda = 0$. The eigenmodes becoming most unstable at larger torsion correspond to non-symmetric flows. Since the two Dean vortices become noticeably different, the instability may set in inside a stronger or weaker vortex, as it was observed for mode 6, so that the eigenmodes can be distinguished also by this feature. This happens, for example, with mode 2, characteristic for small curvatures $\varepsilon \leq 0.1$, and relatively large $\lambda$. This mode is illustrated in Figs. 12 and 13 for $\varepsilon = 0.05, \lambda = 4$. We observe that the most unstable disturbance is located mainly in the clockwise, positive and weaker vortex, and almost does not penetrate in the counter clockwise and stronger one (Fig. 12 and the corresponding animations). As a result, the oscillations of the slightly supercritical flow are visible in the lower part of the snapshots shown in Fig. 13 (see also the corresponding animations), while the upper parts remain almost unchanged. Oscillations of the weaker vortex cause also noticeable oscillations of the zero pseudo streamline.

The zeroing terms numerical experiment carried out for the parameters of Figs. 12 and 13, showed that this instability can be reproduced after neglecting most of bilinear terms. The most interesting is that in Eq. (A5), governing disturbance of the circumferential velocity, we have to leave only the term $-ikV_s\tilde{u}_\xi$ and all the terms containing $V_\xi$. A closer examination of the perturbation snapshots (Fig. 12) and the corresponding animation reveals that along the path of the disturbance wave propagation, the perturbations of $\tilde{u}_\xi$ appear slightly earlier than the two others. Assuming now that the circumferential motion is a source of instability, we look at the isolines of $V_\xi$, where we observe boundary layers that possibly can be unstable. Surprisingly, the largest value of $V_\xi$ in the clockwise Dean vortex is 0.28, while its smallest value in the more intensive counter clockwise vortex is -0.067. Thus, in the lower clockwise vortex the circumferential velocity reaches about a four times larger value, which can explain why the instability sets in inside this vortex only. The radial profile of $V_\xi$ plotted for the value of $\xi$, corresponding to the location of the maximum of $V_\xi$ near the lower boundary, is shown in Fig. 32 and exhibits a steep increase of the circumferential velocity magnitude from the pipe wall inwards. It should be stressed that the advection along the pipe center line still cannot be neglected. Therefore we observe here an instability of swirling motion, but not of some locally rotational flow.

Another example for disturbance located inside only one of the vortices is shown in Figs. 14 and 15. This perturbation mode, denoted as mode 4, replaces the symmetric mode 3 for $0.05 \leq \varepsilon \leq 0.2$ (Fig. 3b,c), and is located inside the weaker clockwise lower vortex. Similarly to mode 2, it arises in the regions of large cross flow velocities that are located now close to the pipe border, where boundary layers form at large curvatures and Reynolds numbers. Compared to the mode 2, this mode forms noticeably smaller structures around its maximal and minimal values, which are advected along the pseudo streamlines (Fig. 14). Note, that the stronger counter-clockwise vortex is almost not disturbed and its oscillations are very weak compared to oscillations of the clockwise vortex and the zero pseudo streamline (Fig. 15 and the corresponding animation). This mode can be characterized as a small circumferential wavenumber downstream wave.

Assuming that mode 4 triggers the instability inside the boundary layer formed by the circumferential velocity, we notice again that the minimal value of the circumferential velocity in the stronger counter clockwise vortex is $\approx -0.13$, while the maximal value in the weaker clockwise vortex is 0.51. Then we examine the radial profile of $V_\xi$ passing through the maximum of $V_\xi$ located near the lower border (Fig. 32). We observe that in the current case the circumferential velocity grows steeper and reaches a smaller minimum value, which can be a reason for smaller-scale wavy structures. The zeroing terms numerical experiment shows that bilinear terms proportional to the dimensionless torsion $\varepsilon\lambda$ play no role in the instability onset. To keep the leading eigenvalue and eigenvector almost unchanged we also can neglect the terms $\left(u_r \frac{\partial V_r}{\partial r} + \frac{u_\xi}{r}\frac{\partial V_r}{\partial \xi}\right)\boldsymbol{e}_r$ and $\frac{u_\xi}{r}\frac{\partial V_\xi}{\partial \xi}\boldsymbol{e}_\xi$. This means that among all the terms describing advection of the base flow velocities $V_r$ and $V_\xi$ by their disturbances, only advection of $V_\xi$ in the radial direction contributes to the instability onset.

Following the neutral curve of $\varepsilon = 0.2$, we arrive to mode 5 illustrated in Figs. 16 and 17. This mode appears at relatively large torsions (Table 1), at which the counter clockwise vortex becomes dominant, while the clockwise one is either noticeably weaker, or completely disappears from the base flow (see animation for $\varepsilon = 0.2$, $\lambda = 5$). Thus, for the parameters of Fig. 16, the minimum and maximum values of the pseudo stream function are approximately -0.15 and 0.06, so that the counter clockwise vortex is more than two times more intensive than the clockwise one. As is seen from the figures, this perturbation mode is characterized by two large-scale structures located in the dominant counter clockwise vortex. The perturbation of the centerline velocity $\tilde{v}_s$ rotates in the clockwise direction around the dominant counter clockwise vortex (Fig. 16 and the corresponding animation), while evolution of the disturbances $\tilde{v}_r$ and $\tilde{v}_\xi$ cannot be easily connected to the direction of base flow motion. At larger torsions, however, the clockwise motion of these two perturbation components is clearly seen. The slightly supercritical flow at smaller torsions results in oscillation of both vortices and a weak clockwise rotational motion of the maximum isolines of the centerline velocity $v_s$ (Fig. 17 and the corresponding animations). During the oscillations period the Dean vortices size noticeably changes, which results also in strong oscillations of the zero pseudostreamline. At larger torsions we observe oscillations of the single counter clockwise cross flow vortex, whose center also performs a weak clockwise rotational motion. Similarly to the previously described mode, we can characterize mode 5 as a large circumferential wavenumber upstream wave.

The zeroing terms numerical experiment conducted for mode 5 showed that the instability is caused by interaction between disturbances $u_\xi$ and $u_s$, and the base flow components $V_s$ and $V_\xi$, while the radial velocity and its disturbance play only a passive role. Therefore, in this case we observe an instability of a swirling flow along the pipe centerline.

An example of a similar, but downstream large circumferential wavenumber wave is observed at larger curvatures $0.3 \leq \varepsilon \leq 0.6$, and large $\lambda$, and is characteristic for the rightmost branches of the corresponding neutral curves (Fig. 3). This perturbation branch is denoted as mode 7 and is depicted in Figs. 18 and 19. At the governing parameters of Figs. 18 and 19, the clockwise vortex is fully suppressed by the counter clockwise one, so that the base flow contains only a single Dean vortex. The perturbation consists of the large scale structures similar to those of mode 5 (cf. Fig. 16 and Fig. 18), but contrary to mode 5 they rotate counter clockwise. In the resulting slightly supercritical flow we observe oscillations of the maximal values of both the centerline velocity and pseudo stream function. The maximum of $v_s$ rotates counterclockwise together with the perturbation, while the maximum of $\psi$ remains almost motionless. The critical wavenumber and the critical frequency have the same sign in this case (Fig. 3a,b), so that an above assumed connection between the direction of the wave propagation along the centerline and along the Dean vortex is supported. Contrary to the previous case, we cannot point on either term of the linearized equations that can be neglected without a change of the disturbances pattern or the eigenvalue. In spite of the fact that the instability in this case results in a quite simple streamwise swirling motion, its appearance includes interconnection of all the perturbations with all base flow components.

Modes 7 and 8 are the last one characteristic for relatively large intervals of $\lambda$. The other modes 9-14 appear at rather large curvatures, $\varepsilon \geq 0.3$, at rather short intervals of $\lambda$, and at non-zero, but not very large torsions, so that $0 < \lambda < 3$. Mode 8 develops in a stronger counter clockwise vortex (Figs. 20 and 21) and is observed at $\varepsilon = 0.3$, $1.37 \leq \lambda \leq 2$, and $\varepsilon = 0.4$, $1.33 \leq \lambda \leq 1.7$. In this case the torsion is moderate, so that both the clockwise and counter clockwise vortices of the base flow are well developed, while the counter clockwise vortex is slightly stronger (Fig. 20). Mode 8 appears in the stronger counter clockwise vortex, as it was observed for mode 6 (Fig. 10), but its structures are noticeably larger. Oscillations of this disturbance mode are seen in almost all bulk of the counter rotating vortex, however, as is seen from the animations, they cannot be characterized as a traveling wave. Contrary to mode 6, we observe penetration of the perturbations $\tilde{v}_s$ and $\tilde{v}_r$ into the weaker clockwise vortex, which results in quite visible oscillations of the clockwise vortex and the zero pseudo streamline (Fig. 21 and the corresponding animation).

Similarly to mode 7, neglecting of each term of the linearized equations leads to a noticeable change of the disturbances pattern or the eigenvalue. To estimate relative intensity of the Dean vortices of the flow shown in Fig. 20, we note that the minimum and maximum of the pseudo stream function are approximately -0.071 and 0.047. The minimal value of the circumferential velocity in the stronger counter clockwise vortex is $\approx -0.1$, while in the weaker clockwise vortex the maximal value is $\approx 0.68$. Nevertheless, the instability sets in in the counter clockwise vortex, but not in a weaker clockwise one, as it was observed for the instability of the circumferential velocity boundary layer. Also, contrary to several previous cases we do not observe here clear advection of the disturbances along or against the Dean vortices. An indication on a possible instability mechanism is revealed from the radial profile of $V_\xi$ plotted through a point of its minimum located near the upper pipe border, also shown in Fig. 32. Here we observe that the circumferential velocity magnitude grows from the zero value at the wall to the minimum of $V_\xi$. Then the circumferential velocity steeply increases to, even larger in magnitude, positive maximal value. Furthermore, changing the sign, the profile exhibits a clear inflection point, so that we observe a configuration of the local mixing layer. The viscous mixing layer becomes unstable at rather small Reynolds numbers (Gelfgat & Kit, 2006), so that this locally developing mixing layer configuration can be a source of the instability.

Another upstream wave developing on the stronger counter clockwise vortex is observed in mode 9 (Figs. 22 and 23), appearing in narrow intervals of $\lambda$ at $0.4 \leq \varepsilon \leq 0.6$ (Table 1). This mode has smaller perturbation structures that can be interpreted as a smaller pseudo-streamlines-wise wavenumber, and almost does not penetrate into the weaker vortex. The smaller perturbation structures can be explained by a slightly steeper increase/decrease of the circumferential velocity starting from the boundary (Fig. 32). We observe also that, unlikely in the mode 8, the disturbance structures propagate upstream the dominant counter clockwise Dean vortex. For the flow shown in Fig. 22, the minimal and maximal values of the stream function are $-0.86$ and $0.67$, while the minimal and maximal values of the circumferential velocity are $-0.15$ and $0.88$, where the minimal value is located inside the unstable dominant vortex. Since the circumferential velocity in the unperturbed clockwise vortex is significantly larger, we cannot connect the instability with the boundary layer of $V_\xi$. Taking into account that the radial profile plotted through a point of minimum of $V_\xi$ located near the upper pipe border is similar to the previous case (Fig. 32), and also contains the inflection point, we assume again that the instability sets in due to an unstable local mixing layer flow. The zeroing terms numerical experiment shows that we can neglect the terms proportional to the torsion $\varepsilon\lambda$, but not to the curvature itself. Thus, this instability is affected by the centrifugal forcing resulting from the pipe bending, which also shows that the disturbances of all the three velocity components interact on the route to instability onset.

The perturbation mode 10 shown in Figs. 24 and 25 is characteristic for base flows in which the maximum of the centerline velocity is shifted towards the inner pipe boundary (located on the left) due to advection by the dominant counter clockwise vortex (Fig. 24). The perturbation pattern is a counter clockwise downstream traveling wave, similar to those observed for mode 7. Nevertheless, these modes are separated, as is clearly seen in Figs. 3b and 3c, possibly because of the different size of the counter clockwise vortex (cf. Fig. 18 and Fig. 24). Also, slightly supercritical regimes of mode 10 involve oscillations of the weaker clockwise vortex (Fig. 25 and the corresponding animations), which is either too small or does not exist in the flows subject to instability due to mode 7. The zeroing terms numerical experiment shows that we cannot neglect any of the terms in the linearized equations. Also, the instability sets in at relatively low Reynolds number, so that no thin boundary layers are observed. Noticing that perturbations of the centerline velocity rotate around the largest maximum of $V_s$ (Fig. 24 and the corresponding animation), we examined the radial profile $V_s(r)$ passing through the maximum. This profile contains an inflection point, which can be an indication for a destabilization mechanism here.

For $0.4 \leq \varepsilon \leq 0.6$ modes 7 and 10 are separated by short intervals of $\lambda$, where the instability is caused by mode 11 depicted in Figs. 26 and 27. Mode 11 seems to be similar to the mode 10, but arising at larger torsions, so that the dominance of the counter clockwise vortex becomes more profound and the vortices become aligned vertically, and not horizontally, as it was for the zero torsion (Fig. 4-10). However, there are several qualitative differences between modes 10 and 11. First, mode 11 sets in with the zero wavenumber, $k_{cr} = 0$, so that the disturbances do not depend on the *s* coordinate. The oscillations are the same in every pipe cross-section, and do not form a wave. The second difference follows from comparison of the perturbation amplitudes that form completely different patterns in the two cases. The third difference is that there is no inflection point on the $V_s(r)$ profile passing through the maximum of the centerline velocity. The zeroing terms numerical experiment shows that we can neglect the terms $\left(\frac{u_\xi}{r}\frac{\partial V_s}{\partial \xi} + \frac{\varepsilon sin(\xi)}{H_s}u_r V_s\right)\boldsymbol{e}_s$ and $\varepsilon\lambda u_s \frac{\partial V_r}{\partial \xi}\boldsymbol{e}_r$, but this does not help to understand the origin of this instability. A closer look at the base flow (Fig. 26) shows that the maximum of $V_s$ and the minimum of the pseudo stream function are located close to each other. In the slightly supercritical regime (Fig. 27 and the corresponding animation) the maximum of $V_s$ makes a round motion around location of these two extremum points. Thus, one can assume that the perturbation of centerline velocity is advected by the base flow around the dominant counter clockwise vortex, affecting in the same way two other velocity components. This assumption is supported by the perturbation snapshots (Fig. 26) and the corresponding animations.

In mode 12, observed only for $\varepsilon = 0.4$ and 0.6 and depicted in Figs. 28 and 29, we again observe an upstream cross flow wave with a relatively large cross flow wavenumber. It is noticeably different from the upstream wave of mode 5, because it develops at relatively small torsions, so that both clockwise and counterclockwise Dean vortices remain well developed. As in the several previous cases, the instability arises in the stronger counter clockwise vortex (Fig. 28). The resulting slightly supercritical flow (Fig.29) exhibits only oscillations of the counter clockwise vortex, while the clockwise one remains almost stationary (see animations listed in Table 1).

Finally, mode 13 (Figs. 30 and 31) observed only for $\varepsilon = 0.4$, exhibits two downstream cross flow waves developing along both Dean vortices. The wave arising along the stronger counter clockwise vortex has a noticeably larger amplitude (Fig. 30), however oscillations of both vortices are clearly seen in the slightly supercritical oscillatory regime (Fig. 31).

The $V_\xi(r)$ profiles of modes 12 and 13 also exhibit inflection points near the upper pipe border, and in the case of mode 13 also near the lower border. Therefore, also in these two cases we assume that the locally developing mixing layer leads to the flow destabilization. Similarly to other flows at large curvatures, we could not find any term of the linearized equations that could be neglected without a qualitative changes in the perturbation patterns. We observe again here that a steeper growth of the circumferential velocity from the pipe wall towards its center, characteristic for mode 13, leads to formation of smaller scale structures in the disturbance patterns, which indicate also on a possible role of the boundary layer instabilities. It is possible also that the two, boundary layer and mixing layer instabilities interact, which results in a self-sustained oscillatory mechanism, yet to be studied and described.

## 6. Concluding remarks

The main result of this study is the map of stability of helical pipe flows reported in Fig. 3. These parametric results showing the dependence of the critical Reynolds number, critical wavenumber and critical frequency on the dimensionless curvature and torsion are presented for the first time. In all the cases considered, the instability sets in as a steady – oscillatory transition via a Hopf bifurcation.

Along with this stability map, a graphical description of 13 distinct disturbance modes that become most unstable at different values of the curvature and torsion is presented. With only one exception, these modes appear as traveling waves propagating downstream or upstream the base through flow. At small curvatures $\varepsilon \leq 0.2$ the instability always set in as a downstream propagating traveling wave. With the increase of the pipe curvature we observe increase of the number of distinct most unstable modes replacing each other in the interval $0 \leq \lambda \leq 5$. These modes sometimes propagate upstream and sometimes are *s*-independent. They are classified by their symmetries, location, and direction of their propagation along the pipe and within its cross-sections. Possible physical mechanisms exciting different perturbations and leading to qualitatively different supercritical oscillatory states are discussed. Based on examination of the perturbation patterns, we offered some explanation of these destabilizing processes that include instability of local viscous boundary, inviscid instability of mixing layers and/or shear through flow, as well as another self-sustained oscillatory process based on the interconnection between the Dean vortices and the through flow.

Two more comments should be made regarding current results on the stability of the helical pipe flow. First, starting from quite small dimensionless curvatures, of the order of 0.01, the instability develops in agreement with the linear stability theory, which is unlikely a straight circular pipe. This difference was mentioned also in Canton et al (2016), where it was confirmed by comparison with the experimental results for the zero torsion. The present results, together with comparisons with experiments made in Gelfgat (2019), confirm this conclusion for a non-zero torsion. At smaller curvatures $\varepsilon < 0.01$, the experimentally observed instability corresponds to a bypass transition similarly to the straight pipe (Sreenivasan & Strykowski, 1983; Cioncilini & Santini, 2006; De Amicis et al, 2014). Thus, studying the flow in a helical pipe with a gradually decreasing/increasing curvature may shed more light in physics of the bypass transition characteristic for the straight pipes and other shear flows. There is also a possibility to study an exchange between linear and bypass transition with variation of the pipe curvature.

The second comment relates to the multiplicity of possible perturbation modes. From the viewpoint of the bifurcation theory, there are points corresponding to the different Hopf bifurcations of codimension 2, e.g. Hopf-Hopf bifurcation (Kuznetsov, 1998) or tangent Hopf bifurcation (Marques et al, 2003). These points can be easily found by the present numerical approach. Using flexible plastic pipes, one can easily alter the curvature and the torsion of a helical pipe in an experimental setup. Therefore, it yields a quite unique possibility to study bifurcations of higher codimension numerically and compare them with the experimental observations done at the prescribed governing parameters values.

Acknowledgments:

This research was supported by Israel Science Foundation (ISF) grant No 415/18 and was enabled in part by support provided by WestGrid (www.westgrid.ca) and Compute Canada (www.computecanada.ca). The author would like to express his thanks to A. Cioncolini, J. Canton and J. Kühnen for kindly agreeing to share their results.

# Figure captions

**Figure 1.** Sketch of a helical pipe (left) and illustration of helical coordinates introduced by Germano (1982). Directions of the coordinate axes *x*, *y*, *r*, and θ correspond to those introduced in Wang (1981) and Germano (1982).

**Figure 2.** Comparison of the friction factors measured by Cioncolini & Santini (2006) and calculated in the present study. The experiment denoted as Coil_02, with $\varepsilon = 0.059256$ and $\lambda = 0.050128$ is chosen for the comparison. Inserts show isolines of the *s*-velocity $v_s$ (color plots) and pseudo stream function $\psi$ (black lines) at several characteristic points. The corresponding values of $R_G$ and $Re$, as well as maximal values of $v_s$ and maximal and minimal values of $\psi$, are shown in the table. All the isolines are equally spaced.

**Figure 3.** (a) Critical Reynolds number versus the torsion to curvature ratio at different dimensionless curvatures; (b) Critical oscillations frequency versus the torsion to curvature ratio at different dimensionless curvatures; (c) Critical wavenumber versus the torsion to curvature ratio at different dimensionless curvatures.

**Figure 3.** Comparison of measured and calculated friction factors. Lines – calculations, symbols – results of De Amicis et al. (2014).

**Figure 4.** Oscillations of the most unstable perturbation at $\varepsilon = 0.01, \lambda = 0, Re_{cr} = 4181$ (mode 1). Left frames – perturbation of the centerline velocity (color) superimposed with isolines of the base flow centerline velocity (lines); center and right frames, respectively, show perturbations of $v_r$ and $v_\xi$ (color) superimposed with the pseudo – streamlines of base flow (lines). All the levels are equally spaced between the minimal and maximal values. Perturbation: $max|\tilde{v}_s| = 0.0192, max|\tilde{v}_r| = 0.00561, max|\tilde{v}_\xi| = 0.00795$. Base flow: $max|v_s| = 1.696, max|\psi| = 0.0101$. Animation files: Perturbation_e=0p01_l=0.avi, Perturbation_e=0p01_l=2.avi.

**Figure5.** Snapshots of a slightly supercritical oscillatory flow at at $\varepsilon = 0.01, \lambda = 0, Re_{cr} = 4181$. The levels are equally spaced between 0 and 1.6 for $v_s$ and between ±0.1 for $\psi$. Animation files: Flow_e=0p01_l=0.avi, Flow_e=0p01_l=2.avi .

**Figure 6.** Oscillations of the most unstable perturbation at $\varepsilon = 0.075, \lambda = 0, Re_{cr} = 3357$ (mode 3). Left frames – perturbation of the centerline velocity (color) superimposed with isolines of the base flow centerline velocity (lines); center and right frames, respectively, show perturbations of $v_r$ and $v_\xi$ (color) superimposed with the pseudo – streamlines of base flow (lines). All the levels are equally spaced between the minimal and maximal values. Perturbation: $max|\tilde{v}_s| = 0.0245, max|\tilde{v}_r| = 0.0646, max|\tilde{v}_\xi| = 0.00567$. Base flow: $max|v_s| = 1.619, max|\psi| = 0.0195$. Animation file: Perturbation_e=0p075_l=0.avi .

**Figure 7.** . Snapshots of a slightly supercritical oscillatory flow at $\varepsilon = 0.075, \lambda = 0, Re_{cr} = 3357$. The levels are equally spaced between 0 and 1.5 for $v_s$ and between ±0.02 for $\psi$. Animation file: Flow_e=0p075_l=0.avi .

**Figure 8.** Oscillations of the most unstable perturbation at $\varepsilon = 0.2, \lambda = 0, Re_{cr} = 3802$ (mode 6). Left frames – perturbation of the centerline velocity (color) superimposed with isolines of the base flow centerline velocity (lines); center and right frames, respectively, show perturbations of $v_r$ and $v_\xi$ (color) superimposed with the pseudo – streamlines of base flow (lines). All the levels are equally spaced between the minimal and maximal values. Perturbation: $max|\tilde{v}_s| = 0.0531, max|\tilde{v}_r| = 0.0167, max|\tilde{v}_\xi| = 0.0232$. Base flow: $max|v_s| = 1.511$, $max|\psi| = 0.0260$. Animation files: Perturbation_e=0p2_l=0.avi, Perturbation_e=0p3_l=0.avi, Perturbation_e=0p4_l=0.avi, Perturbation_e=0p5_l=0.avi, Perturbation_e=0p6_l=0.avi.

**Figure 9.** Snapshots of a slightly supercritical oscillatory flow at $\varepsilon = 0.2, \lambda = 0, Re_{cr} = 3802$. The levels are equally spaced between 0 and 1.4 for $v_s$ and between ±0.025 for $\psi$. Animation file: Flow_e=0p2_l=0.avi.

**Figure 10** Oscillations of the most unstable perturbation at $\varepsilon = 0.3, \lambda = 1, Re_{cr} = 2488$ (mode 6). Left frames – perturbation of the centerline velocity (color) superimposed with isolines of the base flow centerline velocity (lines); center and right frames, respectively, show perturbations of $v_r$ and $v_\xi$ (color) superimposed with the pseudo – streamlines of base flow (lines). All the levels are equally spaced between the minimal and maximal values. Perturbation: $max|\tilde{v}_s| = 0.0627, max|\tilde{v}_r| = 0.0263, max|\tilde{v}_\xi| = 0.0480$. Base flow: $max|v_s| = 1.415$, $\psi_{min} = -0.0444, \psi_{max} = 0.0358$. Animation file: Perturbation_e=0p3_l=1.avi .

**Figure 11.** Snapshots of a slightly supercritical oscillatory flow at $\varepsilon = 0.3, \lambda = 1, Re_{cr} = 2488$. The levels are equally spaced between 0 and 1.4 for $v_s$ and between -0.045 and 0.30 for $\psi$. Animation file: Flow_e=0p3_l=1.avi .

**Figure 12.** Oscillations of the most unstable perturbation at $\varepsilon = 0.05, \lambda = 4, Re_{cr} = 837$ (mode 2). Left frames – perturbation of the centerline velocity (color) superimposed with isolines of the base flow centerline velocity (lines); center and right frames, respectively, show perturbations of $v_r$ and $v_\xi$ (color) superimposed with the pseudo – streamlines of base flow (lines). All the levels are equally spaced between the minimal and maximal values. Perturbation: $max|\tilde{v}_s| = 0.0268, max|\tilde{v}_r| = 0.00724, max|\tilde{v}_\xi| = 0.0101$. Base flow: $max|v_s| = 1.610$, $\psi_{min} = -0.0451, \psi_{max} = 0.0286$. Animation files: Perturbation_e=0p05_l=4.avi, Perturbation_e=0p05_l=4.avi, Perturbation_e=0p1_l=3.5.avi .

**Figure 13.** Snapshots of a slightly supercritical oscillatory flow at $\varepsilon = 0.05, \lambda = 4, Re_{cr} = 837$. The levels are equally spaced between 0 and 1.5 for $v_s$ and between -0.040 and 0.025 for $\psi$. Animation files: Flow_e=0p05_l=4.avi, Flow_e=0p05_l=4.avi, Flow_e=0p1_l=3.5.avi .

**Figure 14.** Oscillations of the most unstable perturbation at $\varepsilon = 0.2, \lambda = 1.5, Re_{cr} = 1904$ (mode 4). Left frames – perturbation of the centerline velocity (color) superimposed with isolines of the base flow centerline velocity (lines); center and right frames, respectively, show perturbations of $v_r$ and $v_\xi$ (color) superimposed with the pseudo – streamlines of base flow (lines). All the levels are equally spaced between the minimal and maximal values. Perturbation: $max|\tilde{v}_s| = 0.0442, max|\tilde{v}_r| = 0.0152, max|\tilde{v}_\xi| = 0.0269$. Base flow: $max|v_s| = 1.463$,

$\psi_{min} = -0.0455$, $\psi_{max} = 0.0340$. Animation files: Perturbation_e=0p2_l=1p5.avi, Perturbation_e=0p1_l=1p5.avi .

**Figure 15.** Snapshots of a slightly supercritical oscillatory flow at $\varepsilon = 0.2, \lambda = 1.5, Re_{cr} = 1904$. The levels are equally spaced between 0 and 1.4 for $v_s$ and between -0.04 and 0.04 for $\psi$. Animation files: Flow_e=0p2_l=1.5.avi, Flow_e=0p1_l=1.5.avi .

**Figure 16.** Oscillations of the most unstable perturbation at $\varepsilon = 0.2, \lambda = 3, Re_{cr} = 449$ (mode 5). Left frames – perturbation of the centerline velocity (color) superimposed with isolines of the base flow centerline velocity (lines); center and right frames, respectively, show perturbations of $v_r$ and $v_\xi$ (color) superimposed with the pseudo – streamlines of base flow (lines). All the levels are equally spaced between the minimal and maximal values. Perturbation: $max|\tilde{v}_s| = 0.0154, max|\tilde{v}_r| = 0.00984, max|\tilde{v}_\xi| = 0.0125$. Base flow: $max|v_s| = 1.904$, $\psi_{min} = -0.148$, $\psi_{max} = 0.0614$. Animation files: Perturbation_e=0p2_l=3.avi, Perturbation_e=0p2_l=4.avi, Perturbation_e=0p2_l=5.avi .

**Figure 17.** Snapshots of a slightly supercritical oscillatory flow at $\varepsilon = 0.2, \lambda = 3, Re_{cr} = 449$. The levels are equally spaced between 0 and 1.9 for $v_s$ and between -0.16 and 0.07 for $\psi$. Animation files: Flow_e=0p2_l=3.avi, Flow_e=0p2_l=4.avi, Flow_e=0p2_l=5.avi .

**Figure 18.** Oscillations of the most unstable perturbation at $\varepsilon = 0.5, \lambda = 4, Re_{cr} = 360$ (mode 7). Left frames – perturbation of the centerline velocity (color) superimposed with isolines of the base flow centerline velocity (lines); center and right frames, respectively, show perturbations of $v_r$ and $v_\xi$ (color) superimposed with the pseudo – streamlines of base flow (lines). All the levels are equally spaced between the minimal and maximal values. Perturbation: $max|\tilde{v}_s| = 0.0102, max|\tilde{v}_r| = 0.00267, max|\tilde{v}_\xi| = 0.00362$. Base flow: $max|v_s| = 2.082$, $\psi_{min} = -1.092$, $\psi_{max} = 0$. Animation files: Perturbation_e=0p3_l=4.avi, Perturbation_e=0p4_l=4.avi, Perturbation_e=0p5_l=4.avi, Perturbation_e=0p6_l=4.avi .

**Figure 19.** Snapshots of a slightly supercritical oscillatory flow at $\varepsilon = 0.5, \lambda = 4, Re_{cr} = 360$. The levels are equally spaced between 0 and 2.0 for $v_s$ and between -1.0 and -0.1 for $\psi$. Animation files: Flow_e=0p3_l=4.avi, Flow_e=0p4_l=4.avi, Flow_e=0p5_l=4.avi, Flow_e=0p6_l=4.avi .

**Figure 20.** Oscillations of the most unstable perturbation at $\varepsilon = 0.3, \lambda = 1.55, Re_{cr} = 1384$ (mode 8). Left frames – perturbation of the centerline velocity (color) superimposed with isolines of the base flow centerline velocity (lines); center and right frames, respectively, show perturbations of $v_r$ and $v_\xi$ (color) superimposed with the pseudo – streamlines of base flow (lines). All the levels are equally spaced between the minimal and maximal values. Perturbation: $max|\tilde{v}_s| = 0.0299, max|\tilde{v}_r| = 0.0124, max|\tilde{v}_\xi| = 0.0136$ Base flow: $max|v_s| = 1.375$, $\psi_{min} = -0.0711$, $\psi_{max} = 0.0475$. Animation files: Perturbation_e=0p3_l=1p55.avi .

**Figure 21.** Snapshots of a slightly supercritical oscillatory flow at $= 0.3, \lambda = 1.55, Re_{cr} = 1384$. The levels are equally spaced between 0 and 1.3 for $v_s$ and between -0.06 and 0.04 for $\psi$. Animation files: Flow_e=0p3_l=1p55.avi .

**Figure 22.** Oscillations of the most unstable perturbation at $\varepsilon = 0.5, \lambda = 1, Re_{cr} = 969$ (mode 9). Left frames – perturbation of the centerline velocity (color) superimposed with isolines of the base flow centerline velocity (lines); center and right frames, respectively, show perturbations of $v_r$ and $v_\xi$ (color) superimposed with the pseudo – streamlines of base flow (lines). All the levels are equally spaced between the minimal and maximal values. Perturbation: $max|\tilde{v}_s| = 0.0279, max|\tilde{v}_r| = 0.0161, max|\tilde{v}_\xi| = 0.0368$ Base flow: $max|v_s| = 1.334$, $\psi_{min} = -0.0861, \psi_{max} = 0.0667$. Animation files: Perturbation_e=0p5_l=1.avi, Perturbation_e=0p4_l=1.avi, Perturbation_e=0p6_l=0p6.avi.

**Figure 23.** Snapshots of a slightly supercritical oscillatory flow at $= 0.5, \lambda = 1, Re_{cr} = 969$. The levels are equally spaced between 0 and 1.3 for $v_s$ and between -0.08 and 0.06 for $\psi$. Animation files: Flow_e=0p5_l=1p1.avi, : Flow_e=0p4_l=1p1.avi, : Flow_e=0p6_l=0p6.avi.

**Figure 24.** Oscillations of the most unstable perturbation at $\varepsilon = 0.5, \lambda = 1.5, Re_{cr} = 391$ (mode 10). Left frames – perturbation of the centerline velocity (color) superimposed with isolines of the base flow centerline velocity (lines); center and right frames, respectively, show perturbations of $v_r$ and $v_\xi$ (color) superimposed with the pseudo – streamlines of base flow (lines). All the levels are equally spaced between the minimal and maximal values. Perturbation: $max|\tilde{v}_s| = 0.0189, max|\tilde{v}_r| = 0.0104, max|\tilde{v}_\xi| = 0.0242$ Base flow: $max|v_s| = 1.729$, $\psi_{min} = -0.200, \psi_{max} = 0.0969$. Animation files: Perturbation_e=0p5_l=1p5.avi .

**Figure 25.** Snapshots of a slightly supercritical oscillatory flow at $\varepsilon = 0.5, \lambda = 1.5, Re_{cr} = 391$. The levels are equally spaced between 0 and 1.8 for $v_s$ and between -0.2 and 0.08 for $\psi$. Animation files: Flow_e=0p5_l=1p1.avi, Perturbation_e=0p6_l=1p7_k=0.avi.

**Figure 26.** Oscillations of the most unstable perturbation at $\varepsilon = 0.4, \lambda = 2.3, Re_{cr} = 171$ (mode 11). Left frames – perturbation of the centerline velocity (color) superimposed with isolines of the base flow centerline velocity (lines); center and right frames, respectively, show perturbations of $v_r$ and $v_\xi$ (color) superimposed with the pseudo – streamlines of base flow (lines). All the levels are equally spaced between the minimal and maximal values. Perturbation: $max|\tilde{v}_s| = 0.0120, max|\tilde{v}_r| = 0.00483, max|\tilde{v}_\xi| = 0.00817$ Base flow: $max|v_s| = 2.184$, $\psi_{min} = -0.325, \psi_{max} = 0.0475$. Animation files: Perturbation_e=0p4_l=2p3_k=0.avi.

**Figure 27.** Snapshots of a slightly supercritical oscillatory flow at $= 0.4, \lambda = 2.3, Re_{cr} = 171$. The levels are equally spaced between 0 and 2.0 for $v_s$ and between -0.3 and 0.04 for $\psi$. Animation files: Flow_e=0p4_l=2p3_k=0.avi, Flow_e=0p6_l=1p7_k=0.avi.

**Figure 28.** Oscillations of the most unstable perturbation at $\varepsilon = 0.4, \lambda = 1.1, Re_{cr} = 1658$ (mode 12). Left frames – perturbation of the centerline velocity (color) superimposed with isolines of the base flow centerline velocity (lines); center and right frames, respectively, show perturbations of $v_r$ and $v_\xi$ (color) superimposed with the pseudo – streamlines of base flow (lines). All the levels are equally spaced between the minimal and maximal values. Perturbation: $max|\tilde{v}_s| = 0.0267, max|\tilde{v}_r| = 0.0241, max|\tilde{v}_\xi| = 0.0394$ Base flow: $max|v_s| = 1.357$, $\psi_{min} = -0.0646, \psi_{max} = 0.0491$. Animation files: Perturbation_e=0p4_l=1p1.avi, Perturbation_e=0p6_l=0p5.avi.

**Figure 29.** Snapshots of a slightly supercritical oscillatory flow at $= 0.4, \lambda = 1.1, Re_{cr} = 1658$. The levels are equally spaced between 0 and 1.3 for $v_s$ and between -0.065 and 0.045 for $\psi$. Animation files: Flow_e=0p4_l=1p1.avi, Flow_e=0p6_l=0p5.avi.

**Figure 30.** Oscillations of the most unstable perturbation at $\varepsilon = 0.6, \lambda = 0.4, Re_{cr} = 2219$ (mode 13). Left frames – perturbation of the centerline velocity (color) superimposed with isolines of the base flow centerline velocity (lines); center and right frames, respectively, show perturbations of $v_r$ and $v_\xi$ (color) superimposed with the pseudo – streamlines of base flow (lines). All the levels are equally spaced between the minimal and maximal values. Perturbation: $max|\tilde{v}_s| = 0.0438, max|\tilde{v}_r| = 0.0251, \ max|\tilde{v}_\xi| = 0.0617$  Base flow: $max|v_s| = 1.273, \psi_{min} = -0.0491, \psi_{max} = 0.0462$. Animation file: Perturbation_ Perturbation_e=0p6_l=0p4.avi.

**Figure 31.** Snapshots of a slightly supercritical oscillatory flow at $= 0.6, \lambda = 0.4, Re_{cr} = 2219$. The levels are equally spaced between 0 and 1.2 for $v_s$ and between -0.050 and 0.047 for $\psi$. Animation file: Flow_e=0p6_l=0p4.avi.

**Figure 32.** Radial profiles of the base flow circumferential velocity passing through its maximum located inside the clockwise Dean vortex (modes 2, 3, 4, 9) or the minimum located in the counter clockwise Dean vortex (modes 10, 13, 14).

**Appendix**

For the "two-dimensional" flow depending only on the coordinates $r$ and $\xi$, the momentum equation in an alternative form with the eliminated mixed second derivatives reads

$$\frac{\partial v_r}{\partial t} + v_r \frac{\partial v_r}{\partial r} + \frac{v_\xi}{r}\frac{\partial v_r}{\partial \xi} - \varepsilon\lambda \frac{v_s}{H_s}\frac{\partial v_r}{\partial \xi} - \frac{v_\xi^2}{r} - \frac{\varepsilon\sin(\xi)}{H_s}v_s^2 =$$

$$= -\frac{\partial p}{\partial r} + \frac{1}{R_G}\frac{1}{rH_s}\left\{\frac{\partial^2}{\partial r^2}[rH_s v_r] + \frac{1}{r}\frac{\partial}{\partial \xi}\left[H_s\frac{\partial v_r}{\partial \xi}\right] + \varepsilon^2\lambda^2 r \frac{\partial}{\partial \xi}\left[\frac{1}{H_s}\frac{\partial v_r}{\partial \xi}\right] - \frac{1}{r}\frac{\partial v_\xi}{\partial \xi} - \varepsilon\lambda\frac{\partial}{\partial \xi}\left[\frac{v_s}{H_s}\right]\right\} \quad (A1)$$

$$\frac{\partial v_\xi}{\partial t} + \frac{v_r}{r}\frac{\partial(rv_\xi)}{\partial r} + \frac{v_\xi}{r}\frac{\partial v_\xi}{\partial \xi} - \varepsilon\lambda \frac{v_s}{H_s}\frac{\partial v_\xi}{\partial \xi} + \frac{v_\xi v_s}{r} - \frac{\varepsilon\cos(\xi)}{H_s}v_s^2 = -\frac{1}{r}\frac{\partial p}{\partial \xi} + \frac{1}{R_G}\frac{1}{H_s}\left\{\frac{\partial}{\partial r}\left[\frac{H_s}{r}\frac{\partial(rv_\xi)}{\partial r}\right] + \right.$$

$$\left. +\frac{1}{r^2}\frac{\partial^2}{\partial \xi^2}[H_s v_\xi] + \varepsilon^2\lambda^2\frac{\partial}{\partial \xi}\left[\frac{1}{H_s}\frac{\partial v_\xi}{\partial \xi}\right] + \frac{2}{r^2}\frac{\partial(H_s v_r)}{\partial \xi} + \varepsilon\cos(\xi)\frac{\partial v_r}{\partial r} + \varepsilon^2\lambda\frac{\partial}{\partial \xi}\left[\frac{\cos(\xi)}{H_s}v_s\right]\right\} \quad (A2)$$

$$\frac{\partial v_s}{\partial t} + v_r\frac{\partial v_s}{\partial r} + \frac{v_\xi}{r}\frac{\partial v_s}{\partial \xi} - \varepsilon\lambda\frac{v_s}{H_s}\frac{\partial v_s}{\partial \xi} + \frac{\varepsilon\sin(\xi)}{H_s}v_r v_s + \frac{\varepsilon\cos(\xi)}{H_s}v_\xi v_s = -\frac{1}{H_s}\frac{\partial p}{\partial s} +$$

$$\frac{1}{R_G}\left\{\frac{1}{r}\frac{\partial}{\partial r}\left[\frac{r}{H_s}\frac{\partial(H_s v_s)}{\partial r}\right] + \frac{1}{r^2}\frac{\partial}{\partial \xi}\left[\frac{1}{H_s}\frac{\partial(H_s v_s)}{\partial \xi}\right] + \varepsilon^2\lambda^2\frac{\partial}{\partial \xi}\left[\frac{1}{H_s^2}\frac{\partial v_s}{\partial \xi}\right] + \frac{\varepsilon^2\lambda\cos(\xi)}{r}\frac{\partial}{\partial r}\left[\frac{r^2}{H_s^2}v_r\right] -$$

$$2\varepsilon^2\lambda\frac{\partial}{\partial \xi}\left[\frac{\sin(\xi)}{H_s^2}v_r\right] - \varepsilon^2\lambda\frac{\partial}{\partial \xi}\left[\frac{\cos(\xi)}{H_s^2}v_\xi\right]\right\} \quad (A3)$$

The equations linearized near the steady state flow $\{V_r(r,\xi), V_\xi(r,\xi), V_s(r,\xi), P(r,\xi)\}$ that govern infinitely small disturbances $\{u_r(r,\xi), u_\xi(r,\xi), u_s(r,\xi), p(r,\xi)\}exp[i\sigma t + iks]$ are

$$\lambda u_r + V_r\frac{\partial u_r}{\partial r} + u_r\frac{\partial V_r}{\partial r} + \frac{V_\xi}{r}\frac{\partial u_r}{\partial \xi} + \frac{u_\xi}{r}\frac{\partial V_r}{\partial \xi} - \varepsilon\lambda u_s\frac{\partial V_r}{\partial \xi} + \frac{V_s}{H_s}\left[-\varepsilon\lambda\frac{\partial u_r}{\partial \xi} + iku_r\right] - \frac{2V_\xi u_\xi}{r} -$$

$$\frac{2\varepsilon\sin(\xi)}{H_s}V_s u_s = -\frac{\partial p}{\partial r} + \frac{1}{R_G}\frac{1}{rH_s}\left\{\frac{\partial^2}{\partial r^2}[rH_s u_r] + \frac{1}{r}\frac{\partial}{\partial \xi}\left[H_s\frac{\partial u_r}{\partial \xi}\right] + \varepsilon^2\lambda^2 r\frac{\partial}{\partial \xi}\left[\frac{1}{H_s}\frac{\partial u_r}{\partial \xi}\right] - \frac{1}{r}\frac{\partial u_\xi}{\partial \xi} -\right.$$

$$\left.\varepsilon\lambda\frac{\partial}{\partial \xi}\left[\frac{u_s}{H_s}\right] - k^2\frac{ru_r}{H_s} - ik\varepsilon\lambda r\frac{\partial}{\partial \xi}\left[\frac{u_r}{H_s}\right] - \frac{ikr}{H_s}\varepsilon\lambda\frac{\partial u_r}{\partial \xi} + \frac{ik}{H_s}u_s\right\} \quad (A4)$$

$$\lambda u_\xi + \frac{V_r}{r}\frac{\partial u_\xi}{\partial r} + \frac{u_r}{r}\frac{\partial V_\xi}{\partial r} + \frac{V_\xi}{r}\frac{\partial u_\xi}{\partial \xi} + \frac{u_\xi}{r}\frac{\partial V_\xi}{\partial \xi} - \varepsilon\lambda\frac{V_s}{H_s}\frac{\partial u_\xi}{\partial \xi} - \varepsilon\lambda\frac{u_s}{H_s}\frac{\partial V_\xi}{\partial \xi} + \frac{V_\xi u_s}{r} + \frac{u_\xi V_s}{r} -$$

$$\frac{2\varepsilon\cos(\xi)}{H_s}V_s u_s + \frac{ik}{H_s}V_s u_\xi = -\frac{1}{r}\frac{\partial p}{\partial \xi} + \frac{1}{R_G}\frac{1}{H_s}\left\{\frac{\partial}{\partial r}\left[\frac{H_s}{r}\frac{\partial(ru_\xi)}{\partial r}\right] + \frac{1}{r^2}\frac{\partial^2}{\partial \xi^2}[H_s u_\xi] + \varepsilon^2\lambda^2\frac{\partial}{\partial \xi}\left[\frac{1}{H_s}\frac{\partial u_\xi}{\partial \xi}\right] +\right.$$

$$\left.\frac{2}{r^2}\frac{\partial(H_s u_r)}{\partial \xi} + \varepsilon\cos(\xi)\frac{\partial u_r}{\partial r} + \varepsilon^2\lambda\frac{\partial}{\partial \xi}\left[\frac{\cos(\xi)}{H_s}u_s\right] - k^2\frac{u_\xi}{H_s} - ik\varepsilon\lambda\frac{\partial}{\partial \xi}\left[\frac{u_\xi}{H_s}\right] - \frac{ik}{H_s}\varepsilon\lambda\frac{\partial u_\xi}{\partial \xi} - ik\varepsilon\cos(\xi)\frac{u_s}{H_s}\right\}$$

$$(A5)$$

$$\lambda v_s + V_r \frac{\partial u_s}{\partial r} + u_r \frac{\partial V_s}{\partial r} + \frac{V_\xi}{r}\frac{\partial u_s}{\partial \xi} + \frac{u_\xi}{r}\frac{\partial V_s}{\partial \xi} - \varepsilon\lambda\frac{V_s}{H_s}\frac{\partial u_s}{\partial \xi} - \varepsilon\lambda\frac{u_s}{H_s}\frac{\partial V_s}{\partial \xi} + \frac{\varepsilon \sin(\xi)}{H_s}(V_r u_s + u_r V_s) +$$

$$\frac{\varepsilon \cos(\xi)}{H_s}\left(V_\xi u_s + u_\xi V_s\right) + \frac{ik}{H_s}V_s u_s = \frac{\varepsilon\lambda}{H_s}\frac{\partial p}{\partial \xi} - \frac{ik}{H_s}p + \frac{1}{R_G}\left\{\frac{1}{r}\frac{\partial}{\partial r}\left[\frac{r}{H_s}\frac{\partial(H_s u_s)}{\partial r}\right] + \frac{1}{r^2}\frac{\partial}{\partial \xi}\left[\frac{1}{H_s}\frac{\partial(H_s u_s)}{\partial \xi}\right] +$$

$$\varepsilon^2\lambda^2\frac{\partial}{\partial \xi}\left[\frac{1}{H_s^2}\frac{\partial u_s}{\partial \xi}\right] + \frac{\varepsilon^2\lambda\cos(\xi)}{r}\frac{\partial}{\partial r}\left[\frac{r^2}{H_s^2}u_r\right] - 2\varepsilon^2\lambda\frac{\partial}{\partial \xi}\left[\frac{\sin(\xi)}{H_s^2}u_r\right] - \varepsilon^2\lambda\frac{\partial}{\partial \xi}\left[\frac{\cos(\xi)}{H_s}u_\xi\right] - \frac{ik}{r}\frac{\partial}{\partial \xi}\left[\frac{r}{H_s}u_\xi\right] -$$

$$\frac{ik}{r}\frac{\partial}{\partial r}\left[\frac{r}{H_s}u_r\right]\right\} \quad (A6)$$

Table 1. Ranges of $\lambda$, at which the modes denoted on Figs. 3b and 3c were found.

| mode | $\varepsilon$ | Range of $\lambda$ | Mode characteristics | Perturbation animation | Flow animation |
|---|---|---|---|---|---|
| 1 | 0.01 | $0 \leq \lambda \leq 2.87$ | Antisymmetry-breaking<br>Dean vortices oscillate in counter phase<br>Zero pseudo streamline is strongly perturbed<br>An increase of through flow intensifies the Dean vortices, which slow down the through flow and consequently weaken by themselves | Perturbation_e=0p01_l=0.avi<br>Perturbation_e=0p01_l=2.avi | Flow_e=0p01_l=0.avi<br>Flow_e=0p01_l=2.avi |
| 2 | 0.01<br>0.03<br>0.05<br>0.075<br>0.1 | $2.87 \leq \lambda \leq 5$<br>$1.1 \leq \lambda \leq 5$<br>$1 \leq \lambda \leq 5$<br>$1.9 \leq \lambda \leq 5$<br>$2.8 \leq \lambda \leq 4.2$ | Perturbations are located inside a weaker vortex and are advected along the streamlines<br>Zero pseudo streamline is strongly perturbed<br>Instability of $V_\xi$ boundary layer is assumed | Perturbation_e=0p01_l=4.avi<br>Perturbation_e=0p05_l=4.avi<br>Perturbation_e=0p1_l=3.5.avi | Flow_e=0p01_l=4.avi<br>Flow_e=0p05_l=4.avi<br>Flow_e=0p1_l=3.5.avi |
| 3 | 0.03<br>0.05<br>0.075<br>0.1 | $0 \leq \lambda \leq 1.1$<br>$0 \leq \lambda \leq 1$<br>$0 \leq \lambda \leq 0.9$<br>$0 \leq \lambda \leq 0.6$ | Antisymmetry-preserving<br>Zero pseudo streamline is not perturbed<br>Dean vortices oscillate in phase and do not interact<br>Instability of $V_\xi$ boundary layer is assumed | Perturbation_e=0p03_l=0.avi<br>Perturbation_e=0p03_l=1.avi<br>Perturbation_e=0p05_l=0.avi<br>Perturbation_e=0p075_l=0.avi<br>Perturbation_e=0p1_l=0.avi | Flow_e=0p03_l=0.avi<br>Flow_e=0p03_l=1.avi<br>Flow_e=0p05_l=0.avi<br>Flow_e=0p075_l=0.avi<br>Flow_e=0p1_l=0.avi |
| 4 | 0.075<br>0.1<br>0.2 | $0.9 \leq \lambda \leq 1.9$<br>$0.6 \leq \lambda \leq 2.8$<br>$0.4 \leq \lambda \leq 2.55$ | Small circumferential wavenumber downstream cross-flow wave located inside a weaker vortex<br>Zero pseudo streamline is strongly perturbed | Perturbation_e=0p1_l=1p5.avi<br>Perturbation_e=0p2_l=1p5.avi | Flow_e=0p1_l=1p5.avi<br>Flow_e=0p2_l=1p5.avi |
| 5 | 0.1<br>0.2<br>0.3<br>0.4 | $4.2 \leq \lambda \leq 5$<br>$2.55 \leq \lambda \leq 5$<br>$2.6 \leq \lambda \leq 2.88$<br>$2.2 \leq \lambda \leq 2.35$ | Large circumferential wavenumber upstream cross-flow wave<br>Zero pseudo streamline, if exists, is strongly perturbed | Perturbation_e=0p1_l=4p5.avi<br>Perturbation_e=0p2_l=3.avi<br>Perturbation_e=0p2_l=4.avi<br>Perturbation_e=0p2_l=5.avi<br>Perturbation_e=0p3_l=2p7.avi<br>Perturbation_e=0p4_l=2p3.avi | Flow_e=0p1_l=4p5.avi<br>Flow_e=0p2_l=3.avi<br>Flow_e=0p2_l=4.avi<br>Flow_e=0p2_l=5.avi<br>Flow_e=0p3_l=2p7.avi<br>Flow_e=0p4_l=2p3.avi |
| 6 | 0.2<br>0.3<br>0.4<br>0.5<br>0.6 | $0 \leq \lambda \leq 0.4$<br>$0 \leq \lambda \leq 1.37$<br>$0 \leq \lambda \leq 0.7$<br>$0 \leq \lambda \leq 0.81$<br>$0 \leq \lambda \leq 0.3$ | Antisymmetry-breaking<br>Dean vortices do not interact<br>Zero pseudo streamline is weakly perturbed<br>At larger $\lambda$ perturbations are located inside a stronger vortex<br>An inviscid instability of through flow is assumed | Perturbation_e=0p2_l=0.avi<br>Perturbation_e=0p3_l=0.avi<br>Perturbation_e=0p3_l=1.avi<br>Perturbation_e=0p4_l=0.avi<br>Perturbation_e=0p5_l=0.avi<br>Perturbation_e=0p6_l=0.avi | Flow_e=0p2_l=0.avi<br>Flow_e=0p3_l=0.avi<br>Flow_e=0p3_l=1.avi<br>Flow_e=0p4_l=0.avi<br>Flow_e=0p5_l=0.avi<br>Flow_e=0p6_l=0.avi |
| 7 | 0.3 | $2.9 \leq \lambda \leq 5$ | The disturbance propagates upstream the mean through flow | Perturbation_e=0p3_l=4.avi | Flow_e=0p3_l=4.avi |
| | 0.4 | $2.45 \leq \lambda \leq 5$ | Large circumferential wavenumber downstream cross-flow wave developing in a single Dean vortex flow | Perturbation_e=0p4_l=4.avi | Flow_e=0p4_l=4.avi |
| | 0.5 | $2.05 \leq \lambda \leq 5$ | | Perturbation_e=0p5_l=4.avi | Flow_e=0p5_l=4.avi |
| | 0.6 | $1.75 \leq \lambda \leq 5$ | | Perturbation_e=0p6_l=4.avi | Flow_e=0p6_l=4.avi |
| 8 | 0.3 | $1.37 \leq \lambda \leq 2$ | Oscillations in the bulk of the counter clockwise vortex that cause oscillations in the whole flow. Instability of a locally developing mixing layer is assumed | Perturbation_e=0p3_l=1p55.avi | Flow_e=0p3_l=1p55.avi |
| | 0.4 | $1.33 \leq \lambda \leq 1.7$ | | Perturbation_e=0p4_l=1p4.avi | Flow_e=0p4_l=1p4.avi |
| 9 | 0.4 | $1.17 \leq \lambda \leq 1.34$ | Small circumferential wavenumber downstream cross-flow wave Zero pseudo streamline is strongly perturbed Instability of a locally developing mixing layer is assumed | Perturbation_e=0p4_l=1p1.avi | Flow_e=0p4_l=1p1.avi |
| | 0.5 | $0.81 \leq \lambda \leq 1.24$ | | Perturbation_e=0p5_l=1.avi | Flow_e=0p5_l=1.avi |
| | 0.6 | $0.57 \leq \lambda \leq 1.2$ | | Perturbation_e=0p6_l=0p6.avi | Flow_e=0p6_l=0p6.avi |
| 10 | 0.3 | $2 \leq \lambda \leq 2.6$ | The disturbance propagates upstream the mean through flow Large circumferential wavenumber downstream cross-flow wave An inviscid instability of through flow is assumed | Perturbation_e=0p3_l=2p4.avi | Flow_e=0p3_l=2p4.avi |
| | 0.4 | $1.7 \leq \lambda \leq 2.2$ | | Perturbation_e=0p4_l=1p8.avi | Flow_e=0p4_l=1p8.avi |
| | 0.5 | $1.24 \leq \lambda \leq 1.88$ | | Perturbation_e=0p5_l=1p5.avi | Flow_e=0p5_l=1p5.avi |
| | 0.6 | $1.2 \leq \lambda \leq 1.6$ | | Perturbation_e=0p6_l=1p3.avi | Flow_e=0p6_l=1p3.avi |
| 11 | 0.4 | $2.27 \leq \lambda \leq 2.45$ | $s$ – independent ($k = 0$) large circumferential wavenumber downstream cross-flow wave Zero pseudo streamline is slightly perturbed | Perturbation_e=0p4_l=2p3_k=0.avi | Flow_ e=0p4_l=2p3_k=0.avi |
| | 0.5 | $1.88 \leq \lambda \leq 2.05$ | | | |
| | 0.6 | $1.6 \leq \lambda \leq 1.75$ | | Perturbation_e=0p6_l=1p7_k=0.avi | Flow_ e=0p6_l=1p7_k=0.avi |
| 12 | 0.4 | $0.7 \leq \lambda \leq 1.16$ | large circumferential wavenumber upstream wave Instability of a locally developing mixing layer is assumed Zero pseudo streamline is slightly perturbed | Perturbation_e=0p4_l=1p1avi | Flow_ e=0p4_l=1p1.avi |
| | 0.6 | $0.47 \leq \lambda \leq 0.55$ | | Perturbation_e=0p6_l=0p5avi | Flow_ e=0p6_l=0p5.avi |
| 13 | 0.6 | $0.3 \leq \lambda \leq 0.47$ | Small circumferential wavenumber downstream waves developing in both vortices Instability of a locally developing mixing layer is assumed Zero pseudo streamline is strongly perturbed | Perturbation_e=0p6_l=0p4avi | Flow_ e=0p6_l=0p4.avi |

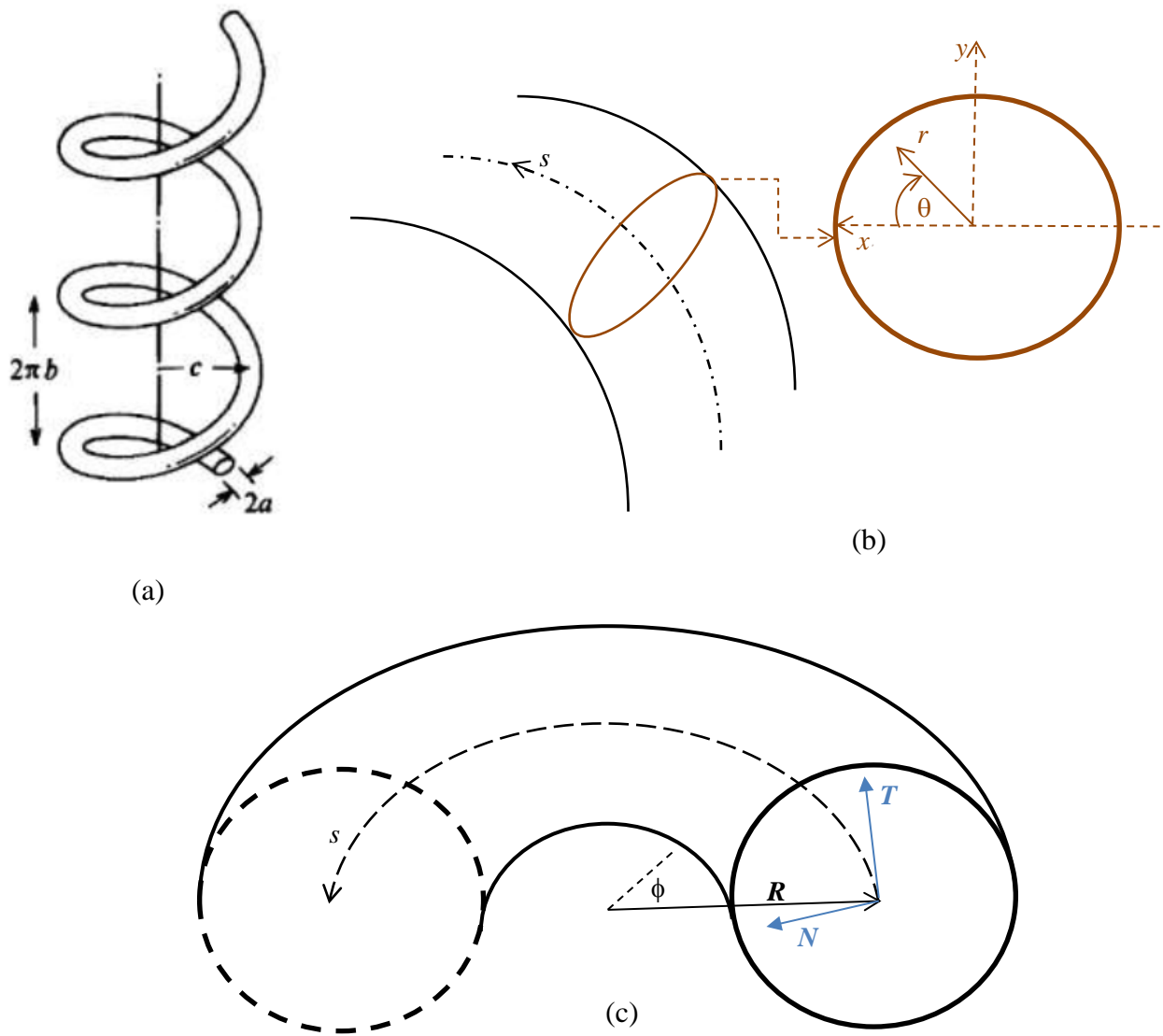

**Figure 1.** Sketch of a helical pipe (left) and illustration of helical coordinates of introduced by Germano (1982). Directions of the coordinate axes *x, y, r,* and θ correspond to those introduced in Wang (1981) and Germano (1982).

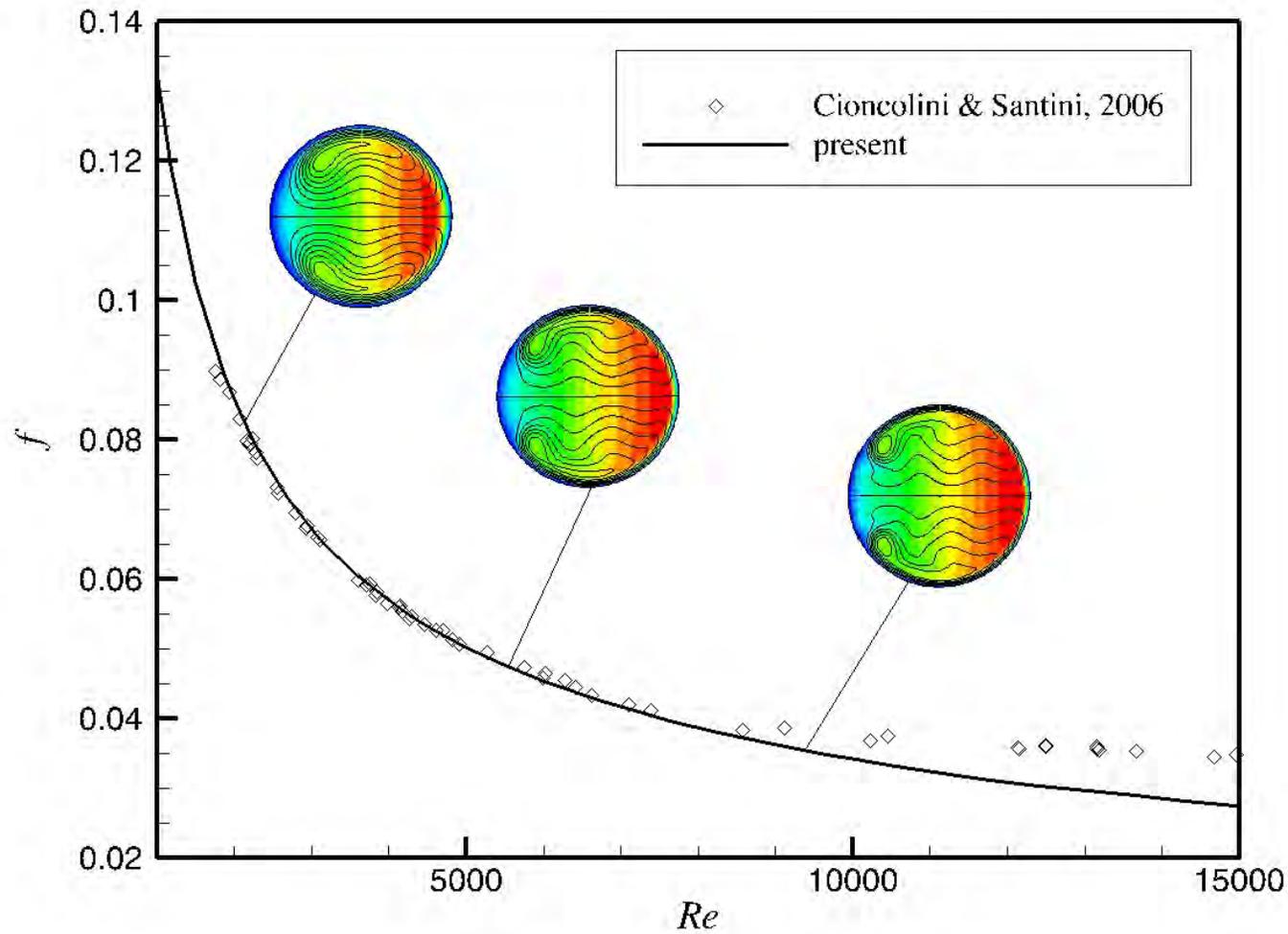

**Figure 2.** Comparison of the friction factors measured by Cioncolini & Santini (2006) and calculated in the present study. The experiment denoted as Coil_02, with $\varepsilon = 0.059256$ and $\lambda = 0.050128$, is chosen for the comparison. Inserts show isolines of the $s$-velocity $v_s$ (color plots) and pseudo stream function $\psi$ (black lines) at several characteristic points. The corresponding values of $R_G$ and $Re$, as well as maximal values of $v_s$ and maximal and minimal values of $\psi$ are shown in the table. All the isolines are equally spaced.

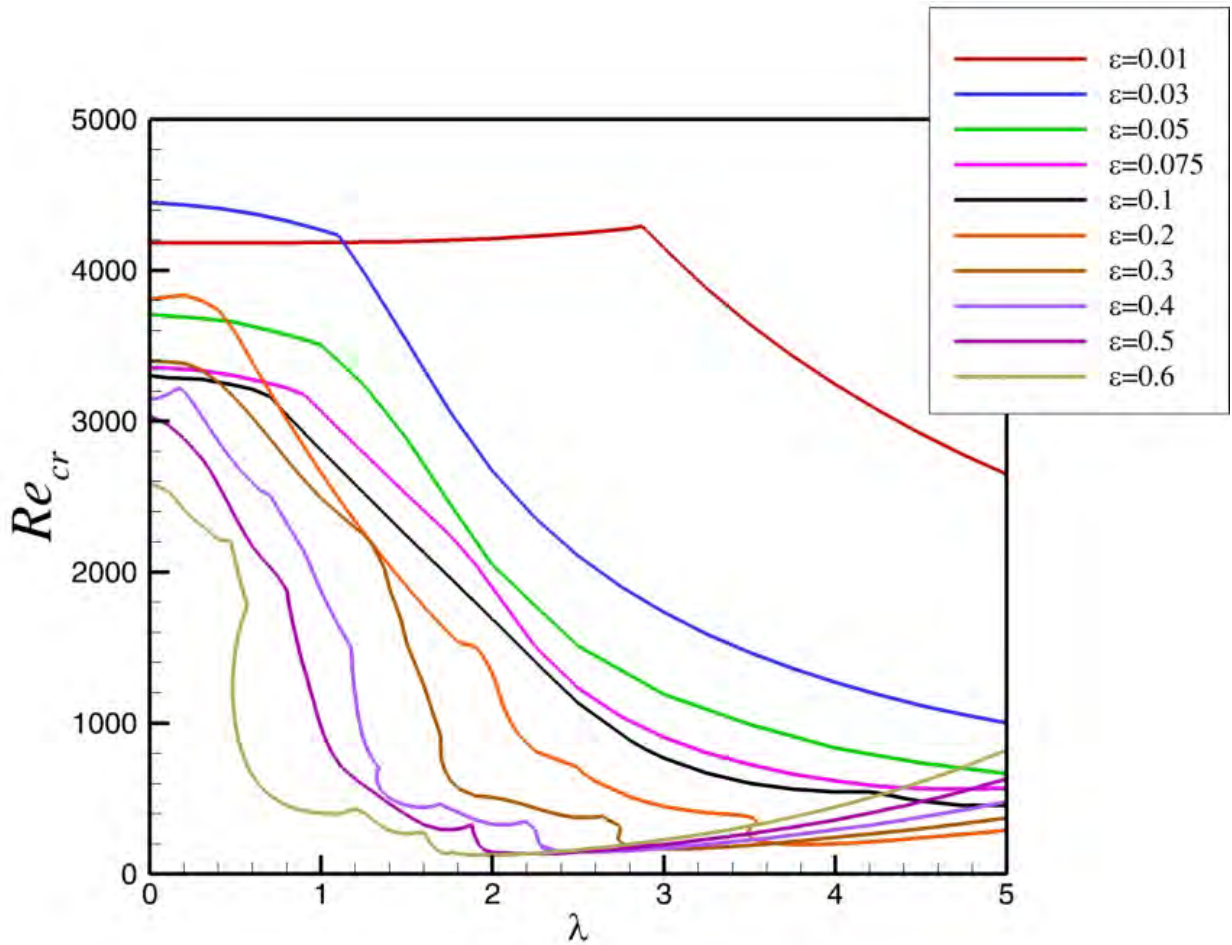

**Figure 3a.** Critical Reynolds number versus the torsion to curvature ratio at different dimensionless curvatures.

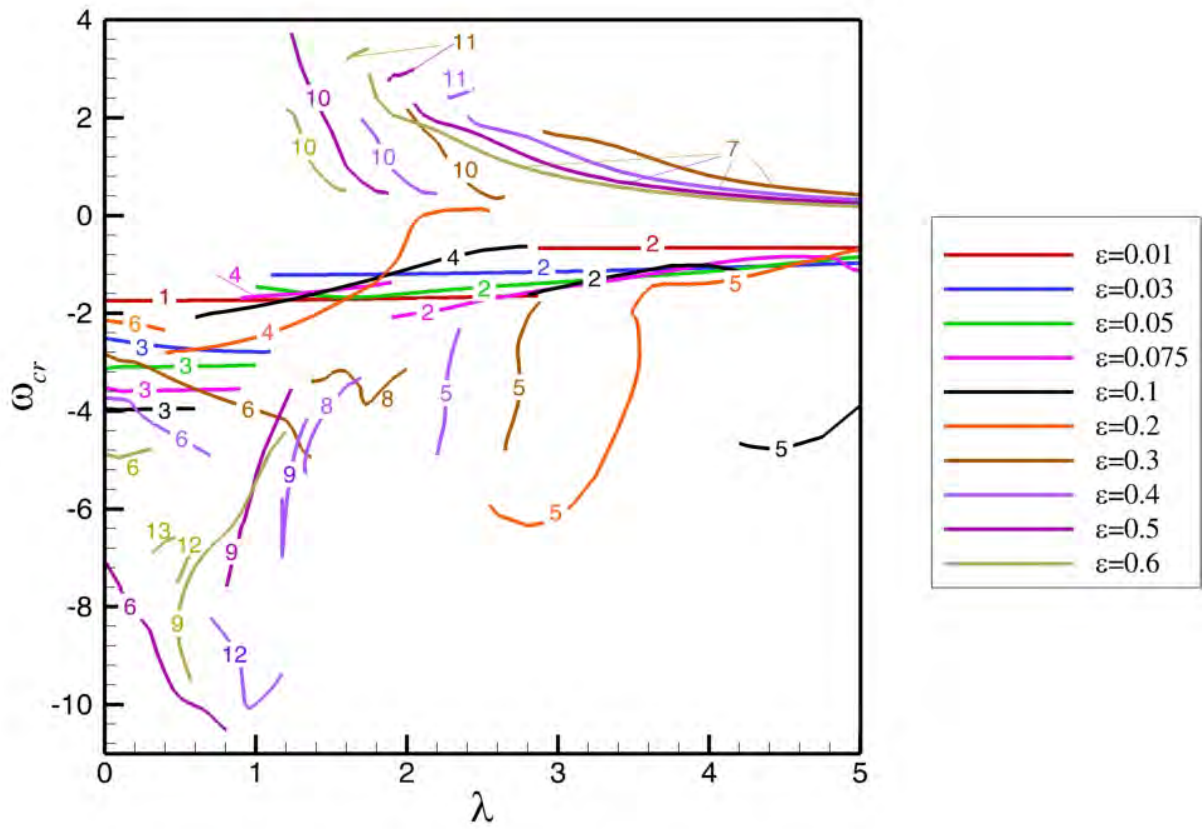

**Figure 3b.** Critical oscillations frequency versus the torsion to curvature ratio at different dimensionless curvatures.

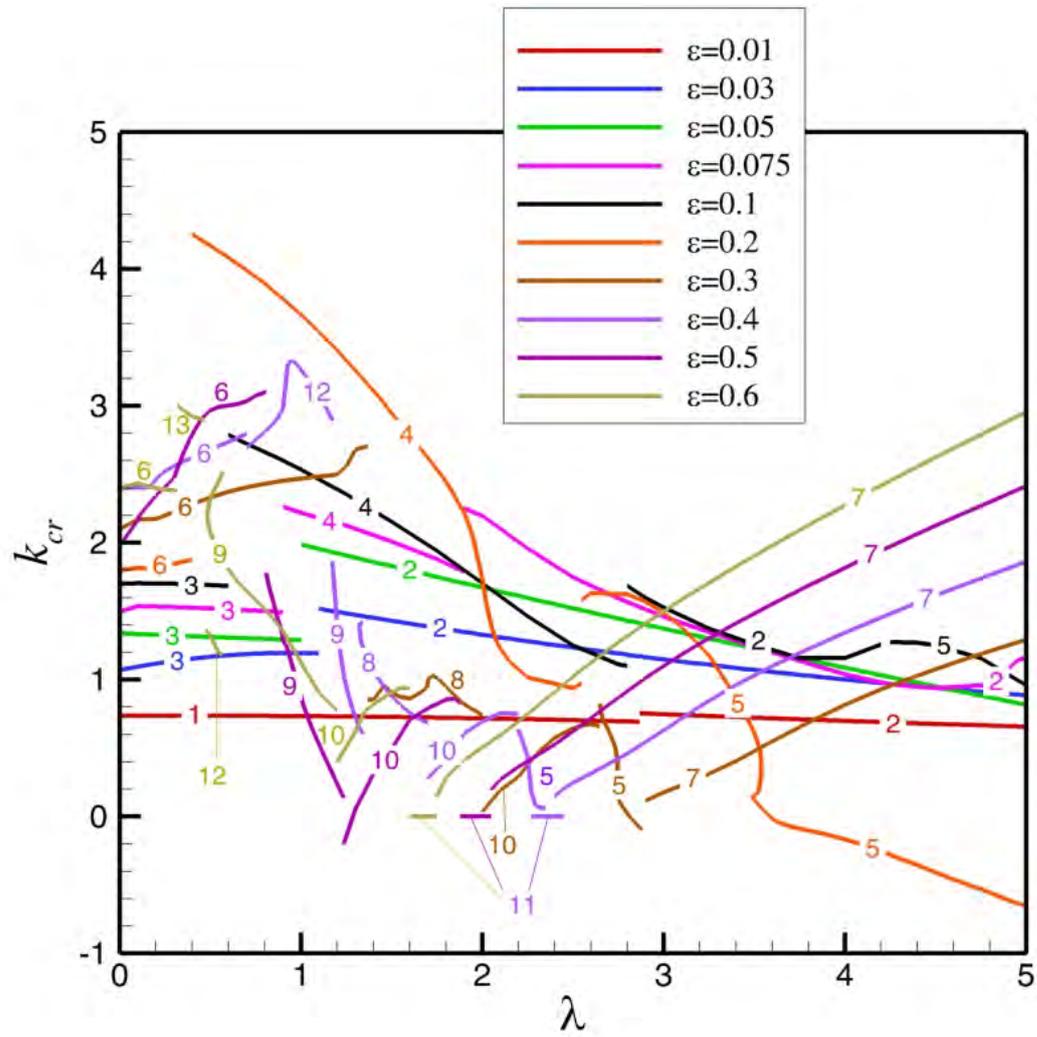

**Figure3c.** Critical wavenumber versus the torsion to curvature ratio at different dimensionless curvatures.

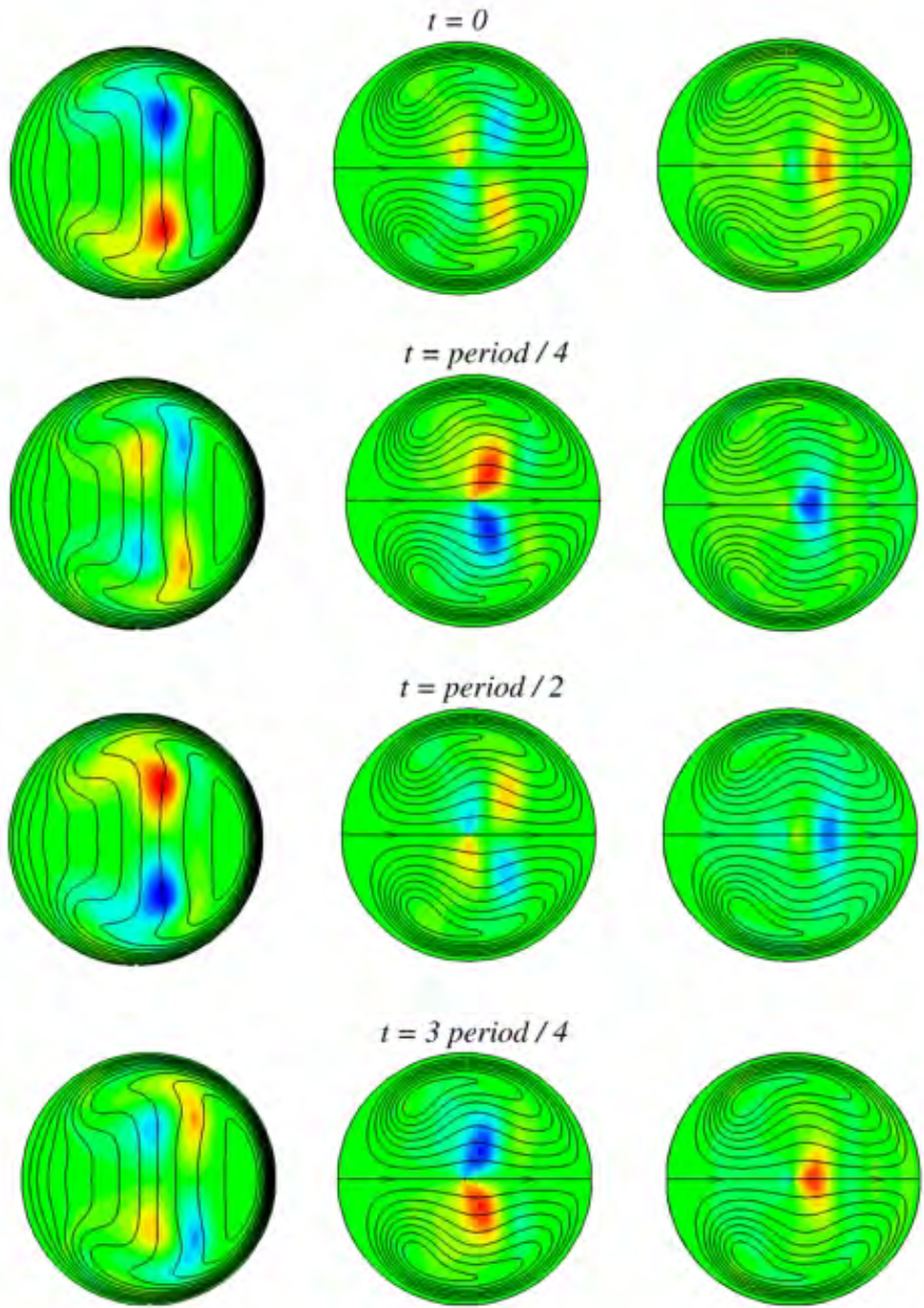

**Figure 4.** Oscillations of the most unstable perturbation at $\varepsilon = 0.01, \lambda = 0, Re_{cr} = 4181$ (mode 1). Left frames – perturbation of the centerline velocity (color) superimposed with isolines of the base flow centerline velocity (lines); center and right frames, respectively, show perturbations of $v_r$ and $v_\xi$ (color) superimposed with the pseudo – streamlines of base flow (lines). All the levels are equally spaced between the minimal and maximal values. Perturbation: $max|\tilde{v}_s| = 0.0192, max|\tilde{v}_r| = 0.00561, max|\tilde{v}_\xi| = 0.00795$. Base flow: $max|v_s| = 1.696, max|\psi| = 0.0101$. Animation files: Perturbation_e=0p01_l=0.avi, Perturbation_e=0p01_l=2.avi.

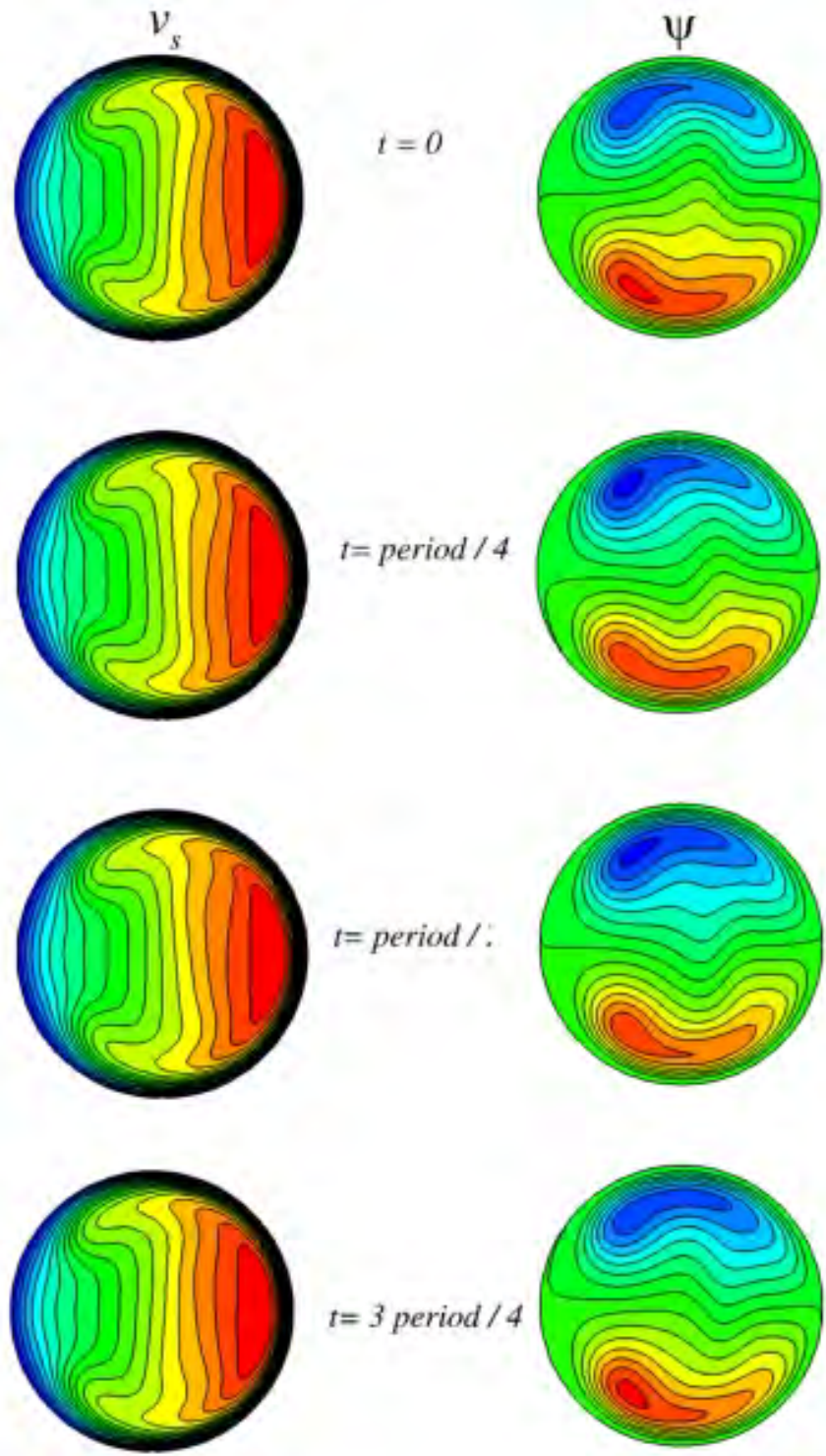

**Figure 5.** Snapshots of a slightly supercritical oscillatory flow at $\varepsilon = 0.01, \lambda = 0, Re_{cr} = 4181$. The levels are equally spaced between 0 and 1.6 for $v_s$ and between ±0.1 for $\psi$. Animation files: Flow_e=0p01_l=0.avi, Flow_e=0p01_l=2.avi.

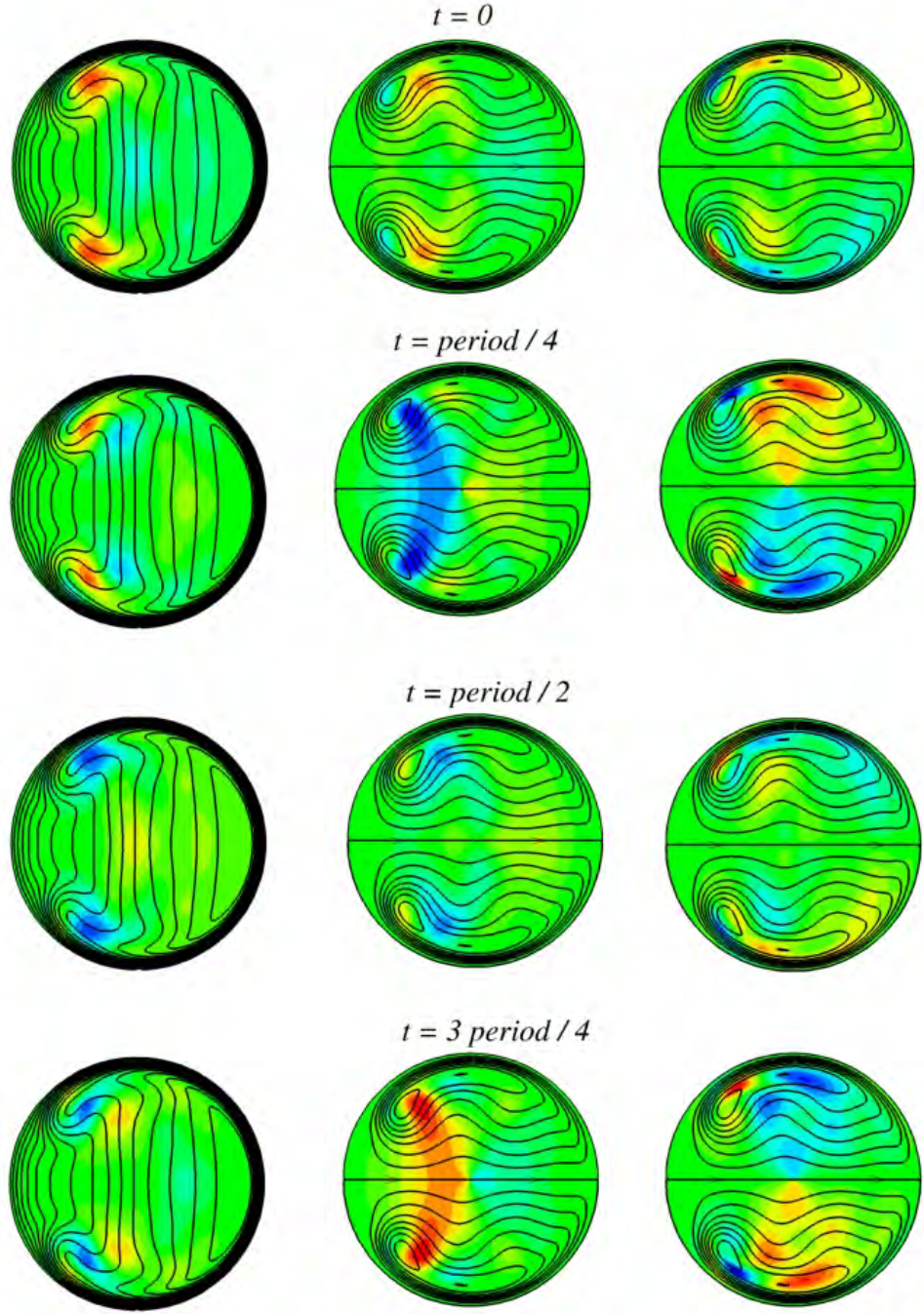

**Figure 6.** Oscillations of the most unstable perturbation at $\varepsilon = 0.075, \lambda = 0, Re_{cr} = 3357$ (mode 3). Left frames – perturbation of the centerline velocity (color) superimposed with isolines of the base flow centerline velocity (lines); center and right frames, respectively, show perturbations of $v_r$ and $v_\xi$ (color) superimposed with the pseudo – streamlines of base flow (lines). All the levels are equally spaced between the minimal and maximal values. Perturbation: $max|\tilde{v}_s| = 0.0245, max|\tilde{v}_r| = 0.0646, max|\tilde{v}_\xi| = 0.00567$. Base flow: $max|v_s| = 1.619, max|\psi| = 0.0195$. Animation file: Perturbation_e=0p075_l=0.avi .

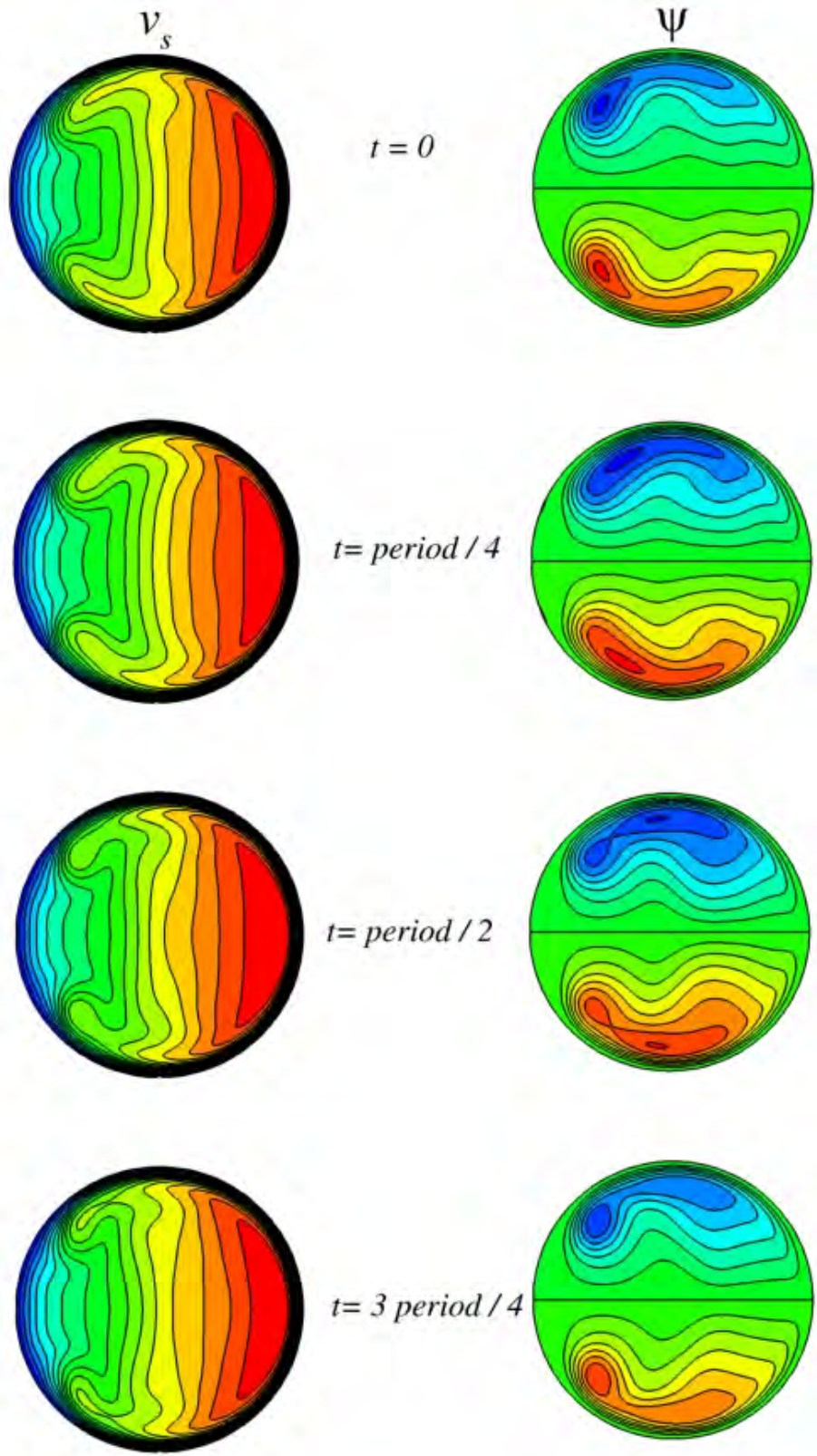

**Figure 7.** Snapshots of a slightly supercritical oscillatory flow at $\varepsilon = 0.075, \lambda = 0, Re_{cr} = 3357$. The levels are equally spaced between 0 and 1.5 for $v_s$ and between $\pm 0.02$ for $\psi$. Animation file: Flow_e=0p075_l=0.avi .

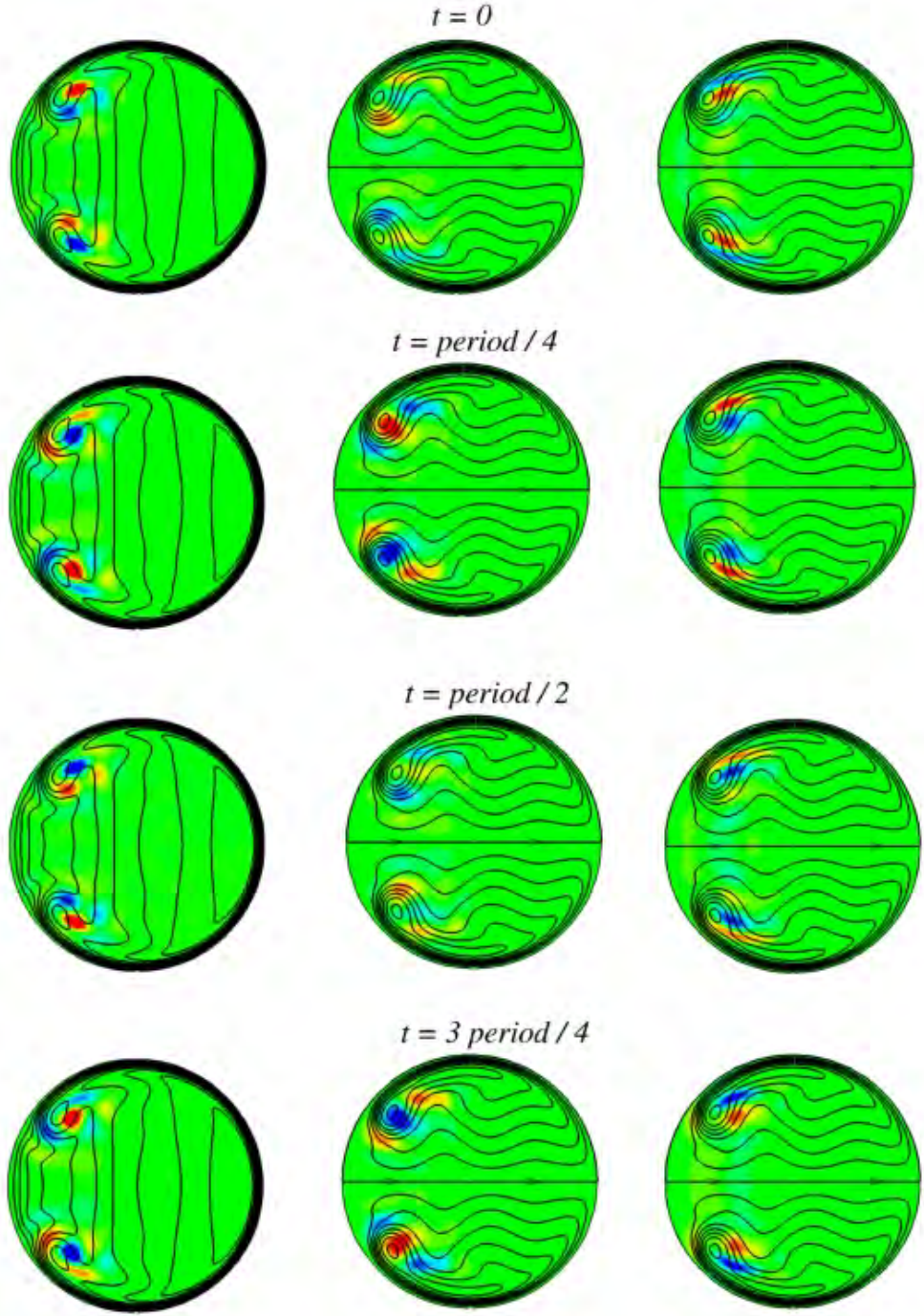

**Figure 8.** Oscillations of the most unstable perturbation at $\varepsilon = 0.2, \lambda = 0, Re_{cr} = 3802$ (mode 6). Left frames – perturbation of the centerline velocity (color) superimposed with isolines of the base flow centerline velocity (lines); center and right frames, respectively, show perturbations of $v_r$ and $v_\xi$ (color) superimposed with the pseudo – streamlines of base flow (lines). All the levels are equally spaced between the minimal and maximal values. Perturbation: $max|\tilde{v}_s| = 0.0531, max|\tilde{v}_r| = 0.0167, max|\tilde{v}_\xi| = 0.0232$. Base flow: $max|v_s| = 1.511, max|\psi| = 0.0260$. Animation files: Perturbation_e=0p2_l=0.avi, Perturbation_e=0p3_l=0.avi, Perturbation_e=0p4_l=0.avi, Perturbation_e=0p5_l=0.avi, Perturbation_e=0p6_l=0.avi .

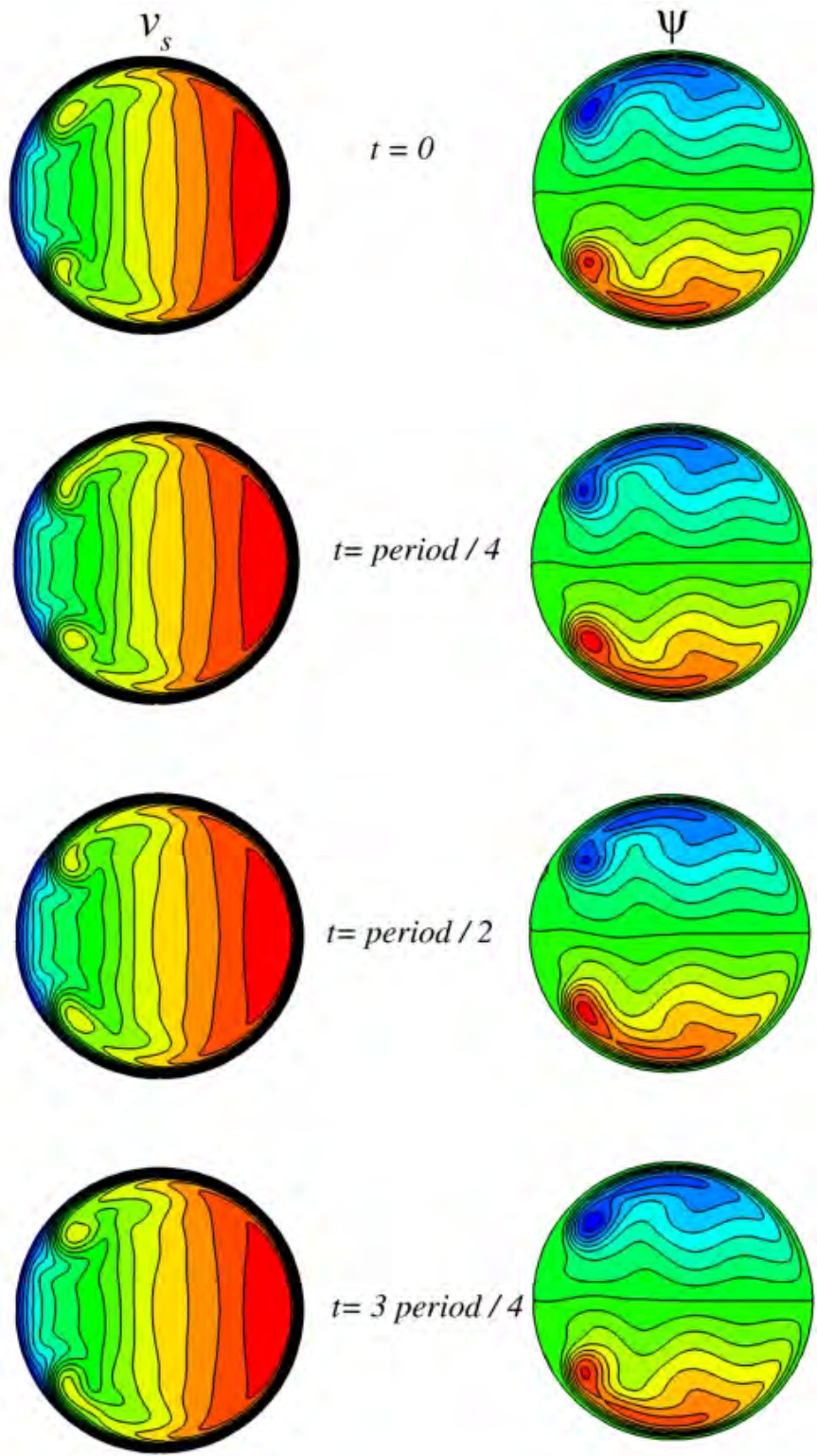

**Figure 9.** Snapshots of a slightly supercritical oscillatory flow at $\varepsilon = 0.2, \lambda = 0, Re_{cr} = 3802$. The levels are equally spaced between 0 and 1.4 for $v_s$ and between $\pm 0.025$ for $\psi$. Animation file: Flow_e=0p2_l=0.avi .

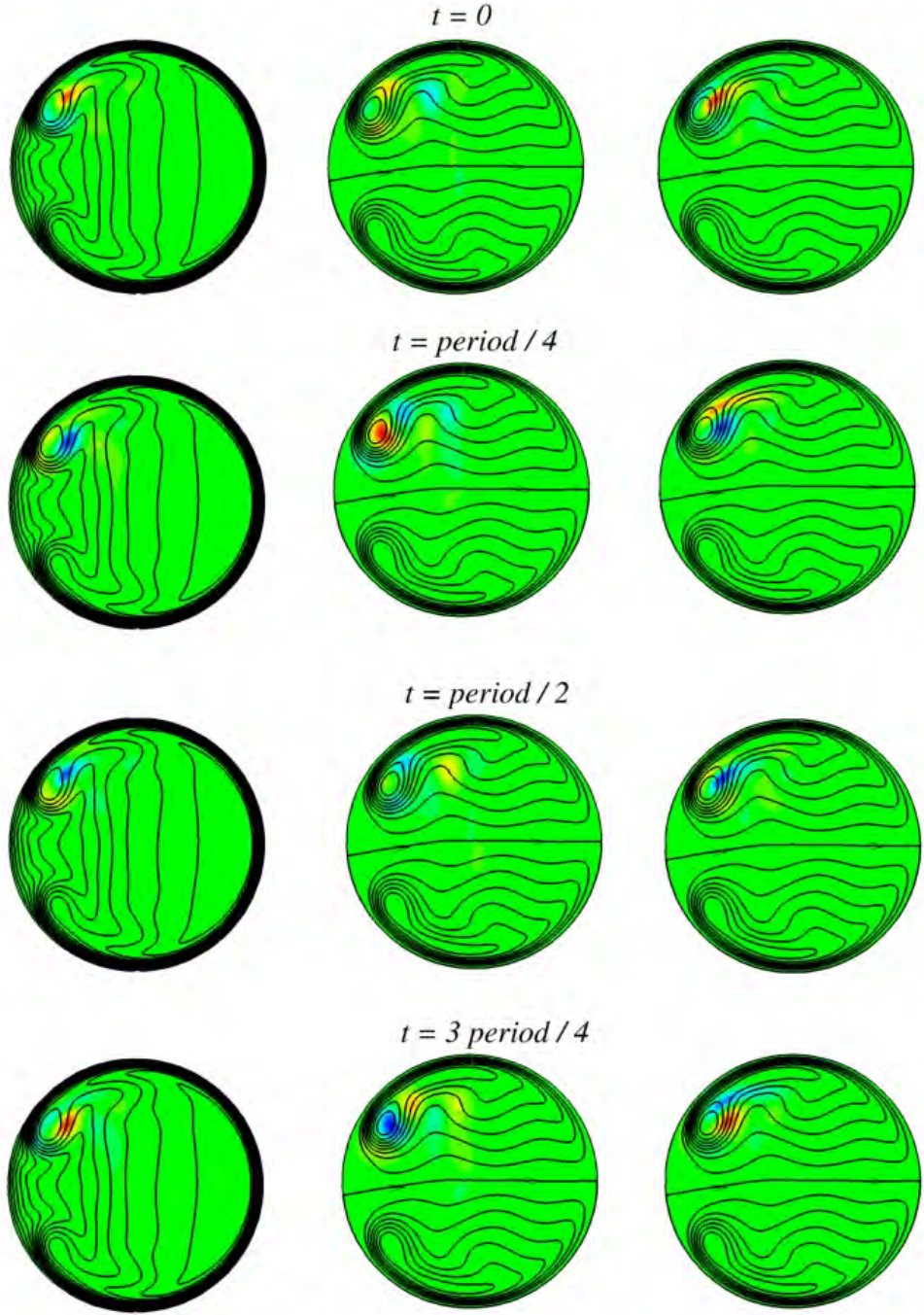

**Figure 10.** Oscillations of the most unstable perturbation at $\varepsilon = 0.3, \lambda = 1, Re_{cr} = 2488$ (mode 6). Left frames – perturbation of the centerline velocity (color) superimposed with isolines of the base flow centerline velocity (lines); center and right frames, respectively, show perturbations of $v_r$ and $v_\xi$ (color) superimposed with the pseudo – streamlines of base flow (lines). All the levels are equally spaced between the minimal and maximal values. Perturbation: $max|\tilde{v}_s| = 0.0627, max|\tilde{v}_r| = 0.0263, max|\tilde{v}_\xi| = 0.0480$. Base flow: $max|v_s| = 1.415, \psi_{min} = -0.0444, \psi_{max} = 0.0358$. Animation file: Perturbation_e=0p3_l=1.avi .

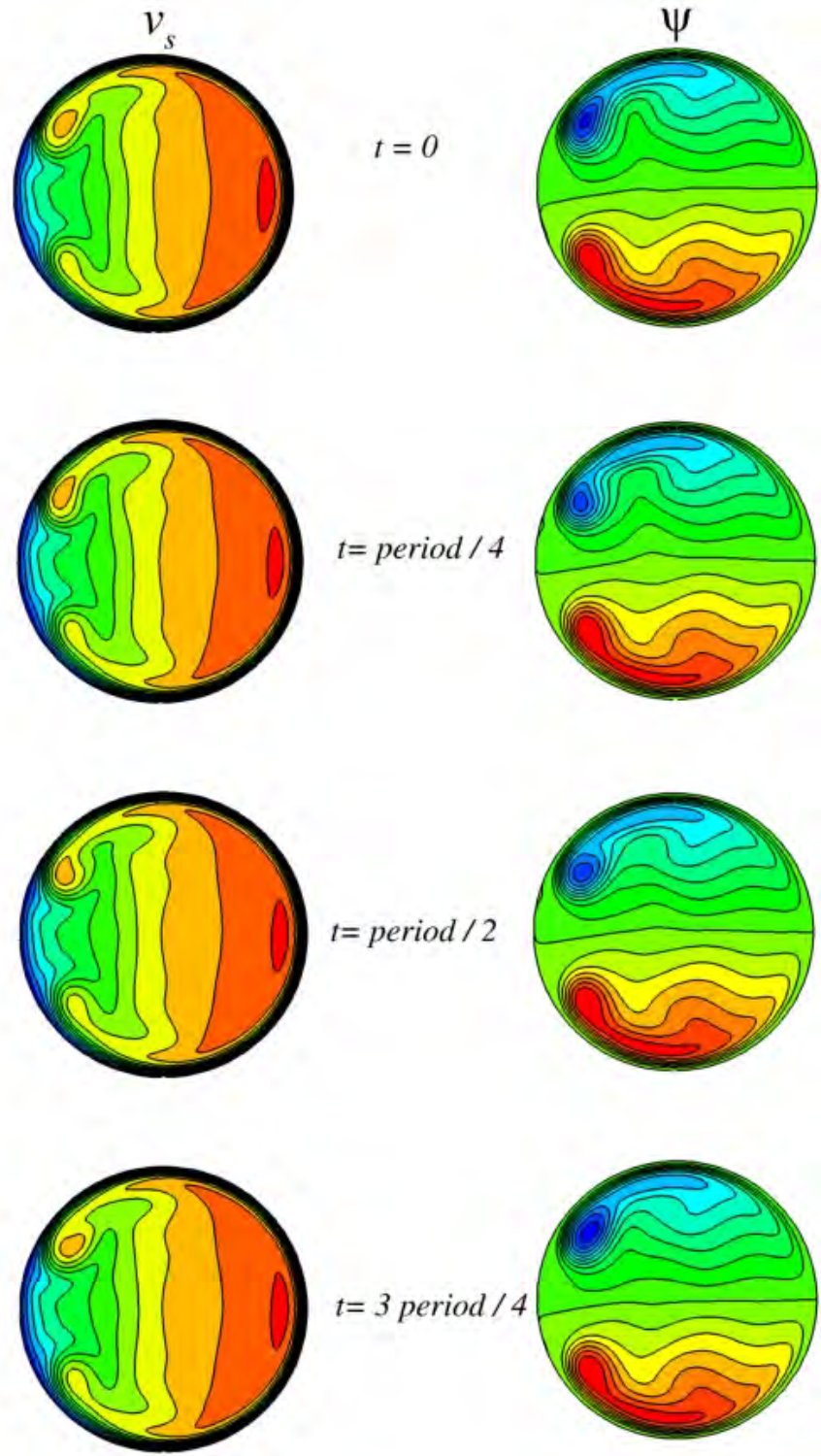

**Figure 11.** Snapshots of a slightly supercritical oscillatory flow at $\varepsilon = 0.3, \lambda = 1, Re_{cr} = 2488$. The levels are equally spaced between 0 and 1.4 for $v_s$ and between -0.045 and 0.30 for $\psi$. Animation file: Flow_e=0p3_l=1.avi .

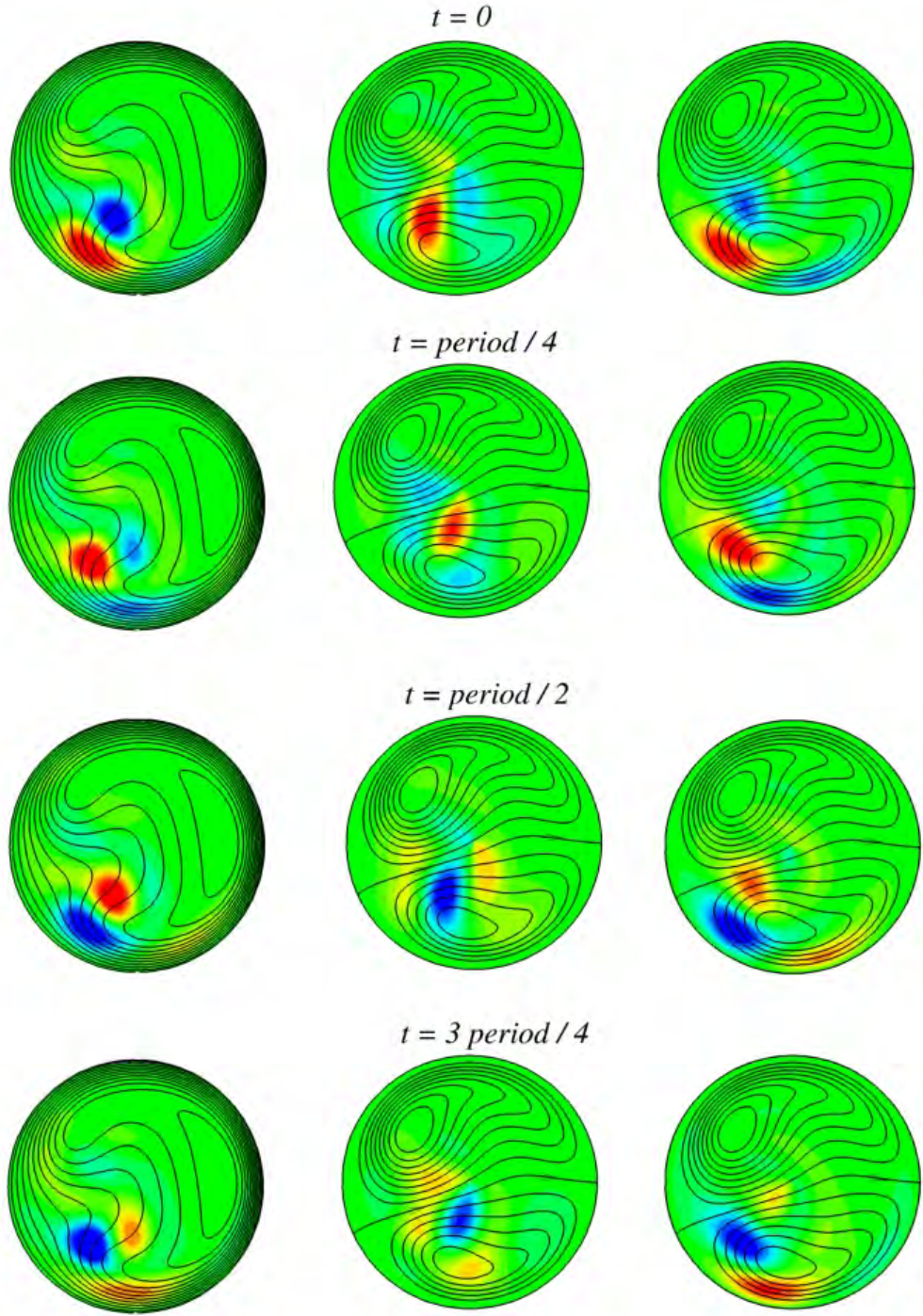

**Figure 12.** Oscillations of the most unstable perturbation at $\varepsilon = 0.05, \lambda = 4, Re_{cr} = 837$ (mode 2). Left frames – perturbation of the centerline velocity (color) superimposed with isolines of the base flow centerline velocity (lines); center and right frames, respectively, show perturbations of $v_r$ and $v_\xi$ (color) superimposed with the pseudo – streamlines of base flow (lines). All the levels are equally spaced between the minimal and maximal values. Perturbation: $max|\tilde{v}_s| = 0.0268, max|\tilde{v}_r| = 0.00724, max|\tilde{v}_\xi| = 0.0101$. Base flow: $max|v_s| = 1.610, \psi_{min} = -0.0451, \psi_{max} = 0.0286$. Animation files: Perturbation_e=0p05_l=4.avi, Perturbation_e=0p05_l=4.avi, Perturbation_e=0p1_l=3.5.avi .

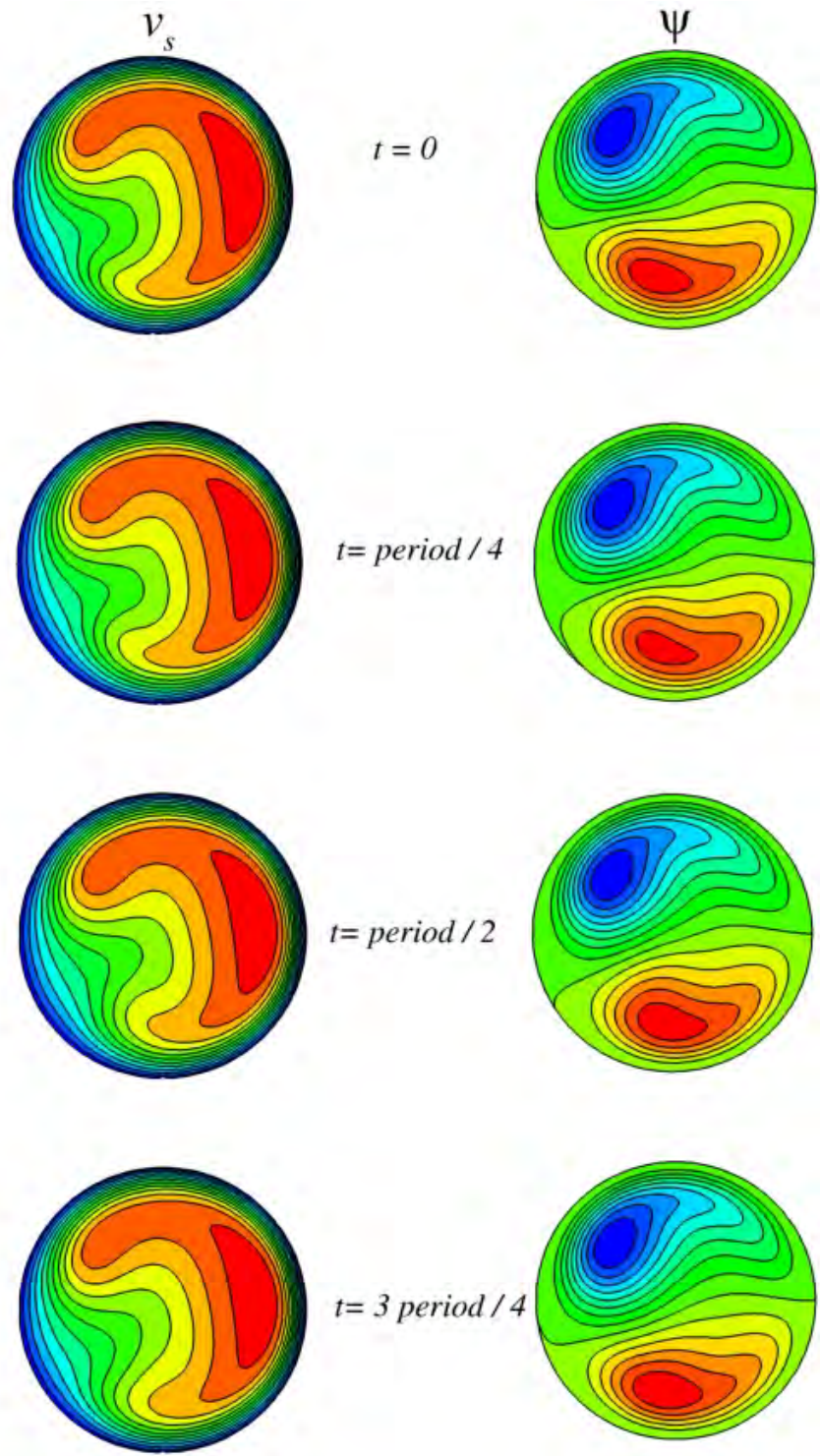

**Figure 13.** Snapshots of a slightly supercritical oscillatory flow at $\varepsilon = 0.05, \lambda = 4, Re_{cr} = 837$. The levels are equally spaced between 0 and 1.5 for $v_s$ and between -0.040 and 0.025 for $\psi$. Animation files: Flow_e=0p05_l=4.avi, Flow_e=0p05_l=4.avi, Flow_e=0p1_l=3.5.avi .

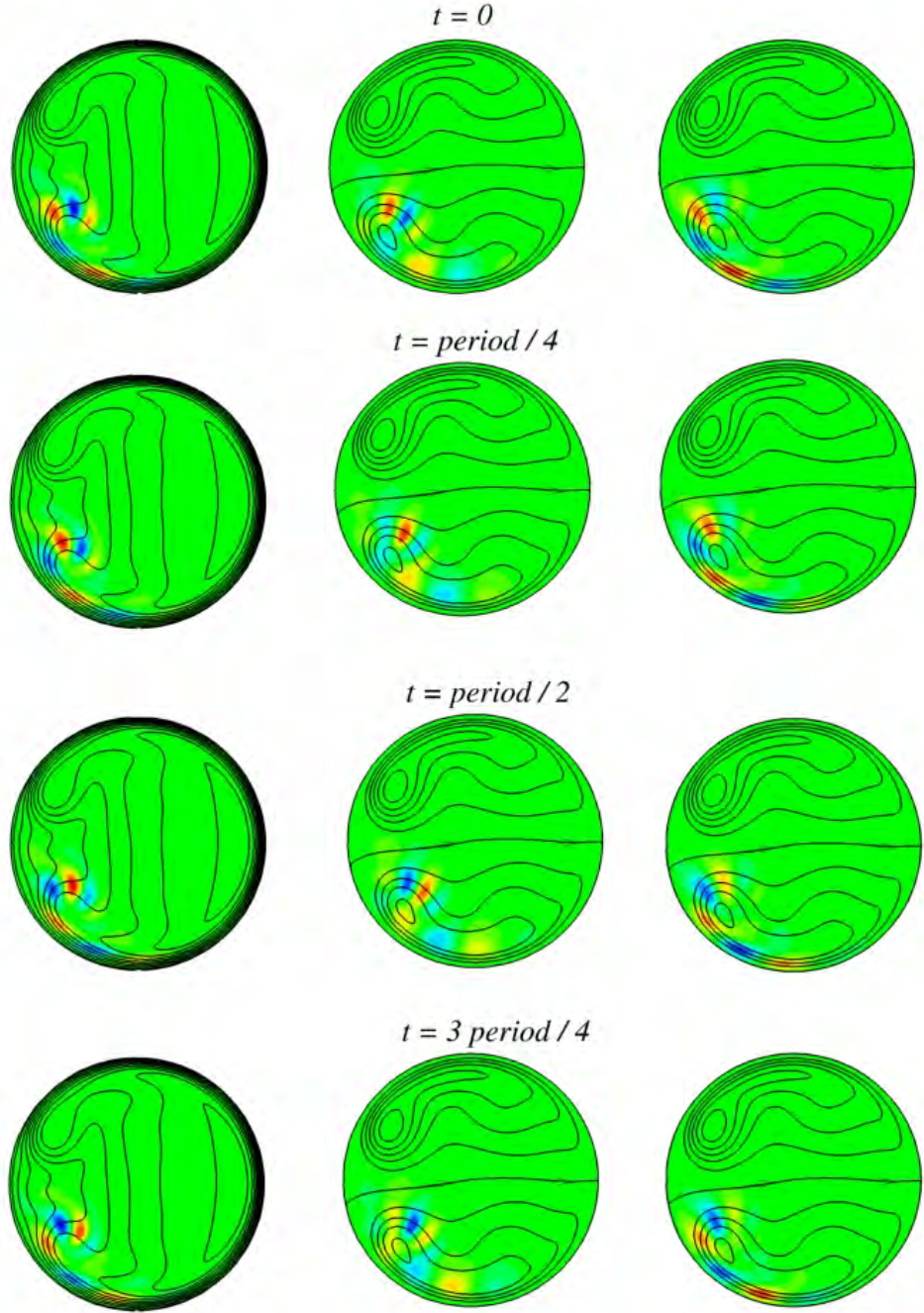

**Figure 14.** Oscillations of the most unstable perturbation at $\varepsilon = 0.2, \lambda = 1.5, Re_{cr} = 1904$ (mode 4). Left frames – perturbation of the centerline velocity (color) superimposed with isolines of the base flow centerline velocity (lines); center and right frames, respectively, show perturbations of $v_r$ and $v_\xi$ (color) superimposed with the pseudo – streamlines of base flow (lines). All the levels are equally spaced between the minimal and maximal values. Perturbation: $max|\tilde{v}_s| = 0.0442, max|\tilde{v}_r| = 0.0152, max|\tilde{v}_\xi| = 0.0269$. Base flow: $max|v_s| = 1.463, \psi_{min} = -0.0455, \psi_{max} = 0.0340$. Animation files: Perturbation_e=0p2_l=1p5.avi, Perturbation_e=0p1_l=1p5.avi.

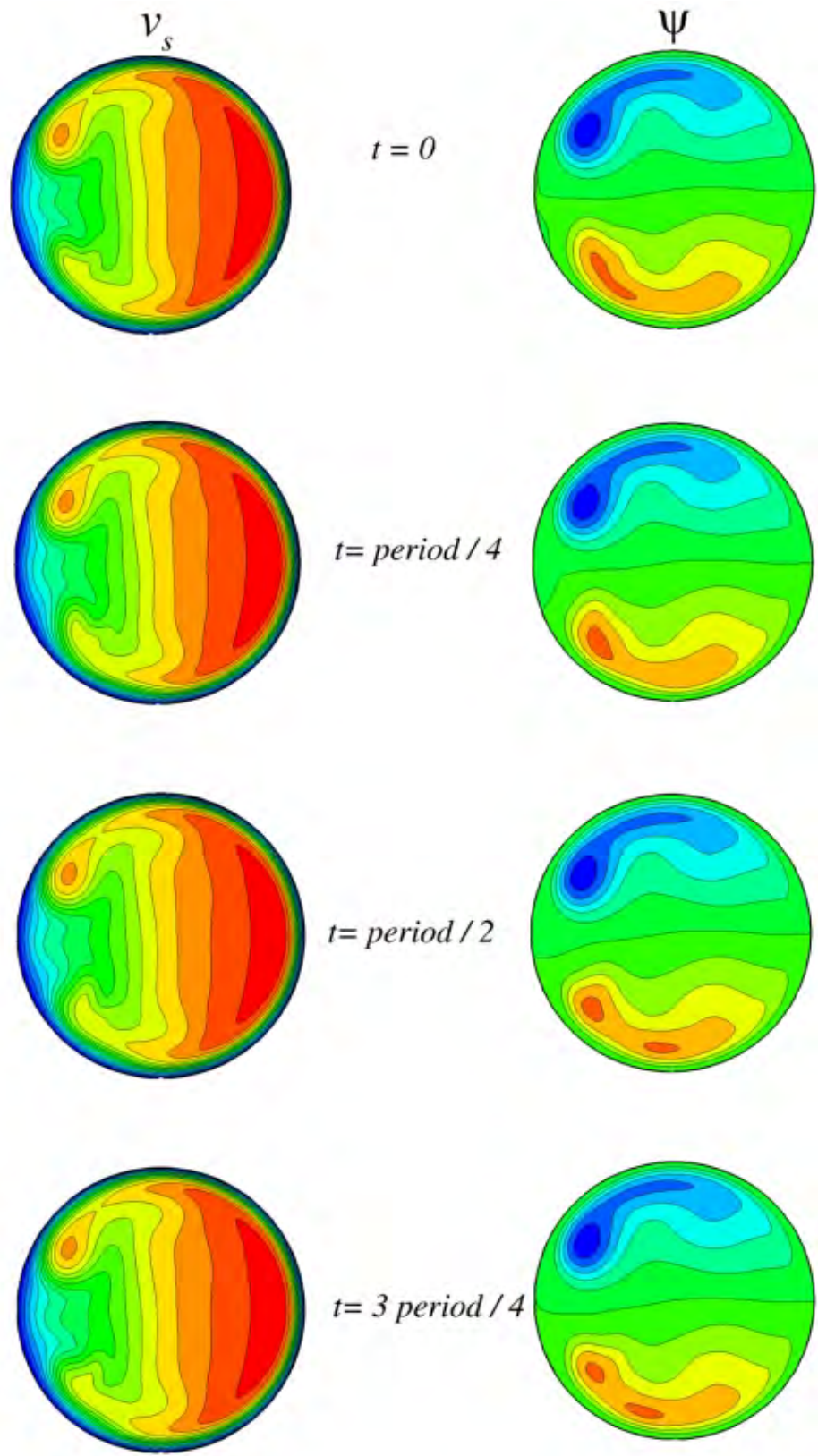

**Figure 15.** Snapshots of a slightly supercritical oscillatory flow at $\varepsilon = 0.2, \lambda = 1.5, Re_{cr} = 1904$. The levels are equally spaced between 0 and 1.4 for $v_s$ and between -0.04 and 0.04 for $\psi$. Animation files: Flow_e=0p2_l=1.5.avi, Flow_e=0p1_l=1.5.avi .

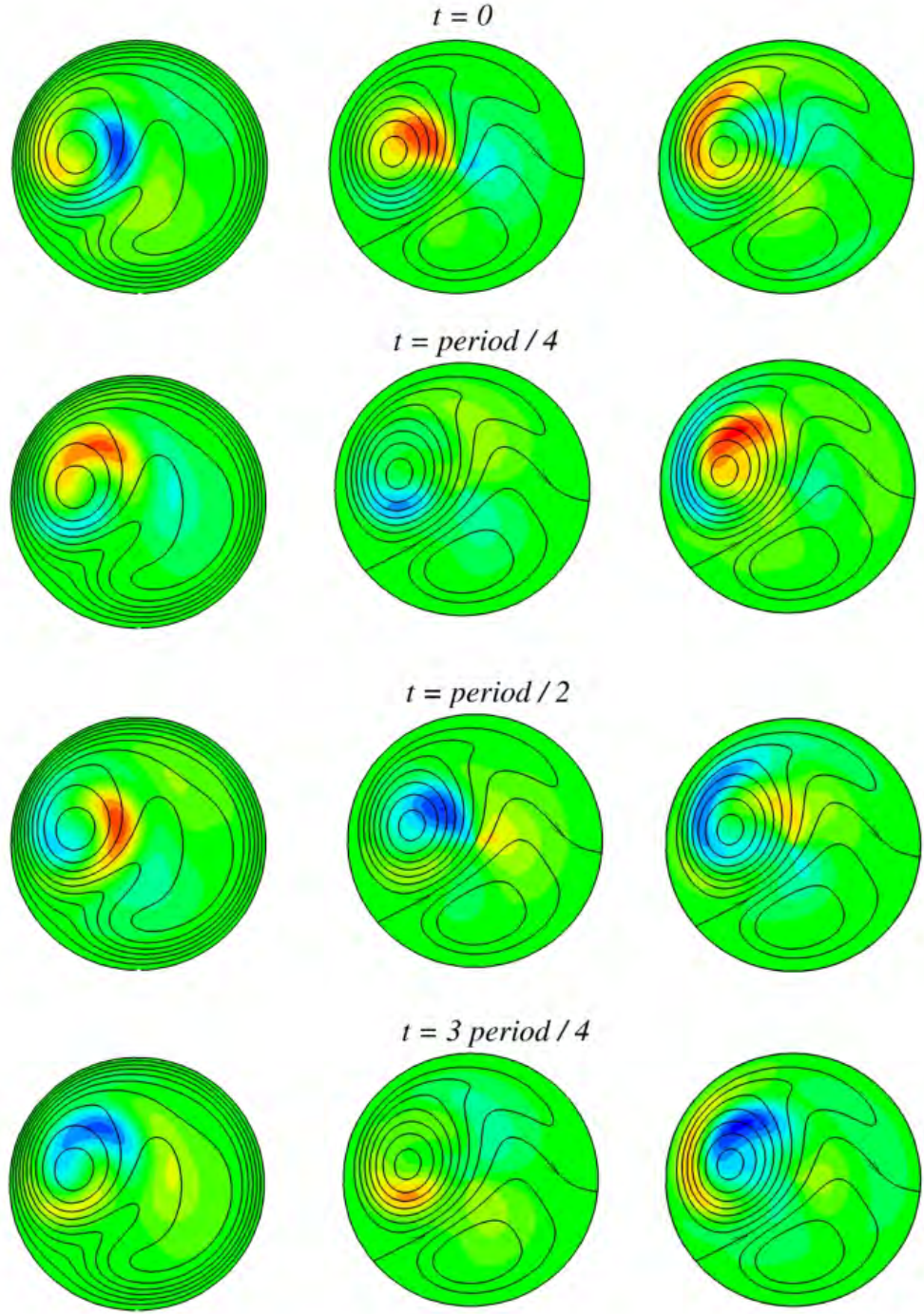

**Figure 16.** Oscillations of the most unstable perturbation at $\varepsilon = 0.2, \lambda = 3, Re_{cr} = 449$ (mode 5). Left frames – perturbation of the centerline velocity (color) superimposed with isolines of the base flow centerline velocity (lines); center and right frames, respectively, show perturbations of $v_r$ and $v_\xi$ (color) superimposed with the pseudo – streamlines of base flow (lines). All the levels are equally spaced between the minimal and maximal values. Perturbation: $max|\tilde{v}_s| = 0.0154, max|\tilde{v}_r| = 0.00984, max|\tilde{v}_\xi| = 0.0125$. Base flow: $max|v_s| = 1.904, \psi_{min} = -0.148, \psi_{max} = 0.0614$. Animation files: Perturbation_e=0p2_l=3.avi, Perturbation_e=0p2_l=4.avi, Perturbation_e=0p2_l=5.avi .

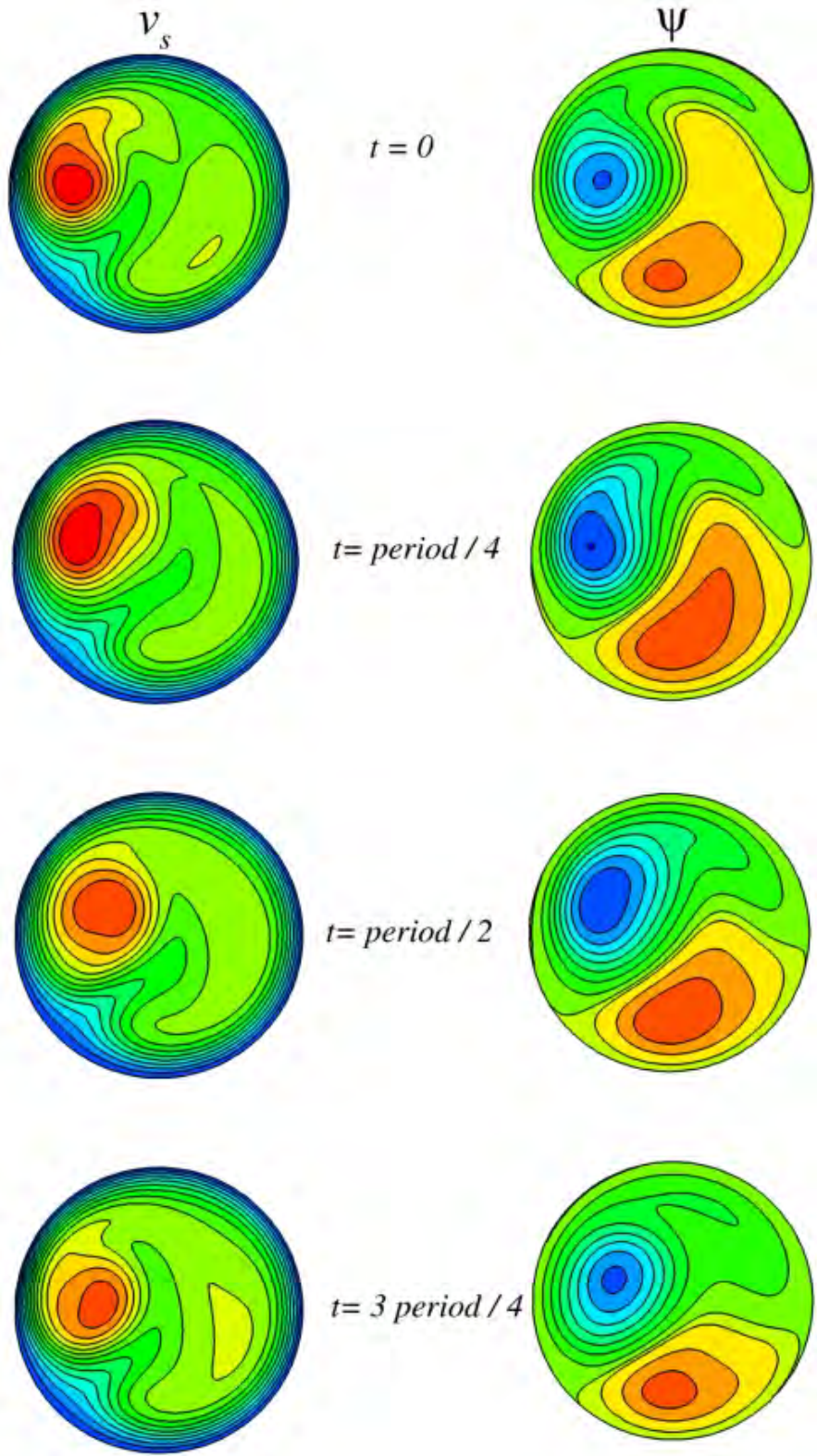

**Figure 17.** Snapshots of a slightly supercritical oscillatory flow at $\varepsilon = 0.2, \lambda = 3, Re_{cr} = 449$. The levels are equally spaced between 0 and 1.9 for $v_s$ and between -0.16 and 0.07 for $\psi$. Animation files: Flow_e=0p2_l=3.avi, Flow_e=0p2_l=4.avi, Flow_e=0p2_l=5.avi.

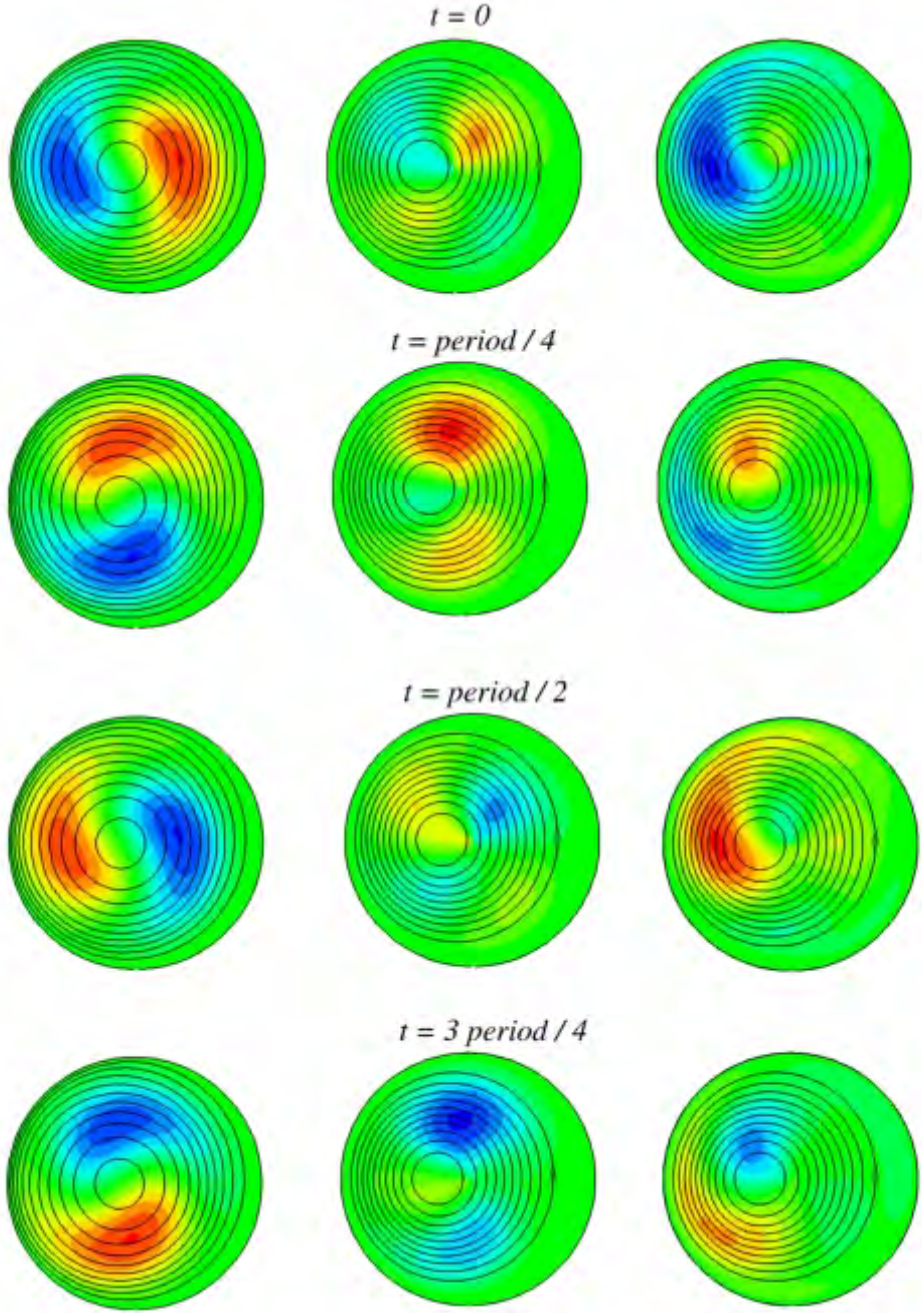

**Figure 18.** Oscillations of the most unstable perturbation at $\varepsilon = 0.5, \lambda = 4, Re_{cr} = 360$ (mode 7). Left frames – perturbation of the centerline velocity (color) superimposed with isolines of the base flow centerline velocity (lines); center and right frames, respectively, show perturbations of $v_r$ and $v_\xi$ (color) superimposed with the pseudo – streamlines of base flow (lines). All the levels are equally spaced between the minimal and maximal values. Perturbation: $max|\tilde{v}_s| = 0.0102, max|\tilde{v}_r| = 0.00267, max|\tilde{v}_\xi| = 0.00362$. Base flow: $max|v_s| = 2.082, \psi_{min} = -1.092, \psi_{max} = 0$. Animation files: Perturbation_e=0p3_l=4.avi, Perturbation_e=0p4_l=4.avi, Perturbation_e=0p5_l=4.avi, Perturbation_e=0p6_l=4.avi.

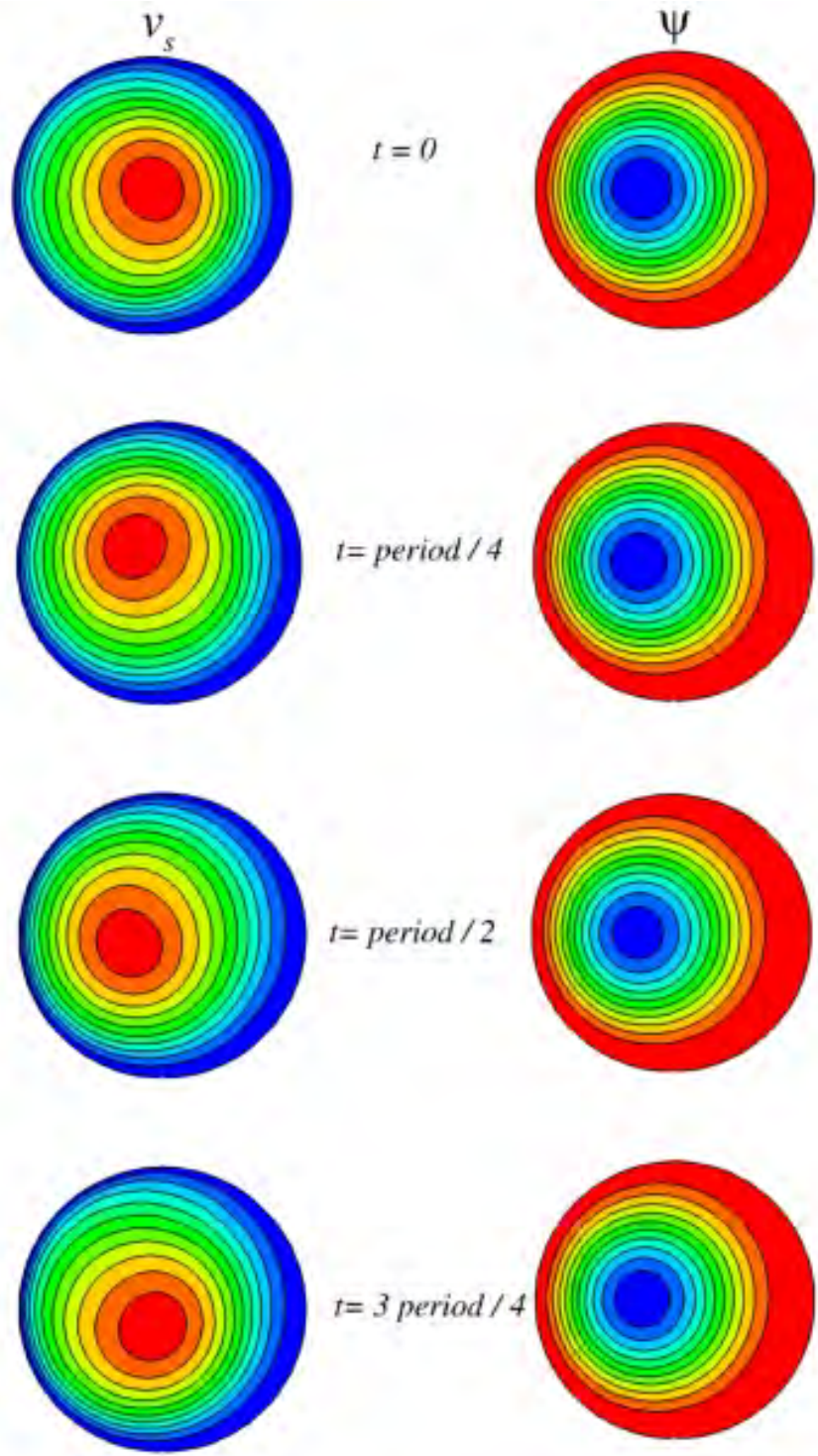

**Figure 19.** Snapshots of a slightly supercritical oscillatory flow at $\varepsilon = 0.5, \lambda = 4, Re_{cr} = 360$. The levels are equally spaced between 0 and 2.0 for $v_s$ and between -1.0 and -0.1 for $\psi$. Animation files: Flow_e=0p3_l=4.avi, Flow_e=0p4_l=4.avi, Flow_e=0p5_l=4.avi, Flow_e=0p6_l=4.avi.

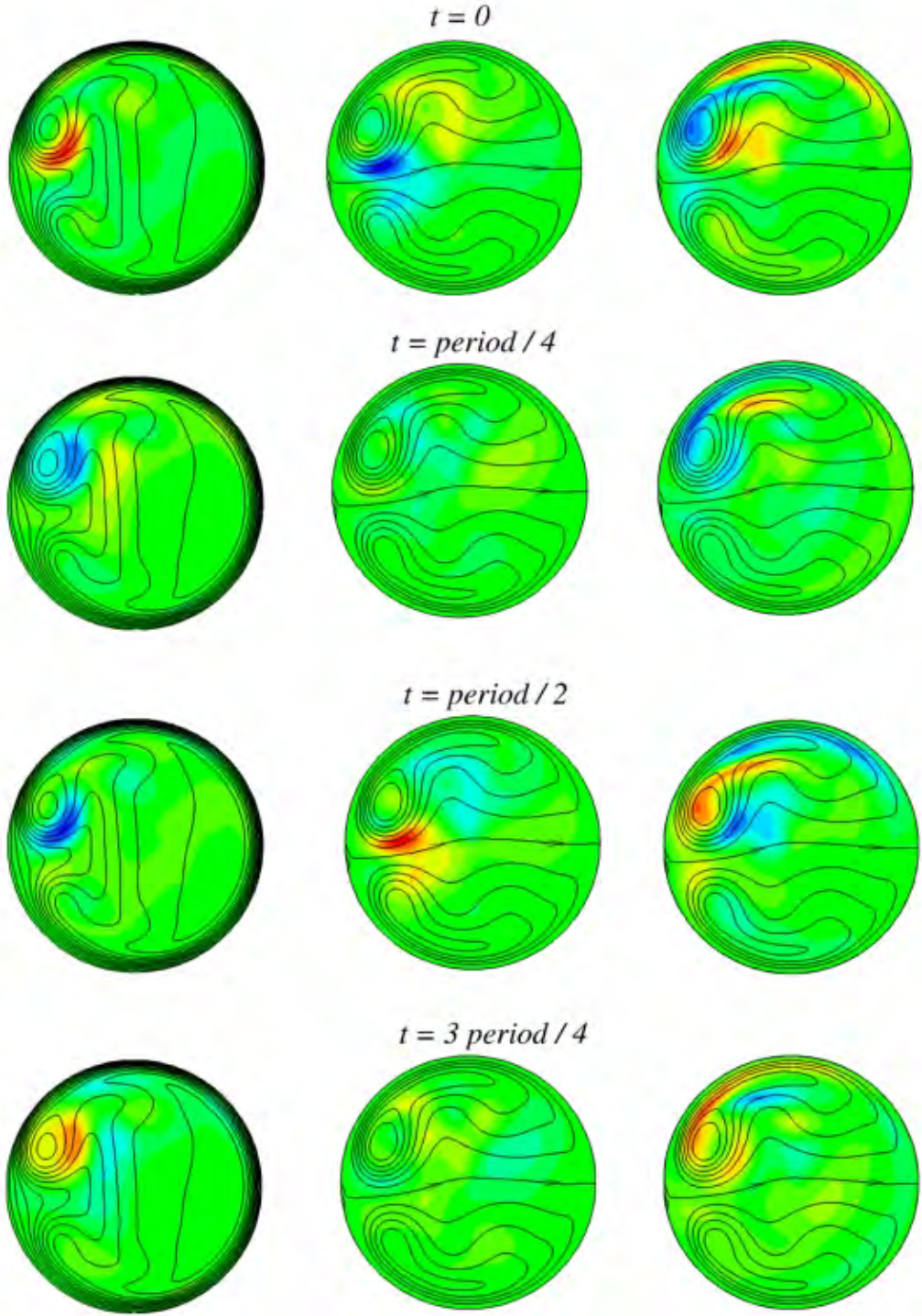

**Figure 20.** Oscillations of the most unstable perturbation at $\varepsilon = 0.3, \lambda = 1.55, Re_{cr} = 1384$ (mode 8). Left frames – perturbation of the centerline velocity (color) superimposed with isolines of the base flow centerline velocity (lines); center and right frames, respectively, show perturbations of $v_r$ and $v_\xi$ (color) superimposed with the pseudo – streamlines of base flow (lines). All the levels are equally spaced between the minimal and maximal values. Perturbation: $max|\tilde{v}_s| = 0.0299, max|\tilde{v}_r| = 0.0124, max|\tilde{v}_\xi| = 0.0136$ Base flow: $max|v_s| = 1.375, \psi_{min} = -0.0711, \psi_{max} = 0.0475$. Animation files: Perturbation_e=0p3_l=1p55.avi.

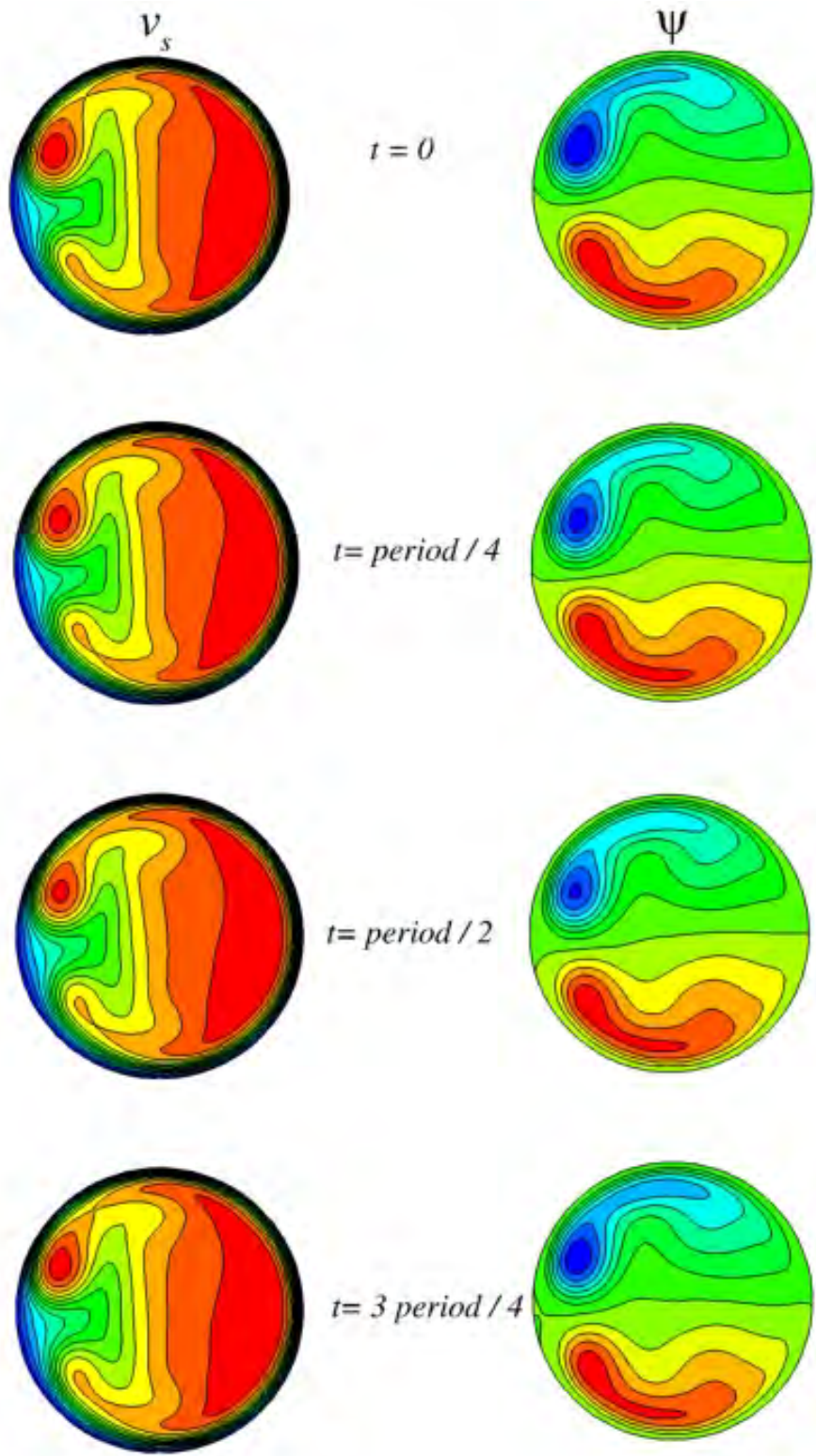

**Figure 21.** Snapshots of a slightly supercritical oscillatory flow at $\varepsilon = 0.3, \lambda = 1.55, Re_{cr} = 1384$. The levels are equally spaced between 0 and 1.3 for $v_s$ and between -0.06 and 0.04 for $\psi$. Animation files: Flow_e=0p3_l=1p55.avi.

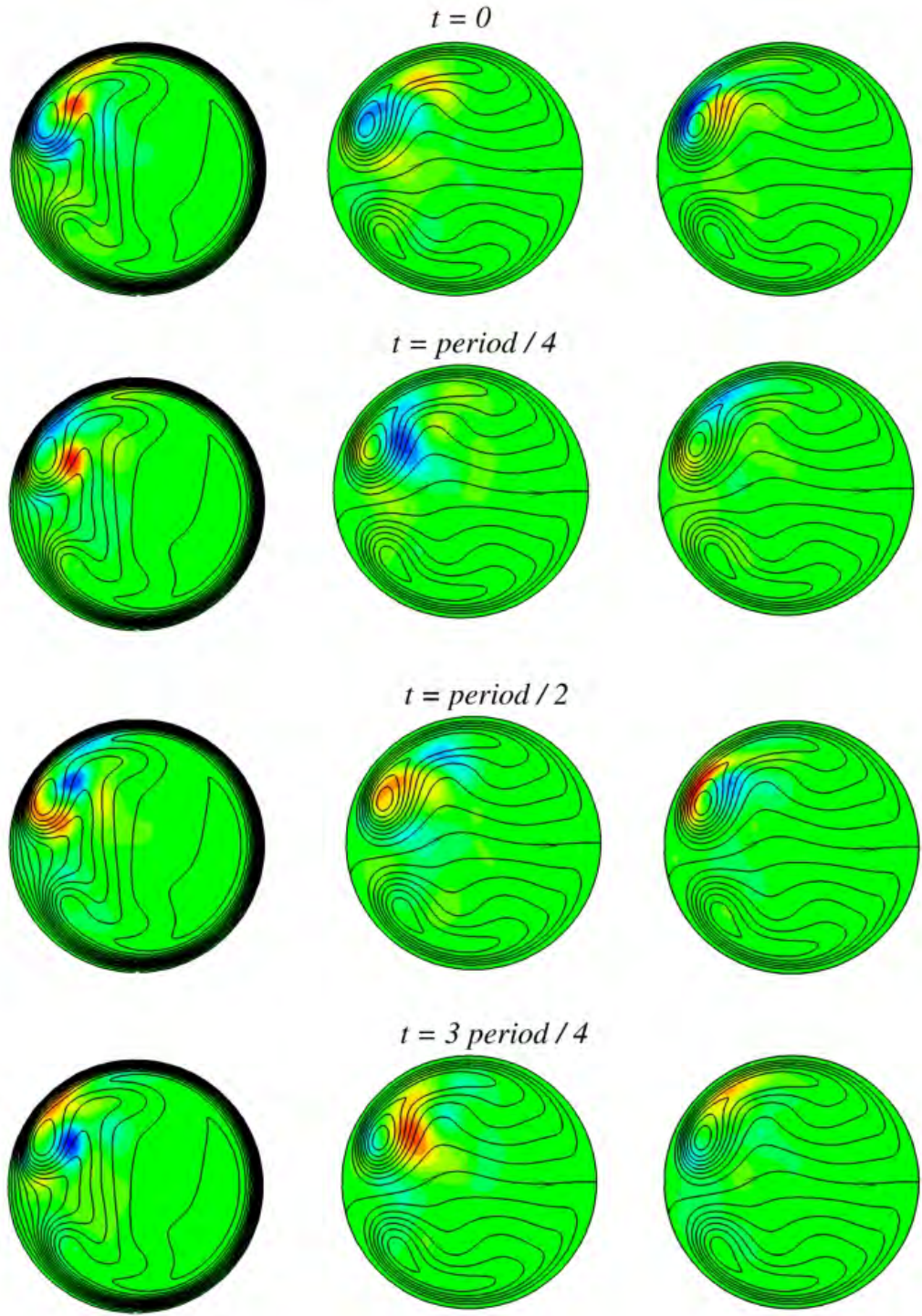

**Figure 22.** Oscillations of the most unstable perturbation at $\varepsilon = 0.5, \lambda = 1, Re_{cr} = 969$ (mode 9). Left frames – perturbation of the centerline velocity (color) superimposed with isolines of the base flow centerline velocity (lines); center and right frames, respectively, show perturbations of $v_r$ and $v_\xi$ (color) superimposed with the pseudo – streamlines of base flow (lines). All the levels are equally spaced between the minimal and maximal values. Perturbation: $max|\tilde{v}_s| = 0.0279, max|\tilde{v}_r| = 0.0161, max|\tilde{v}_\xi| = 0.0368$ Base flow: $max|v_s| = 1.334, \psi_{min} = -0.0861, \psi_{max} = 0.0667$. Animation files: Perturbation_e=0p5_l=1.avi, Perturbation_e=0p4_l=1.avi, Perturbation_e=0p6_l=0p6.avi.

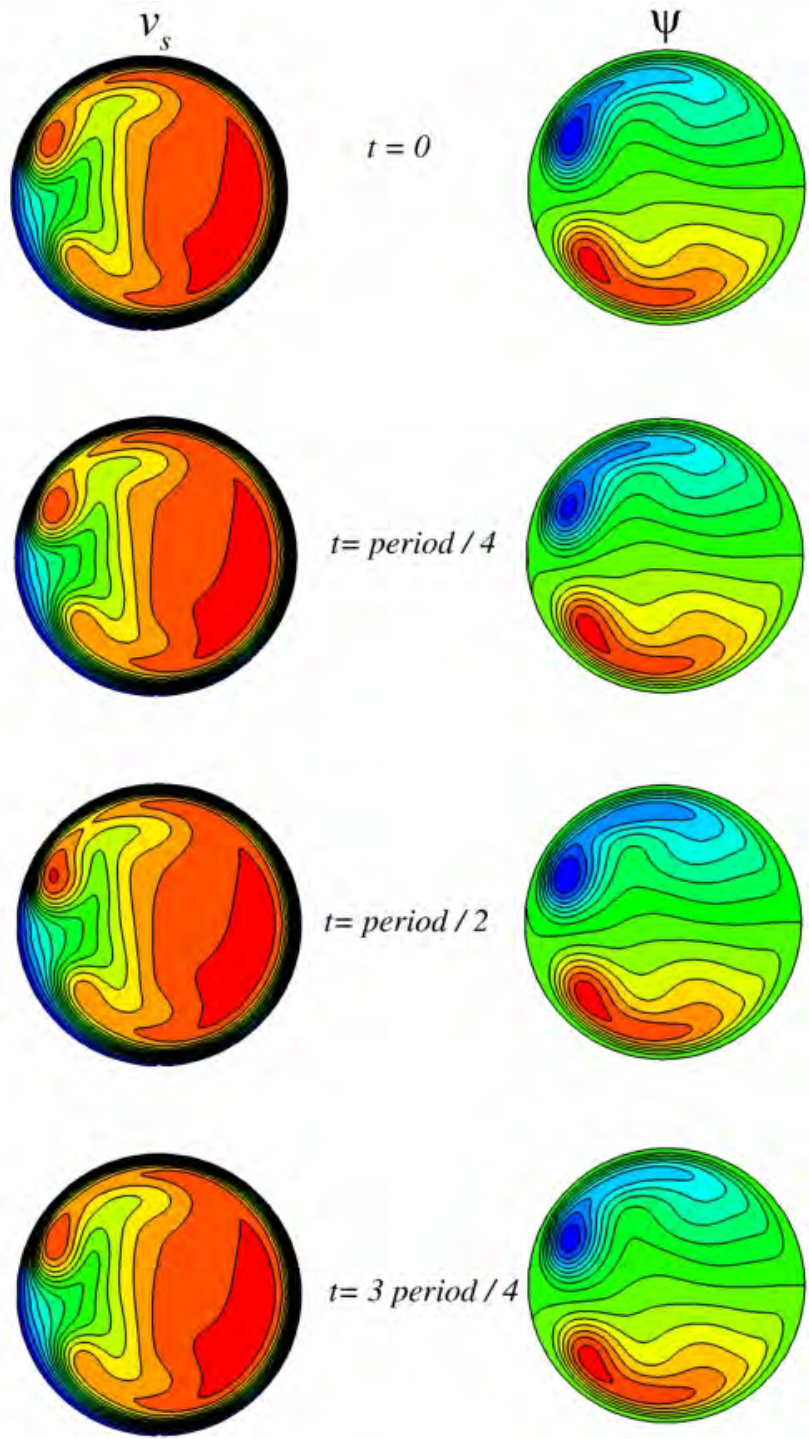

**Figure 23.** Snapshots of a slightly supercritical oscillatory flow at $\varepsilon = 0.5, \lambda = 1, Re_{cr} = 969$ . The levels are equally spaced between 0 and 1.3 for $v_s$ and between -0.08 and 0.06 for $\psi$. Animation files: Flow_e=0p5_l=1p1.avi, : Flow_e=0p4_l=1p1.avi, : Flow_e=0p6_l=0p6.avi.

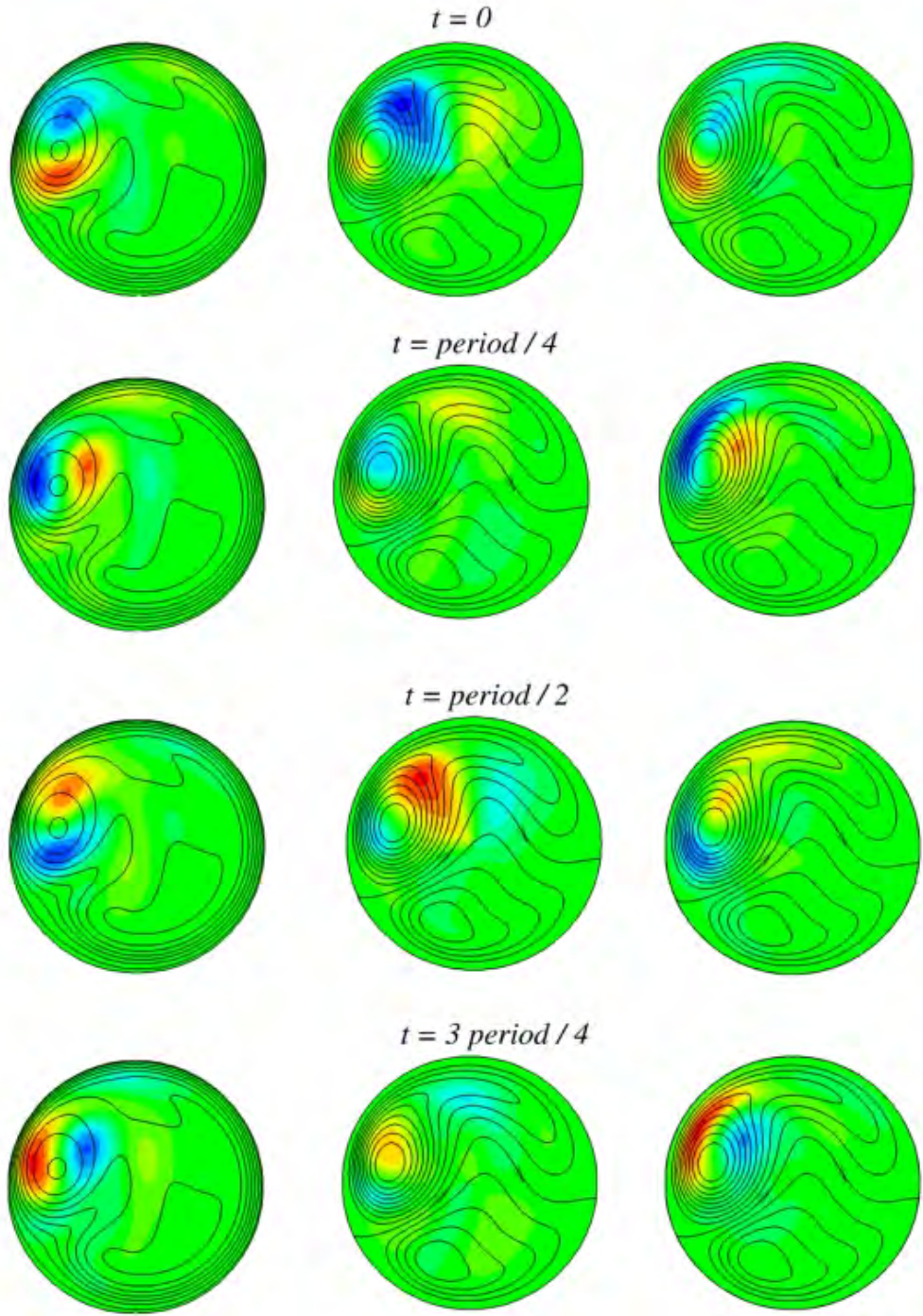

**Figure 24.** Oscillations of the most unstable perturbation at $\varepsilon = 0.5, \lambda = 1.5, Re_{cr} = 391$ (mode 10). Left frames – perturbation of the centerline velocity (color) superimposed with isolines of the base flow centerline velocity (lines); center and right frames, respectively, show perturbations of $v_r$ and $v_\xi$ (color) superimposed with the pseudo – streamlines of base flow (lines). All the levels are equally spaced between the minimal and maximal values. Perturbation: $max|\tilde{v}_s| = 0.0189, max|\tilde{v}_r| = 0.0104, max|\tilde{v}_\xi| = 0.0242$ Base flow: $max|v_s| = 1.729, \psi_{min} = -0.200, \psi_{max} = 0.0969$. Animation files: Perturbation_e=0p5_l=1p5.avi.

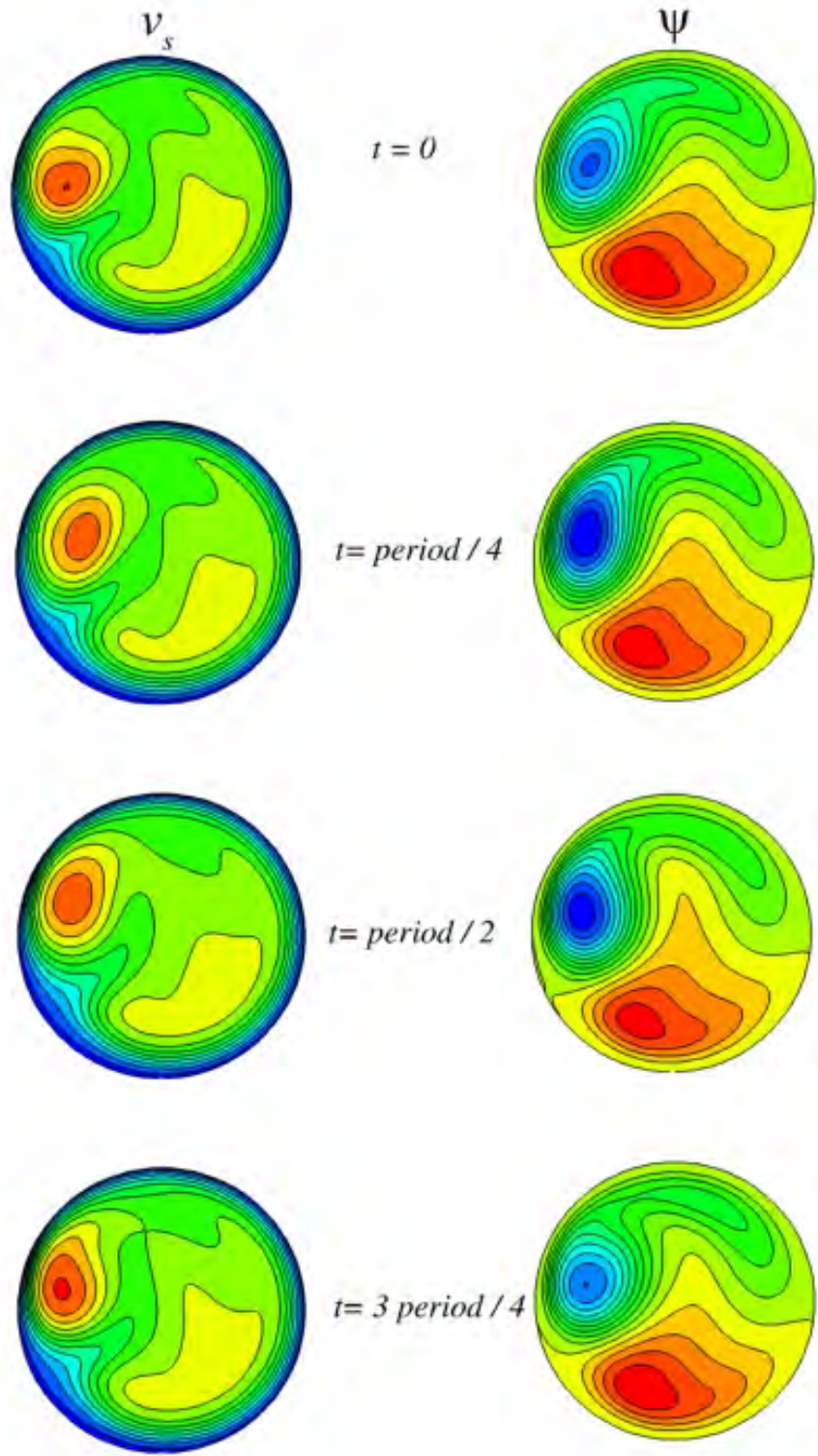

**Figure 25.** Snapshots of a slightly supercritical oscillatory flow at $\varepsilon = 0.5, \lambda = 1.5, Re_{cr} = 391$. The levels are equally spaced between 0 and 1.8 for $v_s$ and between -0.2 and 0.08 for $\psi$. Animation files: Flow_e=0p5_l=1p1.avi.

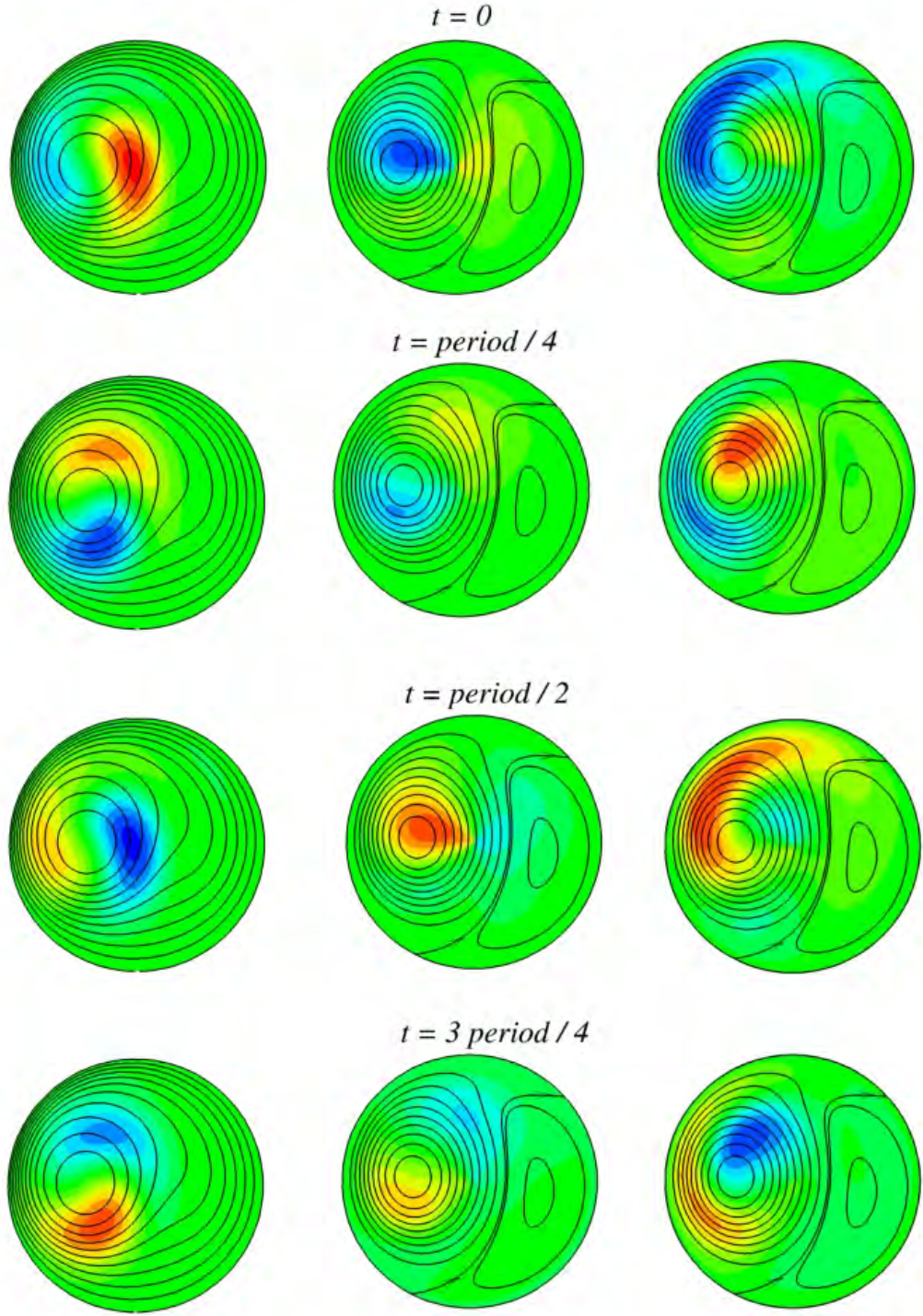

**Figure 26.** Oscillations of the most unstable perturbation at $\varepsilon = 0.4, \lambda = 2.3, Re_{cr} = 171$ (mode 11). Left frames – perturbation of the centerline velocity (color) superimposed with isolines of the base flow centerline velocity (lines); center and right frames, respectively, show perturbations of $v_r$ and $v_\xi$ (color) superimposed with the pseudo – streamlines of base flow (lines). All the levels are equally spaced between the minimal and maximal values. Perturbation: $max|\tilde{v}_s| = 0.0120, max|\tilde{v}_r| = 0.00483, max|\tilde{v}_\xi| = 0.00817$ Base flow: $max|v_s| = 2.184, \psi_{min} = -0.325, \psi_{max} = 0.0475$. Animation files: Perturbation_e=0p4_l=2p3_k=0.avi, Perturbation_e=0p6_l=1p7_k=0.avi.

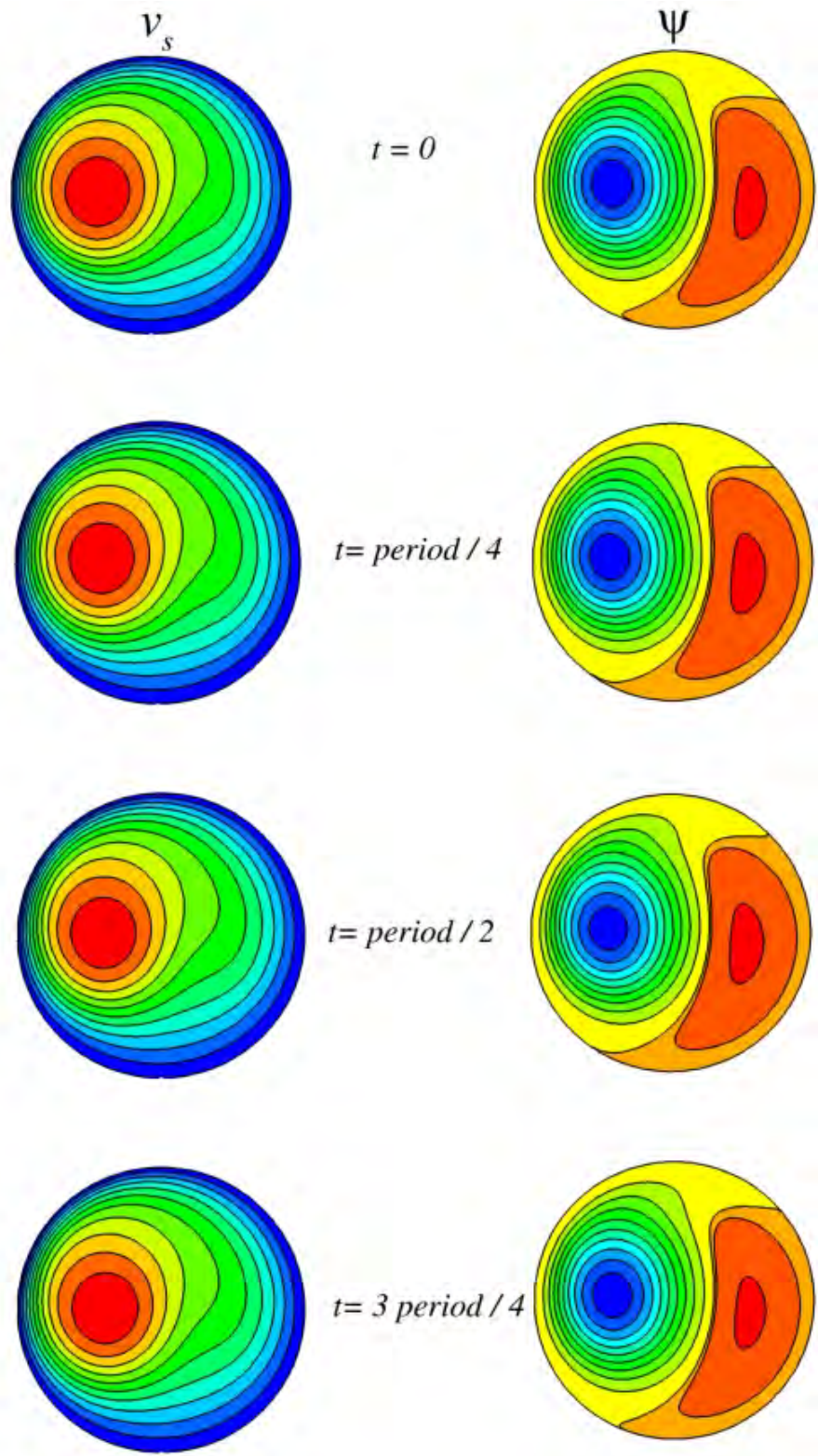

**Figure 27.** Snapshots of a slightly supercritical oscillatory flow at $\varepsilon = 0.4, \lambda = 2.3, Re_{cr} = 171$. The levels are equally spaced between 0 and 2.0 for $v_s$ and between -0.3 and 0.04 for $\psi$. Animation files: Flow_e=0p4_l=2p3_k=0.avi, Flow_e=0p6_l=1p7_k=0.avi.

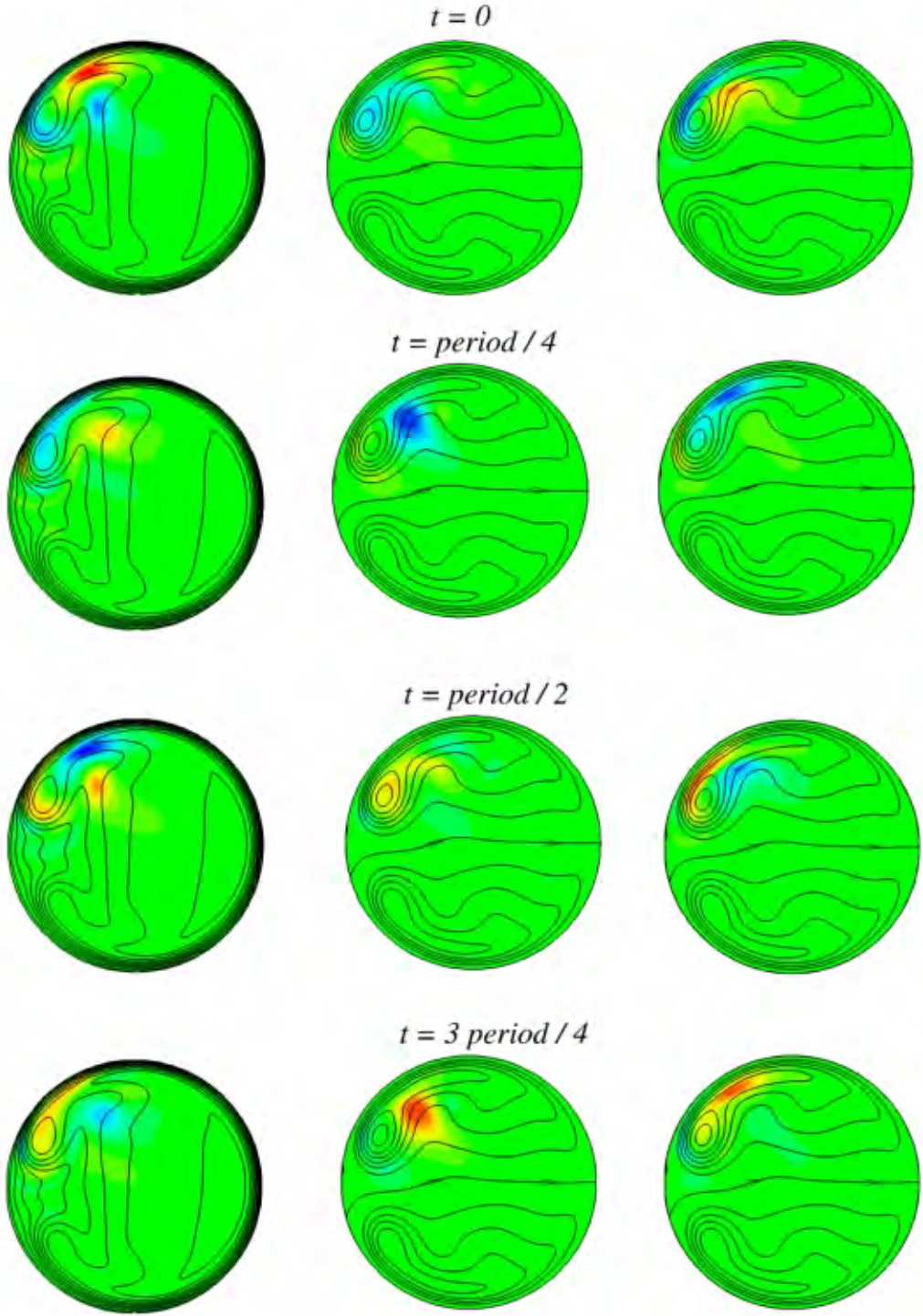

**Figure 28.** Oscillations of the most unstable perturbation at $\varepsilon = 0.4, \lambda = 1.1, Re_{cr} = 1658$ (mode 12). Left frames – perturbation of the centerline velocity (color) superimposed with isolines of the base flow centerline velocity (lines); center and right frames, respectively, show perturbations of $v_r$ and $v_\xi$ (color) superimposed with the pseudo – streamlines of base flow (lines). All the levels are equally spaced between the minimal and maximal values. Perturbation: $max|\tilde{v}_s| = 0.0267, max|\tilde{v}_r| = 0.0241, \ max|\tilde{v}_\xi| = 0.0394$  Base flow: $max|v_s| = 1.357, \ \psi_{min} = -0.0646, \ \psi_{max} = 0.0491$. Animation files: Perturbation_e=0p4_l=1p1.avi, Perturbation_e=0p6_l=0p5.avi.

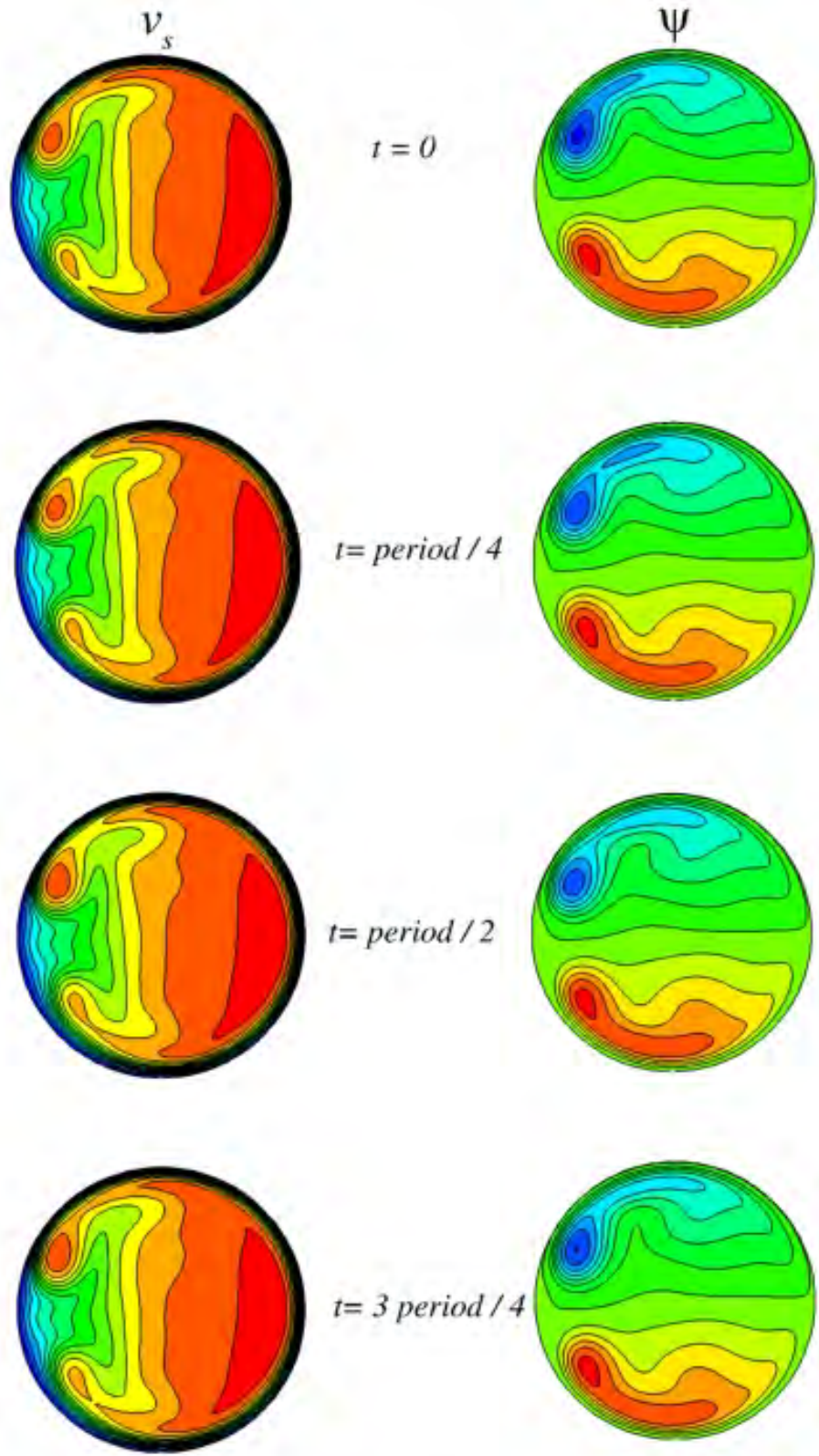

**Figure 29.** Snapshots of a slightly supercritical oscillatory flow at $\varepsilon = 0.4, \lambda = 1.1, Re_{cr} = 1658$. The levels are equally spaced between 0 and 1.3 for $v_s$ and between -0.065 and 0.045 for $\psi$. Animation files: Flow_e=0p4_l=1p1.avi, Flow_e=0p6_l=0p5.avi.

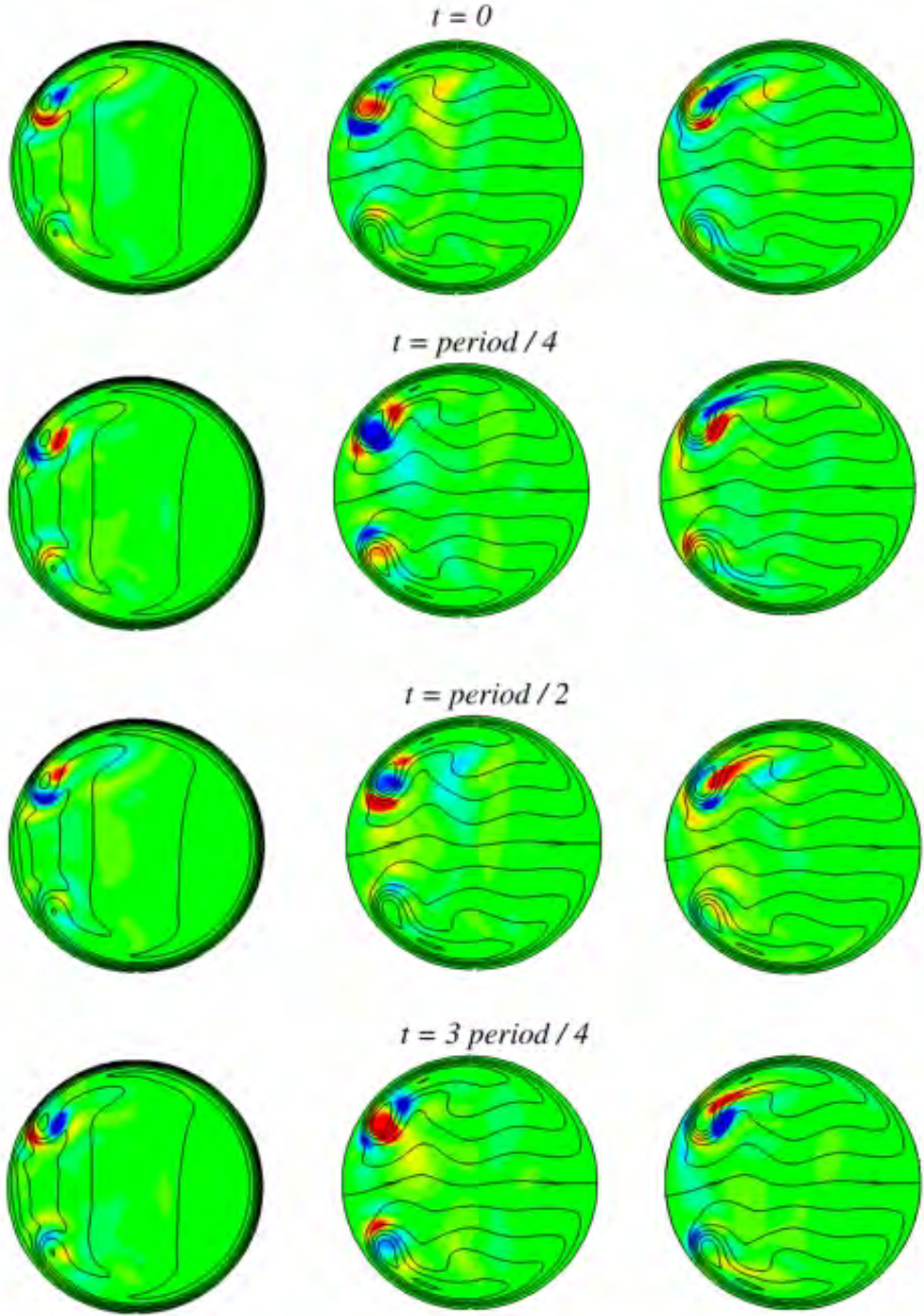

**Figure 30.** Oscillations of the most unstable perturbation at $\varepsilon = 0.6, \lambda = 0.4, Re_{cr} = 2219$ (mode 13). Left frames – perturbation of the centerline velocity (color) superimposed with isolines of the base flow centerline velocity (lines); center and right frames, respectively, show perturbations of $v_r$ and $v_\xi$ (color) superimposed with the pseudo – streamlines of base flow (lines). All the levels are equally spaced between the minimal and maximal values. Perturbation: $max|\tilde{v}_s| = 0.0438, max|\tilde{v}_r| = 0.0251, max|\tilde{v}_\xi| = 0.0617$ Base flow: $max|v_s| = 1.273, \psi_{min} = -0.0491, \psi_{max} = 0.0462$. Animation file: Perturbation_ Perturbation_e=0p6_l=0p4.avi.

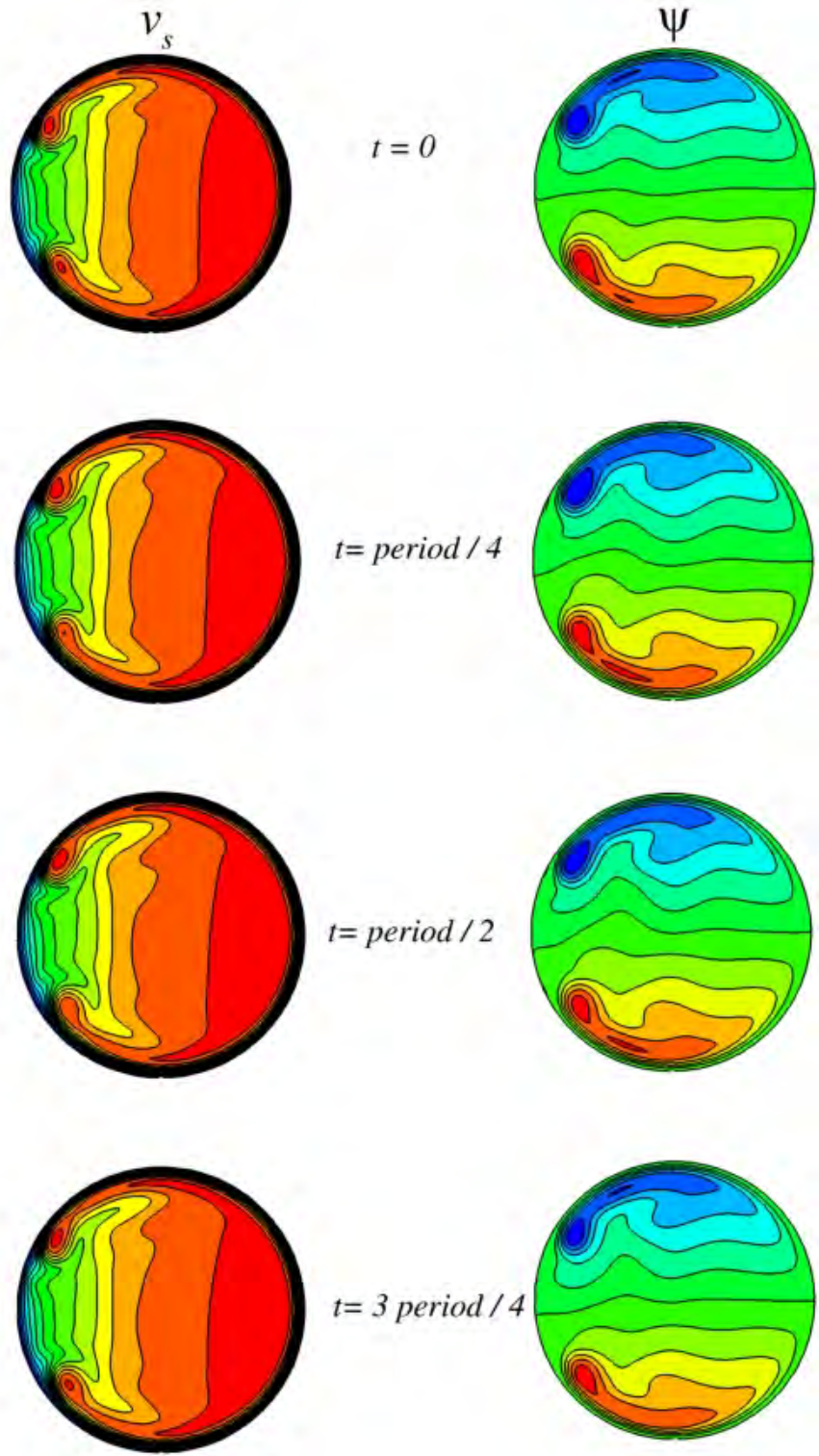

**Figure 31.** Snapshots of a slightly supercritical oscillatory flow at $\varepsilon = 0.6, \lambda = 0.4, Re_{cr} = 2219$. The levels are equally spaced between 0 and 1.2 for $v_s$ and between -0.050 and 0.047 for $\psi$. Animation file: Flow_e=0p6_l=0p4.avi.

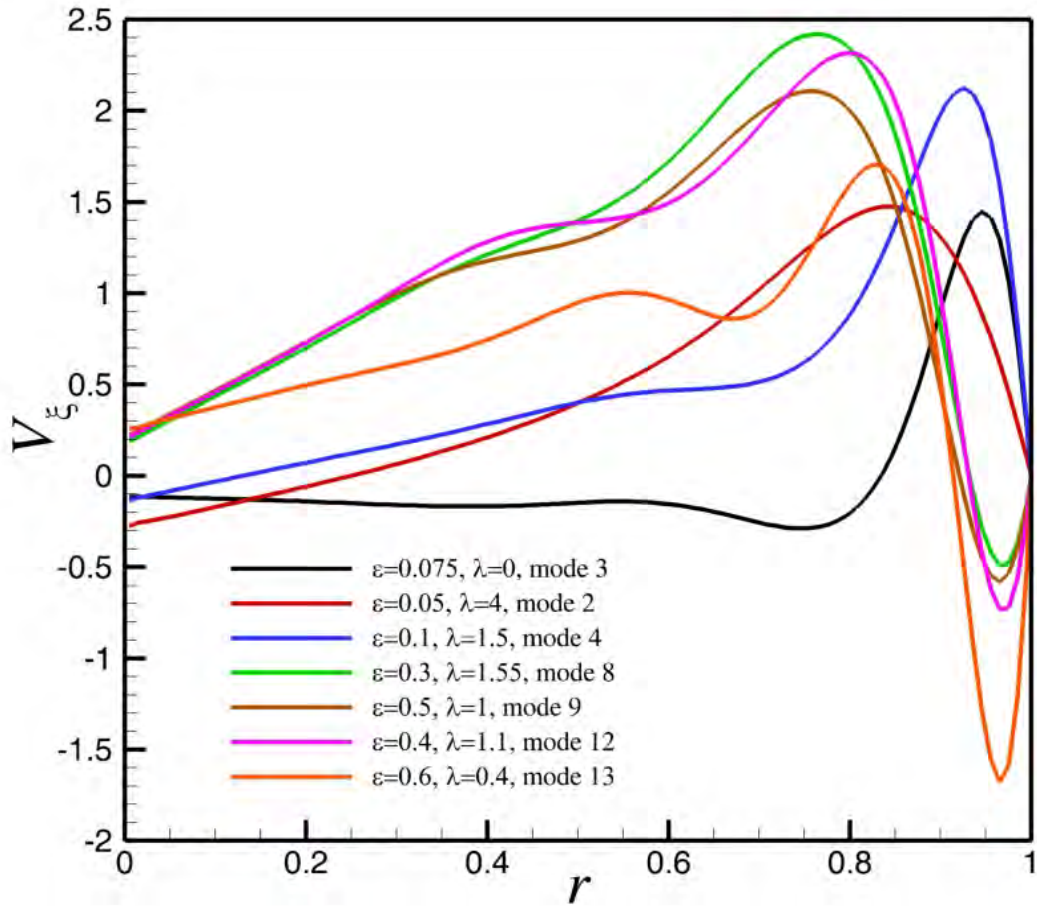

Fig. 32. Radial profiles of the base flow circumferential velocity passing through its maximum located inside the clockwise Dean vortex (modes 2, 3, 4, 9) or the minimum located in the counter clockwise Dean vortex (modes 10, 13, 14).